\documentclass[longauth]{aa}
\usepackage{graphicx}
\usepackage{txfonts}
\usepackage{color}
\usepackage{cellspace}
\newcommand{\printT}{\ensuremath{T_\mathrm{eff} = 760 \pm 20\, \mathrm{K}}}
\newcommand{\printlogg}{\ensuremath{\mathrm{log} \, g=4.26 \pm 0.25}}
\newcommand{\printFEH}{\ensuremath{\mathrm{[Fe/H]} = 1.0 \pm 0.1} dex}
\newcommand{\printf}{\ensuremath{f_{\rm sed} = 1.26^{+0.36}_{-0.29}}}
\newcommand{\printR}{\ensuremath{R = 1.11^{+0.16}_{-0.14}\, \mathrm{R}_\mathrm{J}}}
\newcommand{\printMlogg}{\ensuremath{M_\mathrm{gravity} = 9.1^{+4.9}_{-3.3} \, \mathrm{M}_\mathrm{J}}}
\newcommand{\printLuminosity}{\ensuremath{L/\mathrm{L}_\odot \sim (3.94^{+0.66}_{-0.55}) \times 10^{-6}}}
\newcommand{\printLuminosityRange}{-5.470 and -5.338}

\usepackage{hyperref}

\begin{document}

   \title{Spectral and atmospheric characterization of 51 Eridani b using VLT/SPHERE\thanks{Based on observations made with ESO Telescopes at the Paranal Observatory under programme ID 095.C-0298,  096.C-0241 and 084.C-0739(A)}}

\author{M. Samland\inst{1, 2}
\and
P. Molli{\`e}re\inst{1, 2}
\and
M. Bonnefoy\inst{3, 4}
\and
A.-L. Maire\inst{1}
\and
F. Cantalloube\inst{1, 3, 4, 5}
\and
A. C. Cheetham\inst{6}
\and
D. Mesa\inst{7}
\and
R. Gratton\inst{7}
\and
B. A. Biller\inst{8, 1}
\and
Z. Wahhaj\inst{9}
\and
J. Bouwman\inst{1}
\and
W. Brandner\inst{1}
\and
D. Melnick\inst{10}
\and
J. Carson\inst{10, 1}
\and
M. Janson\inst{11, 1}
\and
T. Henning\inst{1}
\and
D. Homeier\inst{12}
\and
C. Mordasini\inst{13}
\and
M. Langlois\inst{14, 15}
\and
S. P. Quanz\inst{16}
\and
R. van Boekel\inst{1}
\and
A. Zurlo\inst{17, 18, 15}
\and
J. E. Schlieder\inst{19, 1}
\and
H. Avenhaus\inst{17, 18, 16}
\and
A. Boccaletti\inst{20}
\and
M. Bonavita\inst{8, 7}
\and
G. Chauvin\inst{3, 4}
\and
R. Claudi\inst{7}
\and
M. Cudel\inst{3, 4}
\and
S. Desidera\inst{7}
\and
M. Feldt\inst{1}
\and
R. Galicher\inst{20}
\and
T. G. Kopytova\inst{1, 2, 21, 22}
\and
A.-M. Lagrange\inst{3, 4}
\and
H. Le Coroller\inst{15}
\and
D. Mouillet\inst{3, 4}
\and
L. M. Mugnier\inst{5}
\and
C. Perrot\inst{20}
\and
E. Sissa\inst{7}
\and
A. Vigan\inst{15}
}

\institute{Max-Planck-Institut f\"ur Astronomie, K\"onigstuhl 17, 69117 Heidelberg, Germany\\
\email{samland@mpia.de}
\and
International Max Planck Research School for Astronomy and Cosmic Physics at the University of Heidelberg (IMPRS-HD)
\and
Universit\'e Grenoble Alpes, IPAG, 38000 Grenoble, France
\and
CNRS, IPAG, F-38000 Grenoble, France
\and
ONERA, The French Aerospace Lab BP72, 29 avenue de la Division Leclerc, 92322 Ch\^atillon Cedex, France
\and
Geneva Observatory, University of Geneva, Chemin des Mailettes 51, 1290 Versoix, Switzerland
\and
INAF-Osservatorio Astronomico di Padova, Vicolo dell’Osservatorio 5, I-35122 Padova, Italy
\and
Institute for Astronomy, University of Edinburgh, Blackford Hill View, Edinburgh EH9 3HJ, UK
\and
European Southern Observatory, Alonso de Cordova 3107, Casilla 19001 Vitacura, Santiago 19, Chile
\and
Department of Physics \& Astronomy, College of Charleston, 66 George Street, Charleston, SC 29424, USA
\and
Department of Astronomy, Stockholm University, AlbaNova University Center, SE-106 91 Stockholm, Sweden
\and
Zentrum für Astronomie der Universität Heidelberg, Landessternwarte, Königstuhl 12, D-69117 Heidelberg, Germany
\and
Physikalisches Institut, University of Bern, Sidlerstrasse 5, CH-3012 Bern, Switzerland
\and
CRAL,  UMR  5574,  CNRS,  Universit\'e  Lyon  1,  9  avenue  Charles Andr\'e, 69561 Saint Genis Laval Cedex, France
\and
Aix Marseille Univ, CNRS, LAM, Laboratoire d'Astrophysique de Marseille, Marseille, France
\and
Institute for Astronomy, ETH Zurich, Wolfgang-Pauli-Strasse 27, 8093 Zurich, Switzerland
\and
Departamento de Astronom\'ia, Universidad de Chile, Casilla 36-D, Santiago, Chile
\and
N\'ucleo de Astronom\'ia, Facultad de Ingenier\'ia, Universidad Diego Portales, Av. Ejercito 441, Santiago, Chile
\and
NASA Ames Research Center, Moffett Field, CA 94035, USA
\and
LESIA, Observatoire de Paris, PSL Research Univ., CNRS, Univ. Paris Diderot, Sorbonne Paris Cité, UPMC Paris 6, Sorbonne Univ., 5 place Jules Janssen, 92195 Meudon CEDEX, France
\and
School of Earth \& Space Exploration, Arizona State University, Tempe AZ 85287, USA
\and
Ural Federal University, Yekaterinburg 620002, Russia
}

   \date{Received ; accepted }

  \abstract
   {51~Eridani~b is an exoplanet around a young (20 Myr) nearby (29.4 pc) F0-type star, recently discovered by direct imaging. It is one of the closest direct imaging planets in angular and physical separation ($\sim$0.5\arcsec, $\sim$13 au) and is well suited for spectroscopic analysis using integral field spectrographs.}
   {We aim to refine the atmospheric properties of the known giant planet and to further constrain the architecture of the system by searching for additional companions.}
   {We use the extreme adaptive optics instrument SPHERE at the VLT to obtain simultaneous dual-band imaging with IRDIS and integral field spectra with IFS, extending the spectral coverage of the planet to the complete Y- to H-band range and provide additional photometry in the K12-bands (2.11, $2.25\, \muup$m). The object is compared to other known cool and peculiar dwarfs. Furthermore, the posterior probability distributions for parameters of cloudy and clear atmospheric models are explored using MCMC. We verified our methods by determining atmospheric parameters for the two benchmark brown dwarfs Gl~570D and HD~3651B. For probing the innermost region for additional companions, archival VLT-NACO (L') Sparse Aperture Masking data is used.}
   {We present the first spectrophotometric measurements in the Y- and K-bands for the planet and revise its J-band flux to values 40\% fainter than previous measurements. Cloudy models with uniform cloud coverage provide a good match to the data. We derive the temperature, radius, surface gravity, metallicity and cloud sedimentation parameter $f_\mathrm{sed}$. We find that the atmosphere is highly super-solar (\printFEH{}), and the low \printf{} value is indicative of a vertically extended, optically thick cloud cover with small sized particles. The model radius and surface gravity estimates suggest higher planetary masses of \printMlogg{}. The evolutionary model only provides a lower mass limit of $>2\, \mathrm{M}_\mathrm{J}$ (for pure hot-start). The cold-start model cannot explain the planet's luminosity.
   The SPHERE and NACO/SAM detection limits probe the 51~Eri system at solar system scales and exclude brown-dwarf companions more massive than $20$ M$_\mathrm{J}$ beyond separations of $\sim$2.5 au and giant planets more massive than $2$ M$_\mathrm{J}$ beyond 9 au.}
  {}


   \maketitle

%

\section{Introduction}
The number of extrasolar giant planets found with ground-based high-contrast imaging techniques is growing steadily \citep[e.g.,][]{Marois2008b, Marois2010b, Lagrange2010, Rameau2013, 2014ApJ...780L...4B, 2016Sci...353..673W} and the advent of dedicated high-contrast imaging instruments like SPHERE \citep[Spectro-Polarimetric High-contrast Exoplanet REsearch;][]{Beuzit2008} and GPI \citep[Gemini Planet Imager;][]{Macintosh2014} has made it possible to study and characterize these planets and sub-stellar companions in detail with low to mid-resolution spectrometry and/or narrow-band photometry \citep[e.g.,][]{Ingraham2014, Chilcote2015, Apai2016, Maire2016, Vigan2016, Zurlo2016}. At the same time modeling of giant planet and brown dwarf atmospheres has made important progress with the development of cloudy models for colder objects \citep[e.g.,][]{Allard2012, Morley2012, Baudino2015}.\\

\object{51 Eridani b} is the first discovered planet with the GPI-instrument \citep{Macintosh2015} and was characterized using J- and H-band spectra taken with GPI and Keck/NIRC2 photmetry in the L'-band. It  occupies a unique place in parameter space as a young, low-mass ($M<10\, M_{\rm J}$), methane-rich, cold ($\sim$700 K), but seemingly cloudy planet.
This peculiar object is located at an angular separation from its host star ($\rho \sim 0.5$\arcsec) that is suited for spectroscopic characterization within the small field-of-view (FoV) of integral field spectrographs (IFS), but far enough away to achieve good signal-to-noise ratio (SNR) despite its contrast. Given these characteristics, it will become a benchmark object for current and future instruments as well as for the calibration of atmospheric models.\\
Its host star is part of a multiple-system together with an M-dwarf binary \citep{Montet2015} and is located in the well-studied $\beta$ Pictoris moving group \citep{Zuckerman2001}. The age estimates range from 12 to 23 million years (Myr) \citep[e.g.,][]{Simon2011, Binks2014, Mamajek2014, Bell2015}, and we follow the adopted age of the discovery paper as $20 \pm 6$ Myr for all components of the system. A recent dynamical mass estimate of the distant binary M-dwarf companion \object{GJ 3305} predict an older age of the GJ 3305 AB system of $37 \pm 9$ Myr.
An astrometric follow-up paper by \citet{DeRosa2015} confirmed that the planet is co-moving with \object{51 Eri}. The tentative orbital solutions (semimajor axis $a=14^{+7}_{-3}$ au, orbital period $T = 41^{+35}_{-13}$ years, inclination $i = 138^{+15}_{-13}$) suggest that the planet does not share the inclination of the distant M-dwarf companion GJ 3305 \citep{Montet2015}.\\
The host star also has an infrared excess that can be modeled by two components corresponding to a warm belt of debris at 5.5 AU and another colder one at 82 AU \citep{2014ApJS..212...10P, 2014A&A...565A..68R}. As such, the architecture of 51 Eri is reminiscent of our Solar System and of other benchmark systems like HR 8799 and HD 95086.\\
In this work we present new near-infrared (NIR) spectra and photometric data obtained with the SPHERE instrument at the Very Large Telescope (VLT) in Chile, as part of the consortium guaranteed-time exoplanet imaging survey SHINE (SpHere INfrared survey for Exoplanets; Chauvin et al. 2016, in prep).
The SPHERE observations are described in Sect.~\ref{sec:observations} and the data reduction in Sect.~\ref{sec:data_reduction}.
The spectrophotometric analysis of 51~Eri~b is discussed in detail in Sect.~\ref{sec:spectro_phot_analysis}.\\
Finally, we present sensitivity limits to additional closer-in companions in Sect.~\ref{sec:constraints}, extended to the innermost region by archival Sparse Aperture Masking (SAM) data taken with the VLT-NACO instrument in the L'-band, and end with our summary and conclusion in Sect.~\ref{sec:conclusion}. The astrometric analysis of the planet is deferred to a future paper.

\begin{table*}[t]
\caption{\label{tab:observations}Observing log}
\centering
\begin{tabular}{lcccccccccc}
\hline\hline
UT date&Instr. Mode& IRDIS Filter & IFS band& IRDIS DIT\tablefootmark{a}&IFS DIT\tablefootmark{a}& $\mathrm{T}_\mathrm{exp}$\tablefootmark{b}&Field Rot.\tablefootmark{c}& Sr\tablefootmark{d}\\
& & & & (sec, \#) & (sec, \#) &(min)&(deg)& \\
\hline
2015-09-25  &  IRDIFS\_EXT & K12 & YH & $16\times256$ & $16\times256$ & 68.3 & 41.66 & 0.80 -- 0.90 \\
2015-09-26  &  IRDIFS & H & YJ & $4\times918$& $64\times64$& 68.3 & 43.64 & 0.80 -- 0.90 \\
2015-12-26  &  IRDIFS & H23& YJ &$16\times256$ & $32\times128$ & 68.3 (34.7\tablefootmark{b}) & 37.19 & 0.75 -- 0.85\\
2016-01-16  & IRDIFS & H23 & YJ  &$16\times256$ &$64\times64$ & 68.3 & 41.76 & 0.75 -- 0.90 \\
\end{tabular}
\tablefoot{
\tablefoottext{a}{Detector integration time}
\tablefoottext{b}{Exposure time after bad frame removal, about half the frames unusable.}
\tablefoottext{c}{All observation were centered on the meridian passage of the target, with an airmass between 1.08 and 1.10.}
\tablefoottext{d}{Strehl ratio measured at 1.6 $\muup$m as measured by AO system.}
}
\end{table*}

\begin{figure*}
  \centering
  \includegraphics[width=\textwidth]{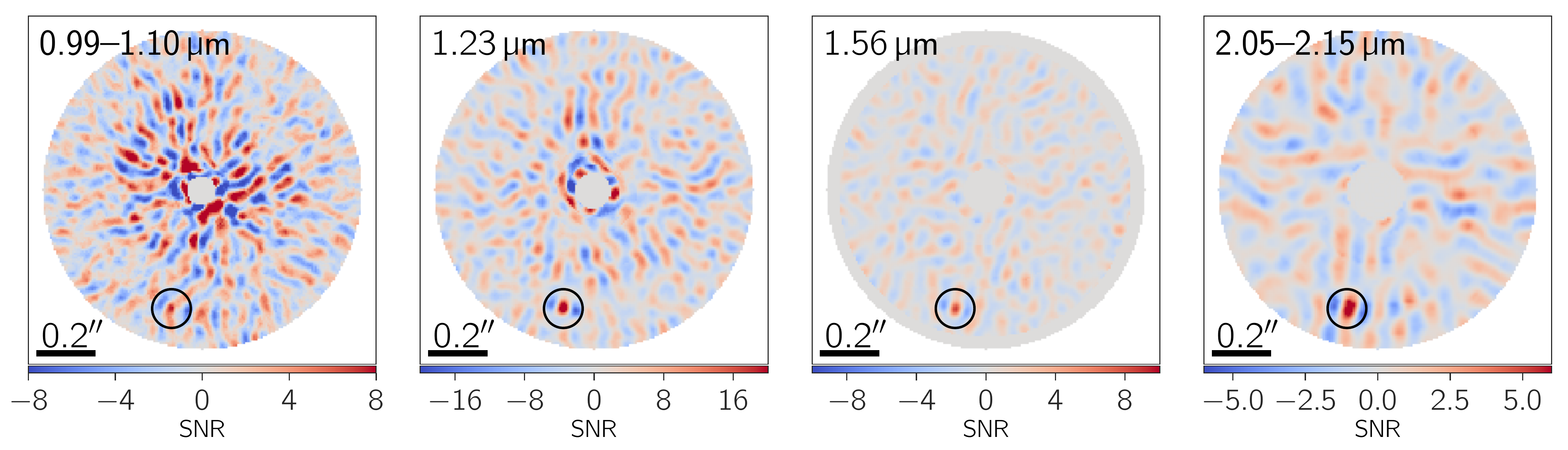}
  \caption{Shown are signal-to-noise maps created by ANDROMEDA for IFS, as well as IRDIS K1. The maps are in order of ascending wavelength. The first two are extracted from the YJ-IFS data and the third from the YH-IFS data. The Y-band image (left) shows the median combined map between 0.99 -- 1.10 $\muup\mathrm{m}$ as signal-to-noise is low, whereas the second and third image, which correspond to the peak in J and H, show single channels. The right panel shows the IRDIS K1 filter. Standard astronomical orientation is used, where up is north and left is east. The black circle marks the position of the planet. Note that the azimuthal negative wings around the planet's signal is the characteristic planet signature that ANDROMEDA is fitting for in ADI data and not undesirable self-subtraction as in the PCA/LOCI approach \citep{Cantaloube2015}.}
  \label{fig:ifs_images}
\end{figure*}

\section{Observations}
\label{sec:observations}
SPHERE \citep{Beuzit2008} is an extreme adaptive optics system \citep[SAXO;][]{Fusco2014} feeding three science instruments: the infrared dual-band imager and spectrograph \citep[IRDIS;][]{Dohlen2008}, an integral field spectrograph \citep[IFS;][]{Claudi2008}, and the visible light imaging polarimeter \citep[ZIMPOL;][]{Thalmann2008}. We observed 51~Eri four times between September 2015 and January 2016 as part of the SPHERE GTO programme using IRDIS in the dual-imaging mode \citep[DBI;][]{Vigan2010} and the IFS operating simultaneously (IRDIFS and IRDIFS\_EXT modes, see Table~\ref{tab:observations}).
The observations were obtained with an apodized pupil Lyot coronagraph \citep{Soummer2005, Boccaletti2008a}, consisting of a focal mask with a diameter of 185 milli-arcsec. The pupil stabilized mode was used close to meridian passage in order to exploit Angular Differential Imaging (ADI) post-processing \citep{Marois2006} with the goal of attenuating residual speckle noise. The usual SPHERE survey observation strategy has been employed: 1) Photometric calibration: imaging of star offset from coronagraph mask to obtain PSF for relative photometric calibration at the beginning and end of the observation sequence; 2) Centering: imaging with star behind the coronagraphic mask with four artificially induced satellite spots using the deformable mirror \citep{Langlois2013} for deriving the star center location directly before and after the science sequence; 3) Coronagraphic sequence; 4) Sky background observation using same configuration as coronagraphic sequence. Finally, North angle offset and pixel scale are determined using astrometric calibrators as part of the SPHERE GTO survey for each run \citep{Maire2016b}. All the other calibration files (e.g., dark, flat, spectral calibration) are obtained during the day following the observation by using the instrument internal calibration hardware. Four IRDIS observations were conducted in three different filter set-ups: once in broadband H (BB\_H), twice in dual-band H23 (H2 $\lambda_{c}=1589\, \rm{nm}$, ${\rm FWHM}=53 \, {\rm nm}$; H3 $\lambda_{c}=1667\, \rm{nm}$, ${\rm FWHM}=56 \, {\rm nm}$), and once in dual-band K12  (K1 $\lambda_{c}=2103\, \rm{nm}$, ${\rm FWHM}=102 \, {\rm nm}$; K2 $\lambda_{c}=2255\, \rm{nm}$, ${\rm FWHM}=109 \, {\rm nm}$, see also Table~\ref{tab:photometry}). The YJ setup (YJ: 0.95 -- 1.35 $\muup\mathrm{m}$, spectral resolution $R \sim 54$) was used three times and the YH mode (YH: 0.95 -- 1.65 $\muup\mathrm{m}$, spectral resolution $R \sim 33$) once.
Observing conditions were variable for the two September data sets, but yielded the best data quality. Both December and January observations were conducted in bad seeing conditions, with strong jet stream that caused saturation near the edge of the coronagraphic mask when using the standard exposure times.

\section{Data reduction and spectrophotometric extraction}
\label{sec:data_reduction}
Basic reduction of both the IRDIS and IFS data (background subtraction, flat fielding, bad pixel removal, centering, spectral calibration for IFS) was performed using the pipeline of the SPHERE data center hosted at OSUG/IPAG in Grenoble using the SPHERE Data Reduction Handling (DRH) pipeline \citep[version 15.0;][]{Pavlov2008}. The calibrated output consists of data cubes for each waveband, re-centered onto a common origin using the satellite spot reference. The unsaturated stellar PSF frames taken before and after the coronagraphic sequence where reduced using the same routines and also corrected for the neutral density filter transmission\footnote{\url{https://www.eso.org/sci/facilities/paranal/instruments/sphere/doc.html}}. The variation of the stellar flux measurement of the host star ($\lesssim$5\% for all used data) is propagated in the uncertainties of the companion flux measurement.
\subsection{IFS data reduction and spectra extraction}
\label{sec:IFS_reduction}
\begin{figure}[t]
  \centering
  \includegraphics[width=\columnwidth]{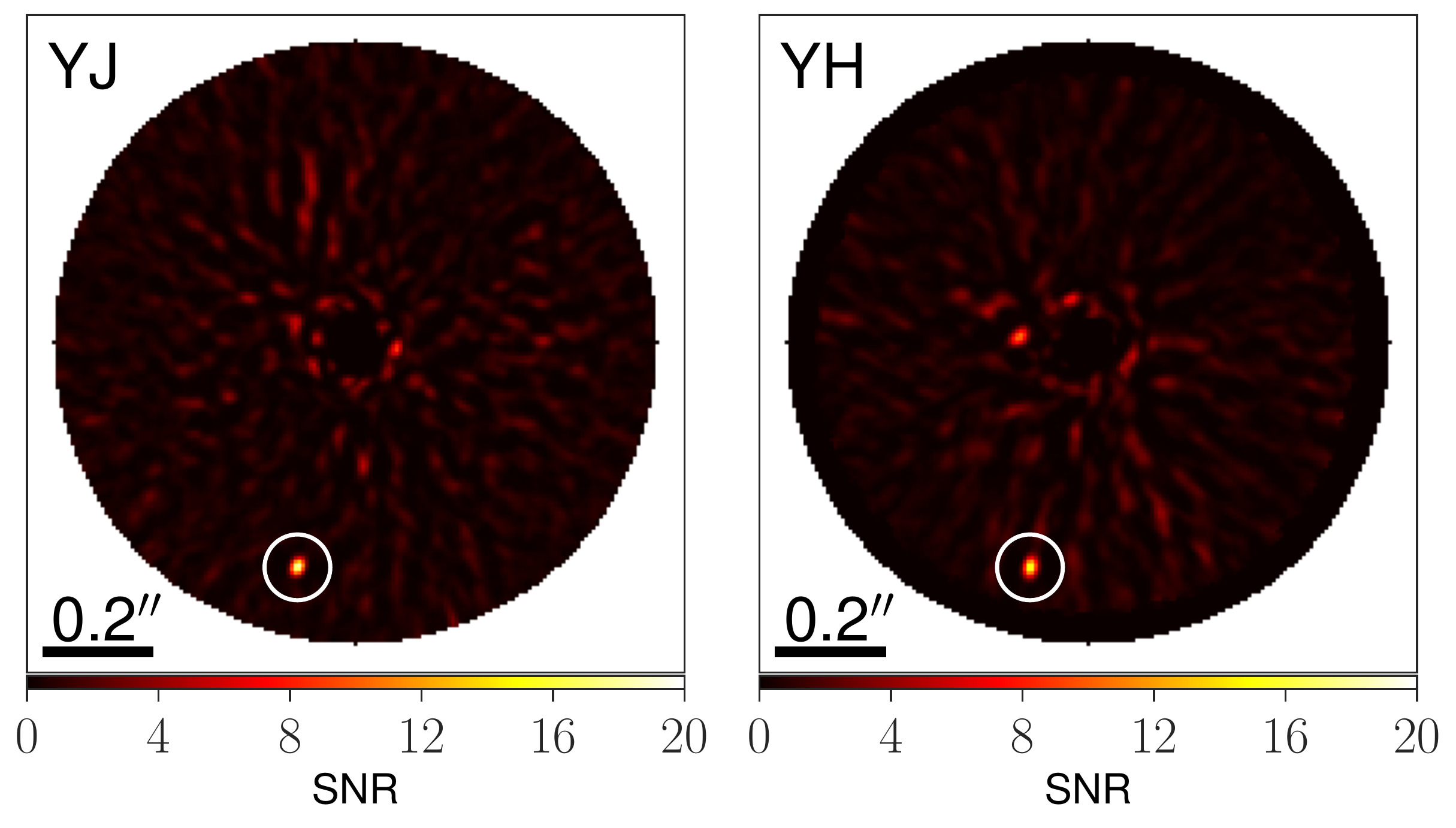}
  \caption{Shown are the detection images made using the method described in \citet{Thiebaut-p-16} with the SNR maps created by ANDROMEDA for IFS-YJ and YH data sets. The left image shows the collapsed image for the YJ-band, the right image YH-band. The circle white marks the position of 51~Eri~b.}
  \label{fig:ifs_collapse}
\end{figure}

In addition to the DRH pipeline, custom IDL routines \citep{Mesa2015} have been used for the basic reduction similar to what has been done in \citet{Maire2016}, with an additional step added to further refine the wavelength calibration by using the shift in satellite spot position.
The analysis of DBI and/or IFS data with aggressive spectral differential imaging \citep[SDI;][]{1999PASP..111..587R} algorithms, such as algorithms that include all other spectral channels as reference to model the speckle pattern, may lead to biases in a planet's signal in a wavelength dependent way which cannot be modeled in a straightforward way \citep{Maire2014}.
In order to avoid biasing the extracted spectrum while still retaining a good signal-to-noise ratio, we opted for a non-aggressive method for the removal of the speckle noise in two steps. We first reduce the data using only ADI post-processing and note the channels which, due to the peak in methane and water absorption, have neither detected flux in the observation, nor expected flux from models. We then go back to the cosmetically reduced data cubes and use these selected channels as reference for a first "classical SDI"-step, i.e. scaled to the same $\lambda / D$ and mean flux for each channel respectively and subtract them from all other channels. For the YJ spectrum, we use the channels between the Y and J-band (1.11 -- 1.17 $\muup\mathrm{m}$) as reference. For the YH spectrum, we use the 1.14 $\muup\mathrm{m}$-channel to reduce all shorter wavelengths (Y-band) and the 1.41 $\muup\mathrm{m}$-channel for the rest of the spectrum (J- and H-band). Note that because the YH spectrum spans a big wavelength range, we use two different reference channels depending on the wavelength, to ensure that the effect of chromatic aberration on the speckle subtraction is minimized.
These SDI pre-processed data cubes with attenuated speckle noise are then used as input for the following ADI reduction using various algorithms. The Specal pipeline (R. Galicher, priv. comm.) developed for the SHINE survey, was used as a first-quick look reduction. For the spectral extraction we test three different reduction approaches: PCA \citep[][Specal implementation used]{Soummer2012, 2012MNRAS.427..948A}, TLOCI \citep[][Specal implementation used]{Marois2014}, and ANDROMEDA \citep{Cantaloube2015}. We chose to focus our analysis on the spectra extracted with the ANDROMEDA algorithm, used for the first time on SPHERE/IFS data. This algorithm provides robust YJ and YH  spectra, and has a number of advantages compared to other reduction methods. In ANDROMEDA the signal is explicitly modeled, therefore no post-processing is necessary to extract an unbiased planetary signal, SNR, and detection limits, i.e. no self-subtraction correction by injection of artificial signals \citep[see e.g.,][]{Lagrange2010, Marois2010} is needed. Furthermore, in contrast to other methods, ANDROMEDA has only one tunable parameter $N_\mathrm{smooth}$ (set to 8 pixels) and it only marginally affects the determined noise level at close separations (thus the SNR of a detection) and could affect the astrometry, but not the signal itself. We confirmed that 51 Eri b is located far enough from the center for none of this to be the case. \citet{Cantaloube2015} lists additional parameters, but these are either set directly by the wavelength of the observation, or can be set to default due to the much higher stability of SPHERE compared to NACO. As such the ANDROMEDA reduction is very reproducible in the sense that it is less prone to subjective choices of parameters which influence the data reduction. Figure~\ref{fig:ifs_images} shows the planet at four different wavelengths, at the Y-, J-, and H-peak in the final ANDROMEDA reduce IFS-YH data cube and in the K1 IRDIS.\\
In addition to the reductions for every spectral channel, we also produce a collapsed "detection image" (see Fig.~\ref{fig:ifs_collapse}) to measure precisely the position of the planet at high SNR and to look for additional companions. For these images, instead of median combining all spectral channels, we follow the method introduced by \citet{Thiebaut-p-16}. We first produce SNR maps for all spectral channels (which are a by-product of the ANDROMEDA algorithm), threshold them above zero, and sum up the squares of the thresholded SNR maps to make a clean, collapsed image. In addition to getting the precise position for a planet, the advantage of this approach is that no assumptions about the exact spectral shape of a potential point source is necessary and it is suitable for visual inspection for further potential candidates.\\
The spectra and photometry extracted using ANDROMEDA and used for the atmospheric characterization are shown in Fig.~\ref{fig:spectrum1} and discussed in Sect.~\ref{sec:spectrum51eri}. Reductions using alternative algorithms are shown and further discussed in Appendix~\ref{sec:alternative_reductions}.

\subsection{Broad and dual-band imaging}
In addition to the basic reduction the cubes are corrected for distortion and for the north angle offset determined from the astrometric calibrations \citep{Maire2016b}. For the unsaturated calibration frames of IRDIS, we used a custom routine that does not interpolate bad pixels in the PSF from the surrounding pixels, but replaces the respective bad pixels with the value obtained by fitting a Moffat function to the PSF in all frames.\\
The cosmetically cleaned and centered data cubes were then used as input for further ADI/SDI post-processing pipelines. We again used three different approaches to reduce the data: PCA, MLOCI \citep{Wahhaj2015}, and ANDROMEDA. All three data reductions were consistent within their respective uncertainties. However, we chose to use ANDROMEDA for the final reduction of the IRDIS photometry presented here to be consistent with the IFS reduction.\\
Additionally we include the broadband L' photometric data point observed with W. M. Keck Observatory’s Near Infrared Camera 2 (NIRC2; L' band, $\lambda_{c}=3780, {\rm nm}$, ${\rm FWHM}=700\, {\rm nm}$)  reported in \citet{Macintosh2015}. The absolute magnitude $\mathrm{L}'=13.82 \pm 0.27$ mag was converted to flux $f_{L'} = (1.82 \pm 0.45) \cdot 10^{-17} \, \mathrm{Wm}^{-2}\muup \mathrm{m}^{-1}$ using the same distance used for the rest of the analysis ($29.4 \pm 0.3 \, \mathrm{pc}$).

\subsection{Conversion of the planet contrasts to physical fluxes}
\begin{figure}
\centering
\includegraphics[width=\columnwidth,clip]{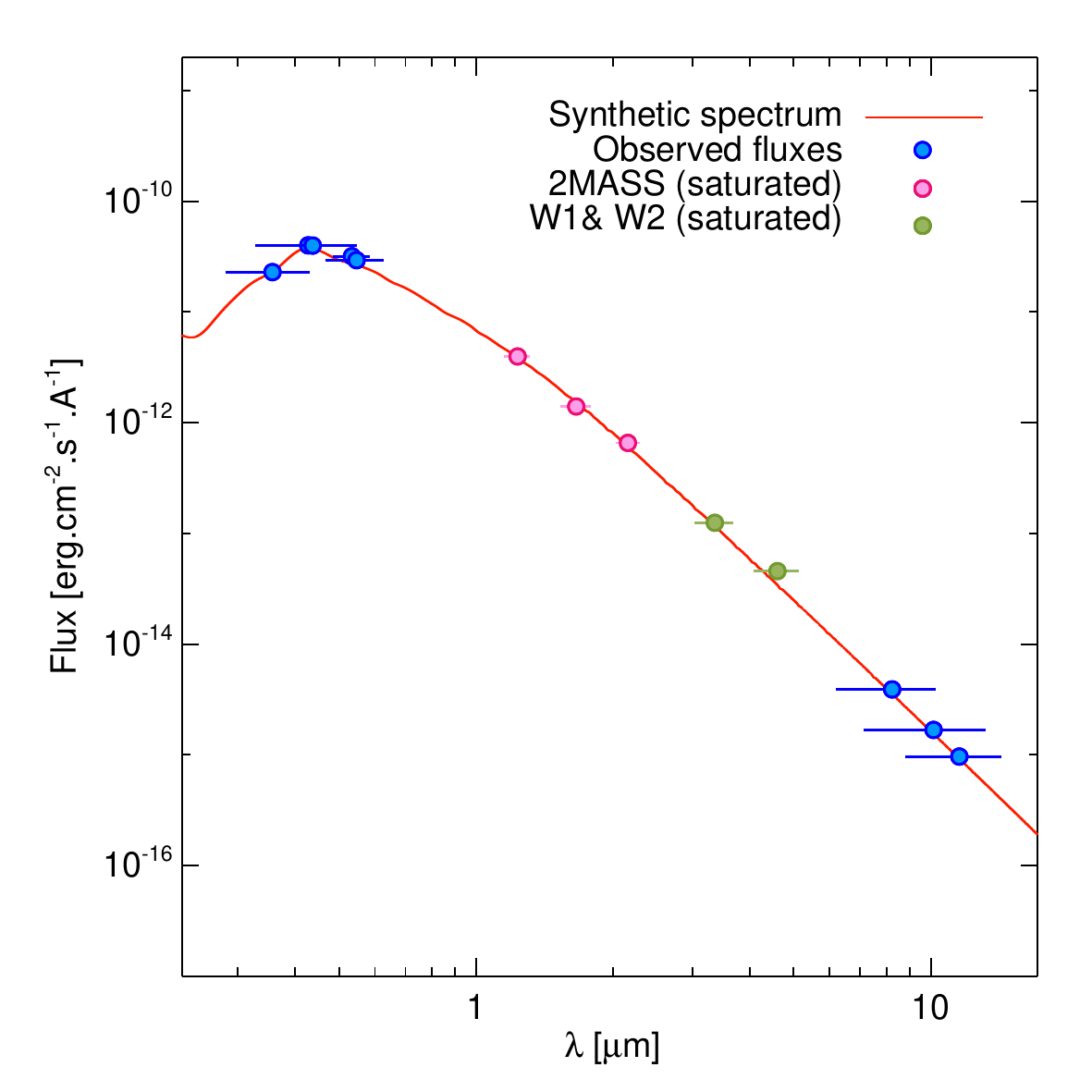}
\caption{BT-NextGen synthetic spectrum of host star 51 Eri, scaled to match SED of optical and mid-infrared photometry. 2MASS $J$, $H$, and $K_s$, as well as W1-W2 were excluded from the fit due to saturation of the star in these bands.}
\label{fig:sed}
\end{figure}

In order to convert the measured star to planet contrast in IFS and IRDIS data to physical fluxes we use a synthetic photometry approach. This can be summarized in three steps:
\begin{enumerate}
\item We build the SED of the star (see Fig.~\ref{fig:sed}) from Tycho $B_T$, $V_T$ \citep{Hoeg1997}, Johnson filter U, V, B \citep{Mermilliod2006}, WISE W3 photometry \citep{Cutri2013}, and IRAS 12 $\muup\mathrm{m}$ photometry \citep{Helou1988}. The 2MASS $J$, $H$, $K_s$ \citep{Cutri2003} as well as W1-W2 photometry could not be used because of saturation of the star's central region. The 2MASS $K_s$-band is not flagged as saturated in the catalog, but can clearly be seen to be saturated in the individual 2MASS $K_s$-band images. On the other hand, W4 had to be excluded due to noticeable infrared excess.
\item We scaled a BT-NextGen model \citep{Allard2012} with $T_\mathrm{eff}=7200$ K, $\mathrm{log}\,\mathrm{g}=4.0$ dex, and $\mathrm{[Fe/H]}= 0.0$ dex, to fit the above mentioned flux values using $\chi^2$-minimization. The chosen model parameters are close to those determined from high-resolution spectra of the star \citep[$T_\mathrm{eff}=7256$ K, $\mathrm{log} \, \mathrm{g}=4.13$ dex, and {[Fe/H]} = 0.0 dex;][]{Prugniel2007}

\item We determine the mean stellar flux in the used SPHERE/IRDIS bandpasses and IFS bins (YJ: 0.95 -- 1.35 $\muup\mathrm{m}$, spectral resolution $R \sim 54$; YH: 0.95 -- 1.65 $\muup\mathrm{m}$, spectral resolution $R \sim 33$), by applying the instrument's spectral response curve, i.e. the normalized wavelength dependent end-to-end transmission including optical elements (e.g., beam-splitters and coronagraph) and filters to the flux-calibrated synthetic spectra. For the IRDIS bands the whole spectral response curve is used. For IFS, because the spectral response is almost flat inside each respective spectral channel, we approximate the spectral response as a Gaussian of a width corresponding to the resolution of the spectrograph in the respective mode.
\end{enumerate}
Our approach differs from that taken in \citet{Macintosh2015}, in that we use a stellar atmosphere model for the flux calibration SED and not a black body spectrum. Comparing the two approaches over the NIR wavelength range of interest, we observe deviations due to spectral features on the order of $\sim3$\%.

\begin{figure*}[!]
\centering
\includegraphics[width=\textwidth,clip]{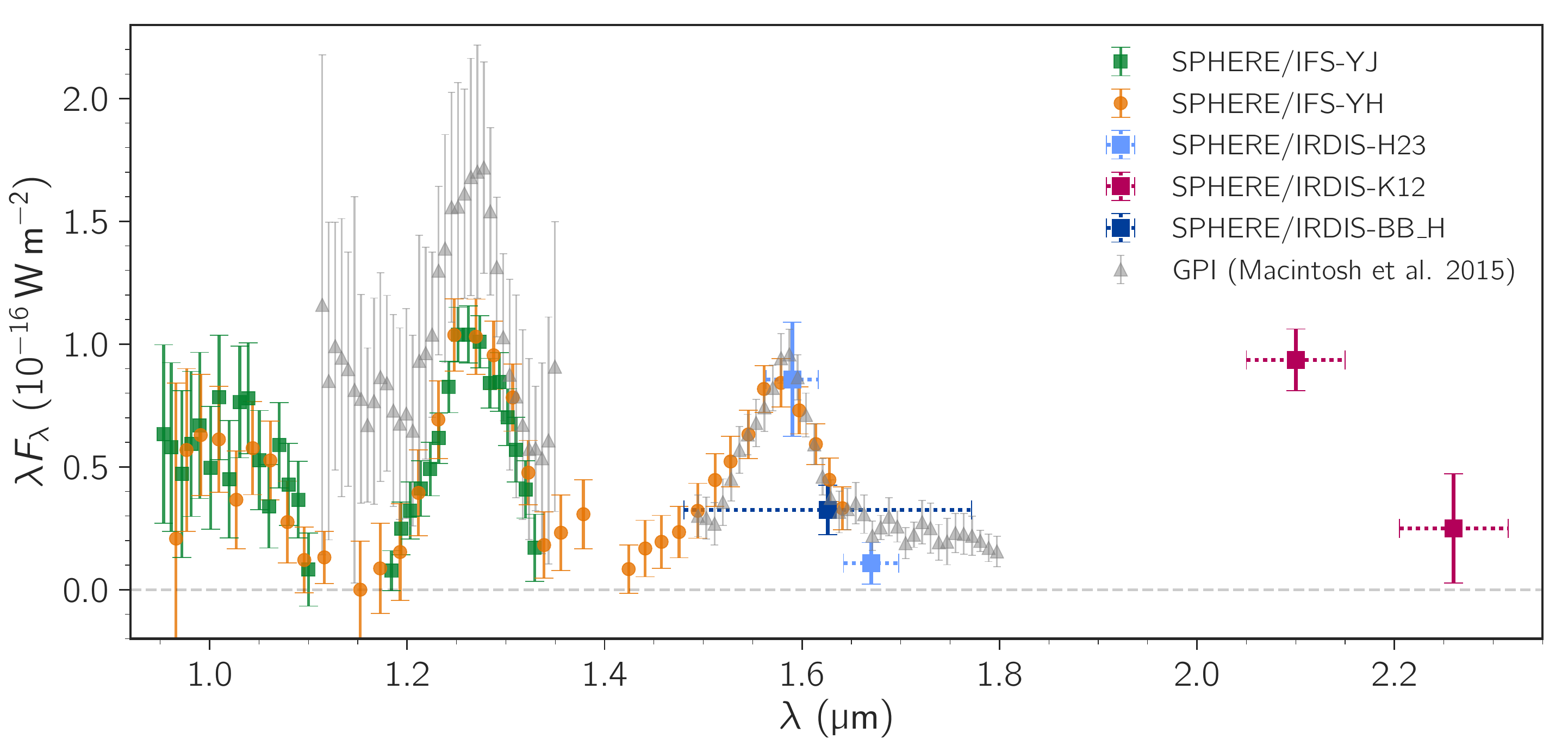}
\caption{SED for 51~Eri~b constructed from our YJ-, YH-IFS spectra and IRDIS photometry in H23, BB\_H, and K12. Channels that have been used as reference for classical SDI have been removed as they are biased. In addition to the SPHERE data we plot the two GPI spectra in J- and H-band respectively. The flux in the J-band is consistent between the two SPHERE spectra, but significantly different compared to GPI. Uncertainties are given as 1$\sigma$ and are assumed to be Gaussian.}
\label{fig:spectrum1}
\end{figure*}

\begin{table*}
\caption{IRDIS photometry}
\centering
\begin{tabular}{lccccccc}
\hline\hline
Filter& $\lambda$ &Width&Contrast& err. Contrast &App. Flux & Err. Flux & Abs. Magnitude\tablefootmark{a}\\
&($\muup$m)&($\muup$m)& & &($\mathrm{Wm}^{-2}\muup \mathrm{m}^{-1}$)&($\mathrm{Wm}^{-2}\muup \mathrm{m}^{-1}$)&\\
\hline
\multicolumn{1}{c}{ADI}\\
\hline
BB\_H &1.626& 0.291&$1.20 \cdot 10^{-6}$& $3.70 \cdot 10^{-7}$ &$2.00 \cdot 10^{-17}$&$6.17 \cdot 10^{-18}$&$17.11 \pm 0.29$\\
H2&1.589&0.048&$3.10\cdot 10^{-6}$&  $8.39 \cdot 10^{-7}$ &$5.39\cdot 10^{-17}$&$1.46\cdot 10^{-17}$&$16.07 \pm0.26$\\
H3\tablefootmark{b}&1.667&0.056&$4.25\cdot 10^{-7}$&  $3.36 \cdot 10^{-7}$ &$6.42 \cdot 10^{-18}$&$5.09 \cdot 10^{-18}$& $>17.59$\\
K1&2.103&0.102&$6.73\cdot 10^{-6}$& $9.02 \cdot 10^{-7}$ &$4.46\cdot 10^{-17}$&$ 5.97\cdot 10^{-18}$&$15.21 \pm 0.14$\\
K2\tablefootmark{b}&2.255&0.109&$2.08 \cdot 10^{-6}$&  $1.64 \cdot 10^{-6}$ &$1.10 \cdot 10^{-17}$&$9.84 \cdot 10^{-18}$& $>15.79$\\
\hline
\multicolumn{1}{c}{SDI+ADI}\\
\hline
H2 - H3&1.589&0.048&$2.32\cdot 10^{-6}$& $3.99 \cdot 10^{-7}$ &$4.04\cdot 10^{-17}$&$6.94\cdot 10^{-18}$&$16.39 \pm 0.17$\\
K1 - K2&2.103&0.102& $5.90\cdot 10^{-6}$& $5.04 \cdot 10^{-7}$ &$3.91\cdot 10^{-17}$&$2.67\cdot 10^{-18}$&$15.35 \pm 0.08$
\end{tabular}
\label{tab:photometry}
\tablefoot{Values under header ADI are obtained using ADI processing only. Values under SDI+ADI are obtained using SDI followed by ADI. Uncertainties are given as 1-$\sigma$. The contrast uncertainties include speckle noise as dominant noise term and the variation in measured host star flux based the two unsaturated stellar images as minor contribution.
\tablefoottext{a}{With distance modulus $\mu = m - M = 2.34$ using Vega magnitude system. Distance uncertainty is negligible in magnitude measurement.}
\tablefootmark{b}{Forced photometry: magnitude 1-$\sigma$ upper-limits obtained by adding the respective flux measurement and uncertainty.}}
\end{table*}

\subsection{Spectrum of 51 Eridani b}
\label{sec:spectrum51eri}
The SED of 51~Eri~b showing all of our observations is presented in Fig.~\ref{fig:spectrum1}. Our IRDIS photometry is summarized in Table~\ref{tab:photometry}, where values are given for ADI and SDI+ADI data reduction. For completeness we also plot the GPI spectra published in the discovery paper \citep{Macintosh2015}. With the SPHERE data, we have extended the spectral coverage of the atmosphere to the Y-band, provide the first photometry in the K-band, and substantially improve the SNR in the J-band. All of these are of tantamount importance for deriving atmospheric parameters and cloud characteristics, as will be discussed in Sect.~\ref{sec:empirical_comparison} and~\ref{sec:atmospheric_modeling}. In further analysis we use both IFS spectra, the four ADI-only narrow-band photometric data points in H- and K-band. Additionally we also use the GPI-H spectrum, as it extends the wavelength coverage of the H-band towards longer wavelengths, as well as the L'-band photometry of \citet{Macintosh2015}. We are not using the broadband H-band observation in our later analysis, because it does not further constrain the spectral shape. The SPHERE YH spectrum is in excellent agreement in the overlapping part with the IRDIS H2 and BB\_H photometry taken on different dates, as well as the previous high S/N H-band spectrum obtained by GPI. The only discrepancy we see is between the J-band flux reported in \citet{Macintosh2015} and our data. The question presenting itself is therefore the origin of the difference observed in the J-band for which there are two possibilities: 1) strong ($\sim$40\%) atmospheric variability in the planet's atmosphere; or 2) systematic offsets in the absolute calibration between the data sets. The J-band is known to be more sensitive to temporal amplitude changes in the atmosphere of L/T-type objects than the H- and K-bands \citep[e.g.,][]{Radigan2012, Biller2015}, but even so, given that we see no significant difference in the H-band, we think it is unlikely that ~40\% variability in the J-band is in agreement with our consistent values for the H-band flux. We therefore believe that the difference in the J-band between our and previous observation is a result of systematics. The reduction for the YH spectrum using different algorithms shows consistent results (Fig.~\ref{fig:alternative_reductions}) giving us confidence in the overall reliability of the data, data reduction, and calibration.\\
Note, while we use the ANDROMEDA method in this paper, \citep{Macintosh2015} uses TLOCI. We also compared different reduction methods in Appendix~\ref{sec:alternative_reductions} and notice a difference in J-band flux between our two data sets depending on the reduction method (e.g., using ANDROMEDA, the flux measured in the two data sets is within 10\%, consistent in their respective uncertainties; using TLOCI it differs by $\sim 40$\%). It is possible that absolute calibration is more difficult for the other algorithms compared here, because additional steps to account for algorithm throughput are necessary.

\section{Spectrophotometric analysis}
\label{sec:spectro_phot_analysis}

\subsection{Empirical comparison to known objects}
\label{sec:empirical_comparison}
Our SPHERE YJ and YH spectra of 51~Eri~b confirm the presence of several deep water and methane absorption bands typical of T-dwarfs from 1.1 to 1.2, 1.3 to 1.5, and longward of 1.6 $\muup$m.
We compared the YH spectrum and K1 photometry of 51~Eri~b to the one of L- and T-type objects from the SpeXPrism library \citep{2014ASInC..11....7B} completed with spectra from \citet{2013ApJS..205....6M} and \citet{2015ApJ...814..118B}. The comparison spectra were smoothed to the resolution of the IFS YH spectrum and their flux was integrated within the wavelength intervals covered by each channel of the IFS and the IRDIS K1 filter. We used the G goodness of fit indicator for the comparison \citep{2008ApJ...678.1372C}, with the implementation following \citet{2016A&A...587A..55V}, accounting for the filter widths and the uncertainty on the 51~Eri~b spectrophotometry.

\begin{figure}[!]
\centering
\includegraphics[width=\columnwidth]{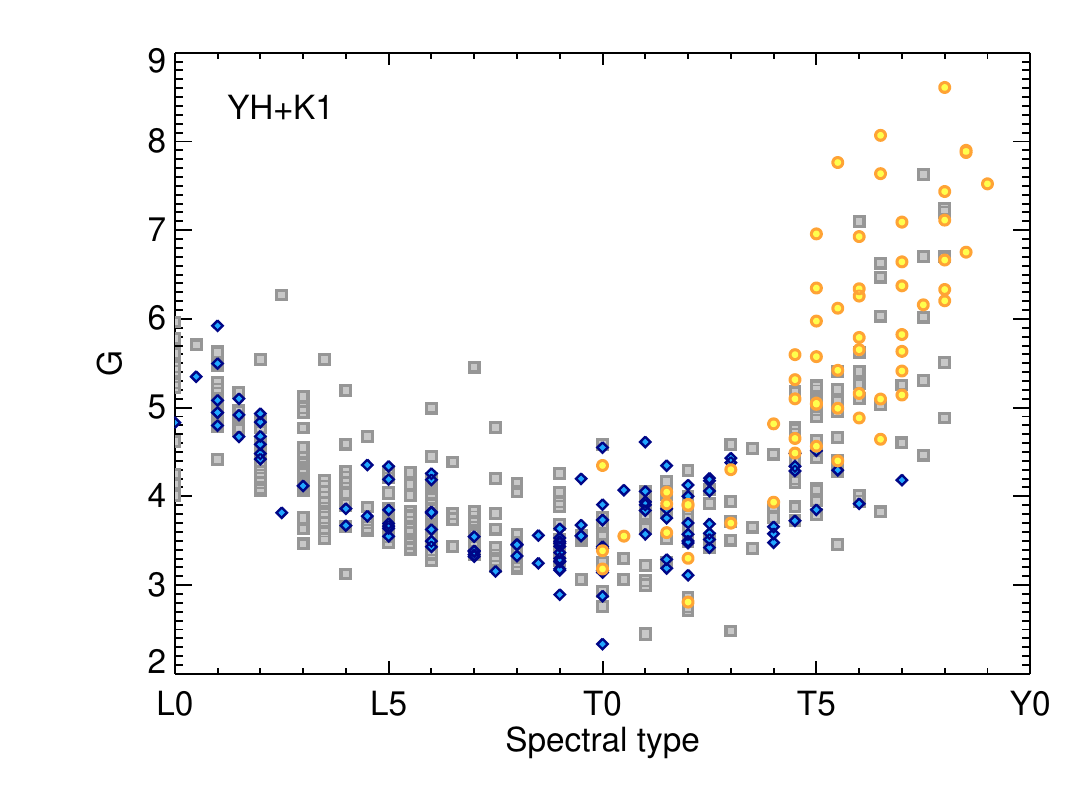}
\caption{Goodness of fit G for the comparison of 51~Eri~b YH+K1 spectrophotometry with the one of template spectra of L and T dwarfs from the SpeXPrism (gray squares), \citet{2013ApJS..205....6M} yellow circles, and \citet{2015ApJ...814..118B} (blue diamonds) libraries.}
\label{fig:G_YH}
\end{figure}

\begin{figure*}[!]
\centering
\begin{tabular}{cc}
\includegraphics[width=8cm]{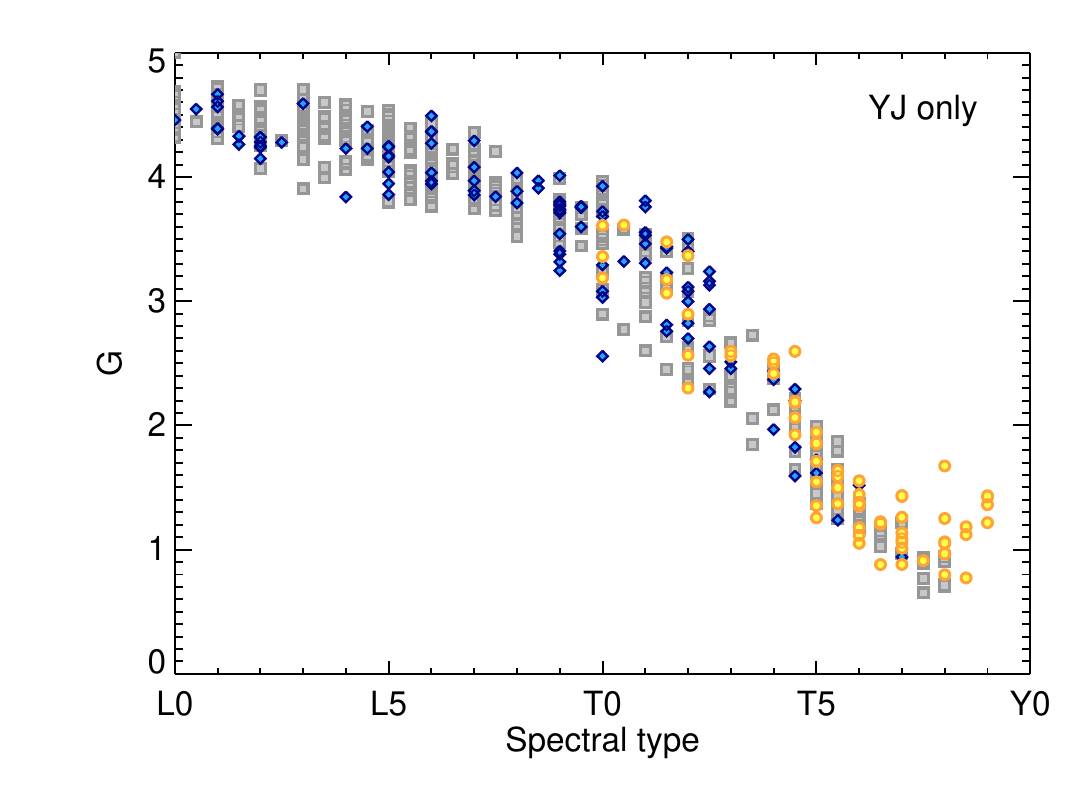} &
\includegraphics[width=8cm]{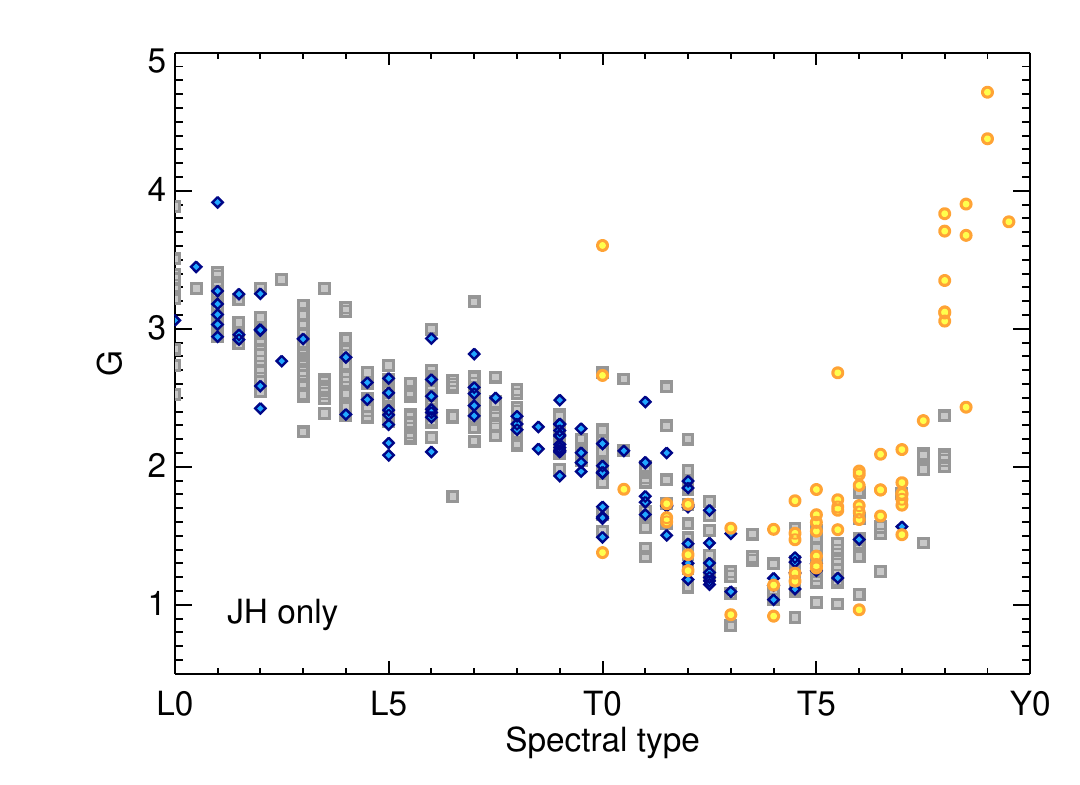} \\
\end{tabular}
\caption{Same as Fig.~\ref{fig:G_YH} but considering the YJ spectrum (left) and JH spectrum (right) of 51~Eri~b only.}
\label{fig:G_YJ}
\end{figure*}

The results are shown in Fig.~\ref{fig:G_YH}. The best fits are obtained for late-L/early-T objects, in agreement with the placement of the planet in color-color and color-magnitude diagrams (see below). The best fitting object is PSO J207.7496+29.4240, a peculiar T0 object, and possibly an unresolved binary, from the \citet{2015ApJ...814..118B} library. A visual inspection of the fit reveals that while the object is able to reproduce the overall JHK spectral slope, 51~Eri~b has deeper methane+water absorptions. The fit of the YH+K1 spectrophotometry is influenced by the strong overluminosity of the K1-band caused by the reduced collision-induced absorption (CIA) of H$_{2}$. The Y-band flux is also known to be modulated by the surface gravity and metallicity \citep{Burgasser2006, Liu2007}. To mitigate this in the comparison to higher log~g objects, we decided to re-run the fit on the YJ  spectrum and on the part of the YH spectrum excluding the Y-band (hereafter JH spectrum). The results are shown in Fig.~\ref{fig:G_YJ}.

T7--T8 objects represent the best match to the planet YJ spectrum only. Among the sample of T5.5--T7 objects, the brown dwarfs SDSSpJ111010.01+011613.1, 2MASSIJ0243137-245329, 2MASSJ12373919+6526148, 2MASSIJ1553022+153236 are minimizing G and therefore represent the best fits to the YJ spectrum. SDSSpJ111010.01+011613.1 and 2MASSIJ0243137-245329 belong to the growing class of red T dwarfs \citep{2014ApJ...783..121G, 2009ApJ...702..154S}.  SDSSpJ111010.01+011613.1 has been proposed as a member of the AB Doradus moving group \citep{2015ApJ...808L..20G}. The two other objects are respectively a binary \citep[unresolved in the SpeX slit,][]{2006ApJS..166..585B} and a magnetically active object with strong H$_{\alpha}$ emission, and displaying some variability in the J-band \citep{2000AJ....120..473B, 2003IAUS..211..451A, 2016ApJ...818...24K}.

The JH spectrum is best represented by SDSSJ141530.05+572428.7, a T3 dwarf from the SpeXPrism libraries, which is again a candidate unresolved binary \citep{2010ApJ...710.1142B}. Therefore, there appears to be a correlation between the spectral type of the best fit template found and the maximum wavelength of the photometric points included in the fit. We interpret it as (i) the consequence of the unusual red slope of the near-infrared spectral energy distribution (SED) of the planet compared to the templates and (ii) the fit limited to the shortest wavelengths becomes more sensitive to the $CH_{4}+H_{2}O$ absorption from 1.1 to 1.2 $\muup$m which is characteristics of late-T dwarfs. To conclude, we notice that the planet SED is often reproduced by candidate unresolved binaries, a class of objects which was also found to provide a good fit to the HR8799b and c planets \citep{2016A&A...587A..58B}.

\begin{table}[h]
\caption{IFS photometry}
\centering
\begin{tabular}{lcccc}
\hline\hline
Filter& $\lambda$ & Width & Contrast & Abs. Magnitude\tablefootmark{a}\\
&($\muup$m)&($\muup$m)& $(10^{-6})$ &\\
\hline
J & 1.245 & 0.240 & $1.03 \pm 0.67$ & $17.40 \pm 0.71$\\
J3 & 1.273 & 0.051 & $2.22 \pm 0.53$ & $16.52 \pm 0.26$\\
H2 & 1.593 & 0.052 & $2.70 \pm 0.70$ & $16.22 \pm 0.28$
\end{tabular}
\label{tab:ifs_phot}
\tablefoot{Photometric magnitudes for IRDIS filters derived from IFS spectra. Uncertainties are given as 1-$\sigma$.
\tablefoottext{a}{With distance modulus $\mu = m - M = 2.34$ using Vega magnitude system. Distance uncertainty is negligible in magnitude measurement.}}
\end{table}

We took advantage of the SPHERE spectra to generate synthetic photometry for the narrow-band filters of SPHERE overlapping with the wavelength range of the IFS spectra (assuming simple top-hat profile): J ($\lambda_{c}=1245\, {\rm nm}$, ${\rm FWHM}=240\, {\rm nm}$), J3 ($\lambda_{c}=1273, {\rm nm}$, ${\rm FWHM}=51\, {\rm nm}$), and H2 ($\lambda_{c}=1593\, \rm{nm}$, ${\rm FWHM}=52 \, {\rm nm}$). Photometric magnitudes in these bands enables a homogeneous comparison of the planet properties with those of known reference objects.

\begin{figure}[!]
\centering
\includegraphics[width=\columnwidth]{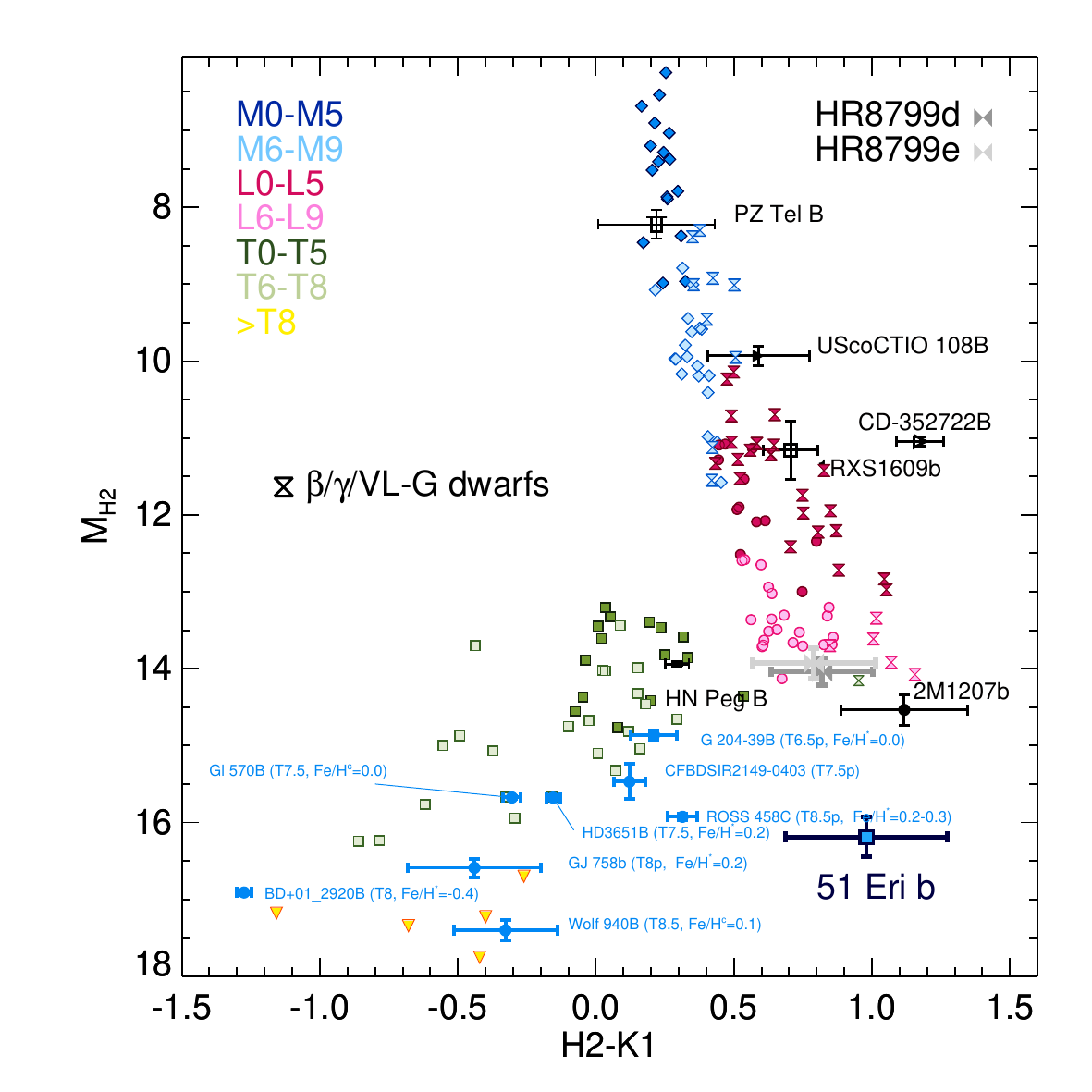}
\caption{Placement of 51~Eri~b in color-magnitude diagram. All magnitudes except for K1 are derived from the IFS spectra.}
\label{fig:CMD}
\end{figure}

\begin{figure}[!]
\centering
\includegraphics[width=\columnwidth]{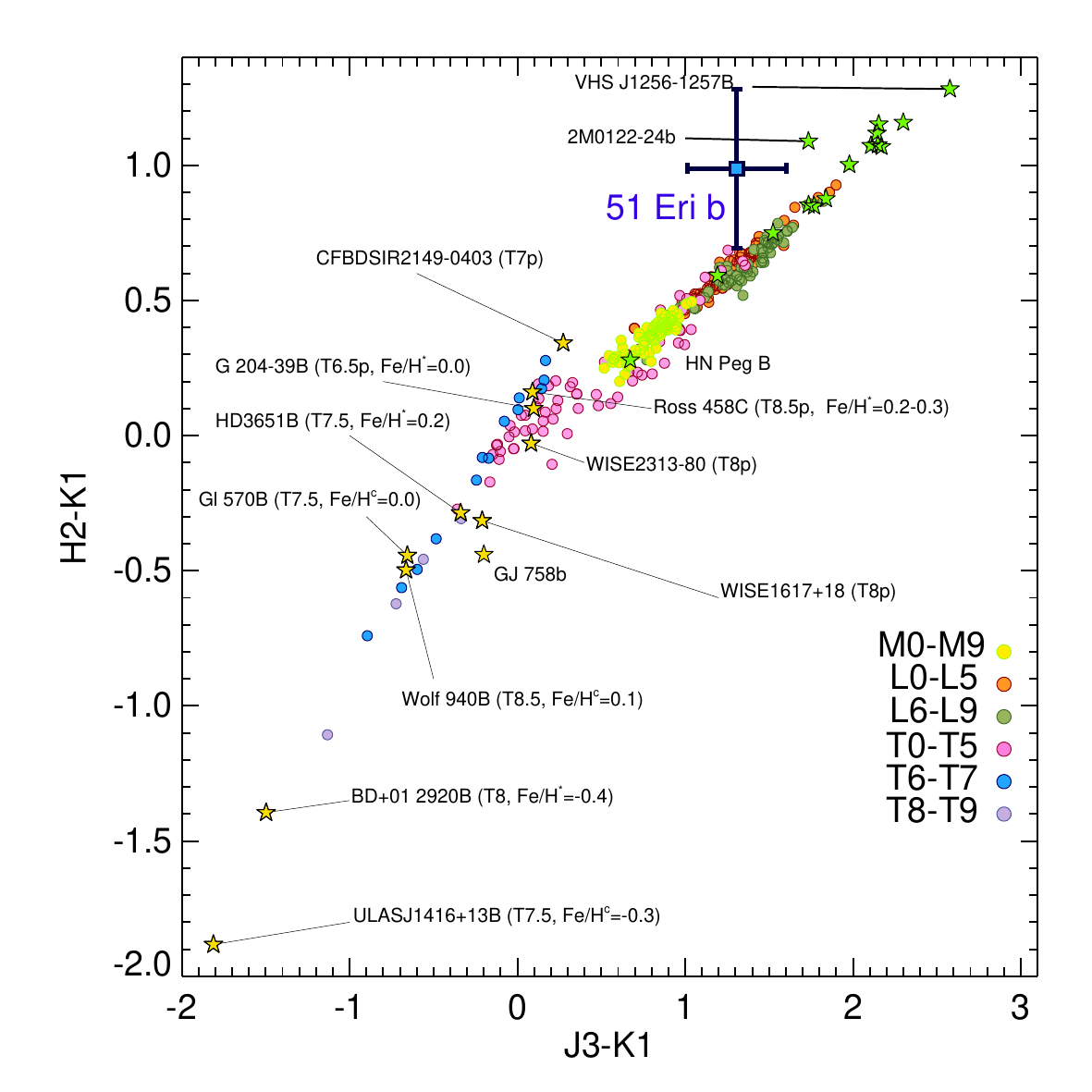}
\caption{Placement of 51~Eri~b in color-color diagram. All magnitudes except for K1 are derived from the IFS spectra.}
\label{fig:CCD}
\end{figure}

The photometry was obtained considering a flux calibrated spectrum of Vega \citep{2007ASPC..364..315B} and ESO \texttt{Skycalc} web application\footnote{http://www.eso.org/obser\-ving/etc/bin/gen/form?INS.MO\-DE=swspectr+INS.NA\-ME=SKYCALC} \citep{2012A&A...543A..92N, 2013A&A...560A..91J}. We find $\mathrm{J}=19.74\pm0.71$ mag, $\mathrm{J3}=18.86\pm0.26$ mag, and $\mathrm{H2}=18.56\pm0.28$ mag (Table~\ref{tab:ifs_phot}). We combined this synthetic photometry with the one obtained in K1 ($17.55\pm0.14$ mag) to show the planet's position in color-color and color-magnitude diagrams (Fig.~\ref{fig:CMD} and \ref{fig:CCD}). The CMD are build using low-resolution spectra taken from the literature and published parallaxes. For the field dwarfs, we used the spectra from \citet{2000ApJ...535..965L} and from the SpeXPrism library \citep{2014ASInC..11....7B}. The SpeXPrism spectra were calibrated in flux using the H-band 2MASS photometry of the targets. We used parallaxes from the literature \citep[mostly from][]{1992AJ....103..638M, 2012ApJ...752...56F} and newly revised values from \citet{Liu2016} where applicable. We repeated this procedure for young, low-gravity and/or dusty M, L, and T dwarfs (spectra taken for the most part from \citet{2013ApJ...772...79A} and parallaxes from \citet{2012ApJ...752...56F} and \citet{2014A&A...568A...6Z}). We added the known T-type companions \citep[and the isolated object CFHTBD2149;][]{Delorme2017} with known distances and with some knowledge of their metallicity (either from the primary star [Fe/H]$^{*}$ or from the companion spectrum [Fe/H]$^{c}$). We defer the reader to Bonnefoy et al. 2016 (in prep., and ref. therein) for a full description.

The planet has the luminosity of T6--T8 dwarfs but much redder colors consistent with those of late-L dwarfs in J3/J3-K1 and H2/H2-K1 color-magnitude diagrams (CMD).  In these diagrams, the benchmark T6.5--T8.5 objects suspected to be metal rich and/or younger than the field \citep[CFBDSIRJ214947.2-040308.9, GJ 758b, ROSS 458C, SDSSJ175805.46+463311.9/G 204-39B,][]{2012A&A...548A..26D, Vigan2016, 2011MNRAS.414.3590B, 2010AJ....139..176F} also have redder colors than the sequence of field dwarfs. Although they are not as red as those of 51~Eri~b.  In color-color diagrams (Figure \ref{fig:CCD}), 51~Eri~b falls at the location of the L/T transition objects in color-color diagrams, although the planet luminosity and the presence of a methane bands in its spectrum is inconsistent with this object being at the L/T transition. Instead, it suggests that the object has a color deviation which is similar to the color deviation seen for young and/or dusty late-L dwarfs (green stars in Fig. \ref{fig:CCD}) with respect to regular late-L dwarfs. The peculiar T7 dwarf CFBDSIRJ214947.2-040308.9 is also deviating from the sequence of T dwarfs but to a lower extent. We interpret the deviation for 51~Eri~b as a consequence of the reduced opacities caused by CIA of $\mathrm{H}_{2}$ \citep{1991Icar...92..273B} that occurs in low gravity and metal-enrich atmospheres and that affect primarily the K-band \citep{2001ApJ...556..357A}. We cannot exclude that it could also be caused by a haze of submicron sized particles as proposed for low-gravity L/T transition objects \citep[see][]{2014MNRAS.439..372M, 2016A&A...587A..58B, 2016arXiv160609485H} and consistent with our atmospheric modeling analysis in Sect~\ref{sec:atmospheric_modeling}.

\begin{figure}[!]
\centering
\includegraphics[width=\columnwidth]{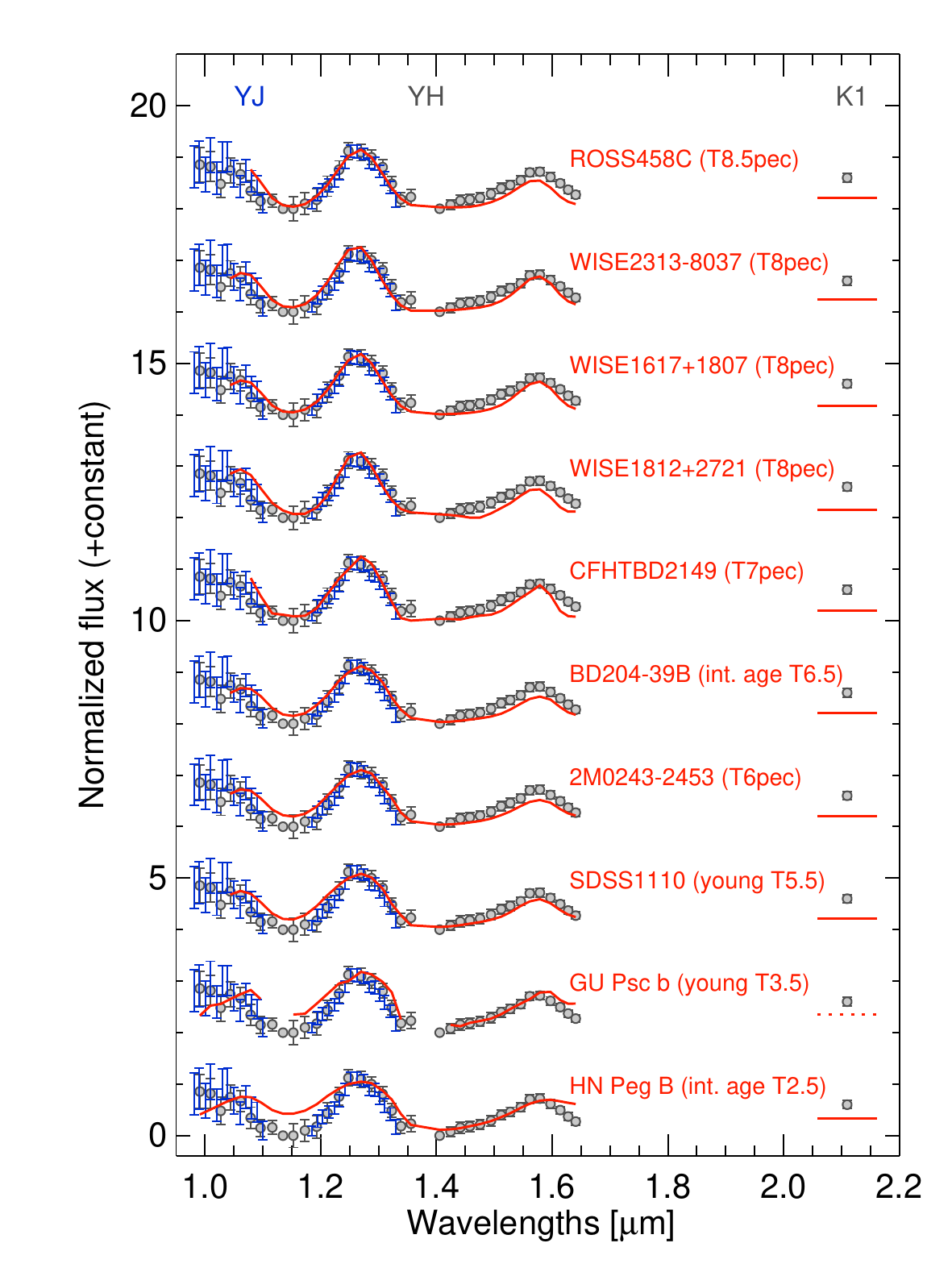}
\caption{Comparison of the flux density $F_\lambda$ of 51~Eri~b to those of selected peculiar T-type objects whose flux density has been normalized to match that of the planet between 1.2 and 1.3 $\mu$m.}
\label{fig:pecTs}
\end{figure}

We compare in Fig.~\ref{fig:pecTs} the spectrophotometry of 51~Eri~b to those of extra T-type objects known to be younger than the field or dusty objects \citep[][]{2011ApJ...735..116B, 2014ApJ...787....5N}. No object could reproduce simultaneously the YH-band features and the K1 flux of 51~Eri~b, in agreement with the previous analysis. We also note a strong departure of the 0.95--1.05 $\muup$m flux of the planet whose origin is unclear, but further discussed in detail in Sect.~\ref{sec:unexplained}. The depth of the 1.1--1.2 and 1.3--1.5 $\muup$m bands  of 51~Eri~b is only reproduced by those of objects later than T7.

In summary, the empirical approach: 1) confirms the peculiarity of 51~Eri~b, 2) further suggests that the planet shares the properties of late-T dwarfs, 3) suggests that some of the properties of the planet are related to low-surface gravity and young age or super-solar metallicity, and 4) is limited by the lack of objects from clusters and young moving groups with spectral types later than T5. These findings are in good agreement with our atmospheric modeling as described in the next section.

\subsection{Atmospheric modeling with petitCODE}
\label{sec:atmospheric_modeling}
In order to characterize 51~Eri~b we carried out dedicated calculations with \emph{petitCODE}, which is a self-consistent 1d radiative-convective equilibrium code, solving for the atmospheric temperature structure and abundances, assuming equilibrium chemistry. For every converged structure the \emph{petitCODE} calculates an emission and transmission spectrum, where the latter is of no importance for studying 51~Eri~b, given that the planet is not transiting. The first version of the code has been reported on in \citet{mollierevanboekel2015} and updates have been shortly described in \citet{mancini2016b, mancinikemmer2016a}. The current version of the code is described in detail in \citet{Molliere2017}.

In its current form the code includes molecular and atomic line and continuum (CIA) opacities, as well as an implementation of the cloud model by \citet{ackermanmarley2001}. The petitCODE treats the non-ideal line shapes of Na and K atoms by using the wing profiles by Nicole Allard, see \citet{mollierevanboekel2015} for a more detailed description. As possible cloud species MgAl$_2$O$_4$, MgSiO$_3$, Mg$_2$SiO$_4$, Fe, KCl and Na$_2$S can be included, with the optical constants taken from \citet{palik2012} for MgAl$_2$O$_4$, \citet{scottduley1996,jaegermolster1998} for MgSiO$_3$, \citet{servoinpiriou1973} for Mg$_2$SiO$_4$, \citet{henningstognienko1996} for Fe, \citet{palik2012} for KCl and \citet{Morley2012} for Na$_2$S.

Finally, we want to note that the implementation of the \citet{ackermanmarley2001} cloud model deviates from the description in the original paper in the sense that the mixing length is set equal to the atmospheric pressure scale height in all cases. This is different from the \citet{ackermanmarley2001} description, where the mixing length can be up to 10 times smaller than the pressure scale height in the radiative regions. In the regions above the cloud deck the cloud mass fraction is proportional to $P^{f_{\rm sed}/\alpha}$, where $f_{\rm sed}$ is the ratio of the mass averaged settling velocity of the cloud particles and the atmospheric mixing speed and $\alpha$ is the ratio between the mixing length of the eddy diffusion process and the atmospheric scale height. In our implementation of the \citet{ackermanmarley2001} model it holds that $\alpha=1$. The given power law can be derived from solving the homogeneous part of the differential equation for the condensate density \citep[Equation 4;][]{ackermanmarley2001}. Therefore, for a given $f_{\rm sed}$ value, clouds in the \emph{petitCODE} implementation will be more extended than in the \citet{ackermanmarley2001} description. Further, the atmospheric mixing speed is equal to $K_{zz}/L$, where $K_{zz}$ is the atmospheric eddy diffusion coefficient and $L$ is the associated mixing length, or mean free path, of the mixing process. Because the \emph{petitCODE} implementation sets $L=H_P$, where $H_P$ is the pressure scale height, the mixing velocity will be smaller than in the \citet{ackermanmarley2001} description, which favors smaller cloud particles at a given $f_{\rm sed}$ value. Therefore, adopting $L=H_P$ results in effectively smaller $f_{\rm sed}$ values when comparing cloud properties of the original \citet{ackermanmarley2001} description to the \emph{petitCODE} at the same $f_{\rm sed}$ value.

Two dedicated grids were calculated for 51~Eri~b (see Table~\ref{tab:model_grids}). The first grid is a clear grid (subsequently PTC-Clear, i.e. cloud free), assuming scaled solar compositions for the planetary abundances. We varied the effective temperature $T_{\rm eff}$ between 500 and 1700~K, the surface gravity by assuming ${\rm log}\, g$ values between 3 and 6 (with $g$ in cgs units) and the metallicities [Fe/H] between -1.0 and 1.4.

The second grid is a cloudy grid (PTC-C, "cloudy"), for which we assumed a mixing coefficient $K_{zz}=10^{7.5}$, similar to the value used in \citet{Macintosh2015}.
Here the varied grid parameters are $T_{\rm eff} = $500--850~K, ${\rm log} \, g=$3-5, [Fe/H]~=~0.0--1.4 and $f_{\rm sed}=$~0.5--2.0 (1--5 for initial exploration). Following \citet{Morley2012}, the opacities of MgAl$_2$O$_4$, MgSiO$_3$, Mg$_2$SiO$_4$ and Fe were neglected for this cool grid, such that for the clouds only KCl and Na$_2$S opacities were considered.

Finally, our calculations were carried out assuming equilibrium chemistry for the gas composition, and for identifying the cloud deck locations within the atmospheres. It is well known that for planets, compared to the higher mass brown dwarfs, non-equilibrium effects, and the associated quenching of CH$_4$ and NH$_3$ abundances may be more important \citep[see, e.g.,][]{zahnlemarley2014}. Due to the fact that we clearly detect methane in 51~Eri~b's atmosphere, and that we find best fit ${\rm log} \, g>4$ (see Sect.~\ref{sec:physical_parameters}) we conclude that CH$_4$ quenching is not very strong in this object, in agreement with the results presented in \citet{zahnlemarley2014} for higher ${\rm log} \, g$ objects.

In addition to the two grids outlined above we compare our results with the cloudy model atmospheres described in \citet{Morley2012}. However, as the grid does not include super-solar metallicity the resulting parameters are skewed (Appendix Fig.~\ref{fig:fit_morley} and Fig.~\ref{fig:corner_morley}). We will focus our discussion on the \emph{petitCODE} models.\\
A summary of the used grids can be found in Table~\ref{tab:model_grids} and our \emph{petitCODE} model grids will be made available online.

\begin{table*}[t]
\caption{Model grids used as input for MCMC exploration. The radius of the planet was included as an additional analytic fit-parameter regardless of the model, ranging from $0.1\,R_\mathrm{J}$ to $2 \,R_\mathrm{J}$. }
\centering
\begin{tabular}{l c c c c c c c c}
\hline\hline
Model& $T_\mathrm{eff}$ & $\Delta T$ & $\mathrm{log}\, g$ & $\Delta \mathrm{log}\, g$ &[Fe/H]& $\Delta \mathrm{[Fe/H]}$ &$f _\mathrm{sed}$ & $\Delta f _\mathrm{sed} $\\
& $(K)$ & $(K)$ & $\log_{10} \mathrm{(cgs)}$ & $\log_{10} \mathrm{(cgs)}$ & (dex) & (dex) & & \\
\hline
petitCODE (clear) & 500 -- 1700 & 50 & 3.0 -- 6.0 & 0.5 & -1.0 -- 1.4 & 0.2 & n/a & n/a \\
petitCODE (cloudy) & 500 -- 850 & 50 & 3.0 -- 5.0 & 0.5 & 0.0 -- 1.4 & 0.2 & 0.5 -- 2.0\tablefootmark{a}& 0.5\tablefootmark{a}\\
Morley+12 & 600 -- 800 & 100 & 4.0 -- 5.5 & 0.5 & 0.0 & n/a & 2 -- 5 & 1
\end{tabular}
\label{tab:model_grids}
\tablefoot{
\tablefoottext{a}{For the initial exploration a wider grid between 1 and 3 was used with a step size of 1. Smaller values were consistently preferred, leading to the final grid values.}
}
\end{table*}

\subsubsection{Determination of the spectral covariance matrices}
\label{sec:correlation}
\begin{figure}[!]
\centering
\includegraphics[width=\columnwidth]{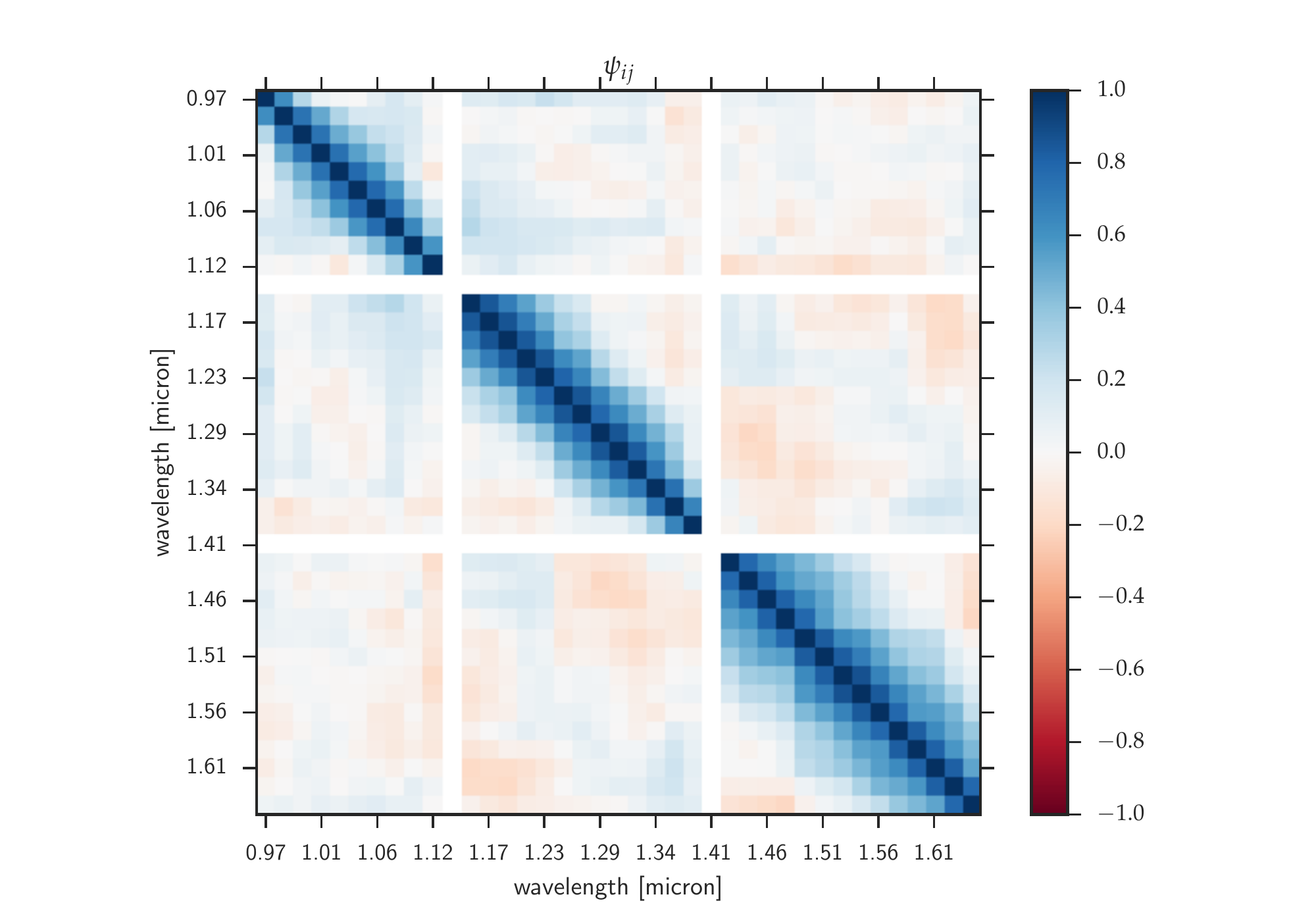}
\caption{Correlation matrix $\psi_{ij}$ showing the correlation between each pair of spectral channels (1: completely correlated; -1: completely anti-correlated; 0: uncorrelated). The 1.14 $\muup\mathrm{m}$ channel was used as reference for SDI for all wavelength channels shorter than this, whereas the 1.41 $\muup\mathrm{m}$ channel was used as reference for all other channels.}
\label{fig:correlation_yh}
\end{figure}

When comparing the spectrum obtained with an IFS-instrument with a model, taking into account the spectral covariance of the residual speckle noise has been shown to be of great importance for assessing the uncertainty of the fitted atmospheric model parameters \citep{Greco2016}. Following the methods presented by these authors, we determine the mean spectral correlation between all spectral channels within an annulus of width $1.5 \lambda / D$ at the planet's separation (with the planet masked out by a $2 \lambda / D$  radius mask):
\begin{equation}
\psi_{ij} =\dfrac{\langle I_i I_j \rangle}{\sqrt{\langle I^{2}_i \rangle \langle I^{2}_j \rangle}},
\end{equation}
where $\langle I_i \rangle$ is the average intensity inside the annulus at wavelength $\lambda_i$. The correlation matrix can then be used to obtain the covariance matrix $C$, which is used in computing the Gaussian log-likelihood $\ln \, \mathcal{L}$ (or $\chi^2$) for the MCMC model fit according to
\begin{equation}
-2 \, \ln \, \mathcal{L} \equiv \chi^2 = (S - F)^T \, C^{-1} \, (S - F),
\end{equation}
where $S$ is the observed spectrum and $F$ the model spectrum. In the case of uncorrelated noise $C$ is equal to the unity matrix and $\chi^2$ reduces to the more familiar sum over the residuals squared, which is not correct for correlated IFS data. The correlation matrix for the YH-spectrum is shown in Fig.~\ref{fig:correlation_yh}. We can see that each channel at the separation of 51~Eri~b is strongly correlated with three to four of its adjacent channels in both directions. Contrary to \citet{Greco2016}, we note that there are also anti-correlations present, which are due to the use of classical SDI and the larger spectral coverage available with the SPHERE IFS, spanning multiple bands and band gaps, unlike GPI spectra.\\
As we do not have access to the reduced GPI data of 51~Eri~b, we assume the fiducial model for a GPI-H spectrum reduced using simultaneous SDI+ADI as given in \citet{Greco2016} to calculate the correlation matrix at the angular separation of 51~Eri~b.

\subsubsection{MCMC exploration of atmospheric models and parameters}
\label{sec:mcmc}
We use the python implementation of the affine-invariant Markov-Chain Monte Carlo algorithm \textit{emcee} \citep{Goodman2010, Foreman2013} to explore the posterior probability distribution of model parameters for various atmospheric model grids (see Table~\ref{tab:model_grids}). Our custom procedure can handle model grids of arbitrary number of parameters, number of photometric data points and/or spectra, as well as their respective covariance matrices. The only restriction is that we require the model grid to be regularly spaced in each individual parameter to allow for efficient N-dimensional linear interpolation, where N is the number of free atmospheric model grid parameters. As the atmosphere of 51~Eri~b is not well characterized yet, we use flat priors over parameter ranges listed in Table~\ref{tab:model_grids}. Planetary radii are fitted as a separate analytic parameter. Evaluation of the log-likelihood for each spectrum with their complete covariance matrix and each photometric data point is done separately. They are then summed to obtain the overall log-likelihood of the model given the data, with no statistical weighting between the data sets done. Rather than defining a wavelength dependent weighting scheme, this is more properly taken into account by using the real covariances between the data. This effectively down-weights the relative importance of the many spectral data points with respect to the fewer but independent photometric data points. Uncertainties are assumed to be uncorrelated between the separate data sets. The likelihood evaluation is done in luminosity space, taking into account the additional uncertainty of the systems distance ($29.4 \pm 0.3 \, \mathrm{pc}$), which can be important for the planet's radius uncertainty, which otherwise would be slightly underestimated. In this case, due to the proximity and brightness of the host star, the distance uncertainty is only on the order of 1\%, and thus does not impact the radius uncertainty much, but for objects at larger distance this can become a significant factor.\\

We follow a different approach when treating upper-limits compared to many previous studies. We treat data points "below the detection limit" not as \textit{non-detection} or \textit{upper-limits} in the fit, because one does not look for a previously undiscovered source. We know where to measure the flux. Knowing the position of the source contains strong prior information, and even data points that are below the formal detection limit contain useful information (as can be seen by the fact that even data points that are technically not $3\sigma$-detections follow the model predictions quite well). These non-detection points can still contain significant flux, and in the "worst case" are consistent with negligible flux within their $1\sigma$ uncertainties. We use this approach of "forced photometry" \citep{Lang2016}, a method successfully used in other fields of astronomy, such as the study of faint galaxies and quasars \citep[e.g.,][]{Venemans2015}, to replace the more common practice of simply excluding data points below the classical detection threshold for point sources of unknown position, because this would mean mixing two unrelated statistical quantities in an unjustified way and effectively leads to throwing away informative data. Also replacing these measurements with an upper-limit as is commonly done -- while seemingly the conservative choice -- is not necessarily the optimal choice. In direct imaging, reporting only the upper-limit is equivalent to just reporting the uncertainty for the measurement without reporting the measurement itself. Applying forced photometry for all measurements means consistently reporting both, measurement and uncertainty. This has the advantage that all the data is treated uniformly and no arbitrary choice about a cut-off value for "detection" has to be chosen. The problem is illustrated by Fig.~\ref{fig:ifs_images}, where one would not claim the discovery of an unknown planet at the position of 51~Eri~b given only the Y-band image, but the fact that the occurrence of a clear excess in flux is located at the exact position of the planet visible in the other bands is much more informative and shows how important prior knowledge of the planet's position is for characterization.\\
However, to put the importance of the Y-band measurement in this particular case into perspective, it should  be pointed out that while it is true that all points included for forced photometry do contain some information on the spectrum of the planet, due to their low signal-to-noise and because other parts of the spectrum already put very strong constraints on the Y-band model fluxes, they do not impact the derived atmospheric parameters significantly. In other cases, where the predictability of the rest of the spectrum on the model in the wavelength range covered by non-significant flux measurements is not strong (e.g., models comparisons that seek to distinguish between the presence or absence of a physical model component, like thermal inversion or significant non-equilibrium chemistry), forced photometry and the inclusion of all the measurements whether statistically significant or not, can help distinguishing between valid models.

For all of these reasons, we also include the measured flux in the methane "non-detection" bands H3 and K2 (see Table~\ref{tab:photometry}). The only data that we do not include in the fit are the spectral channels that have been used as a reference in the SDI step of data reduction as these are biased and the first three IFS channels as they are most affected by degrading overall system performance and telluric lines.

\begin{figure*}[t]
\centering
\includegraphics[width=\textwidth]{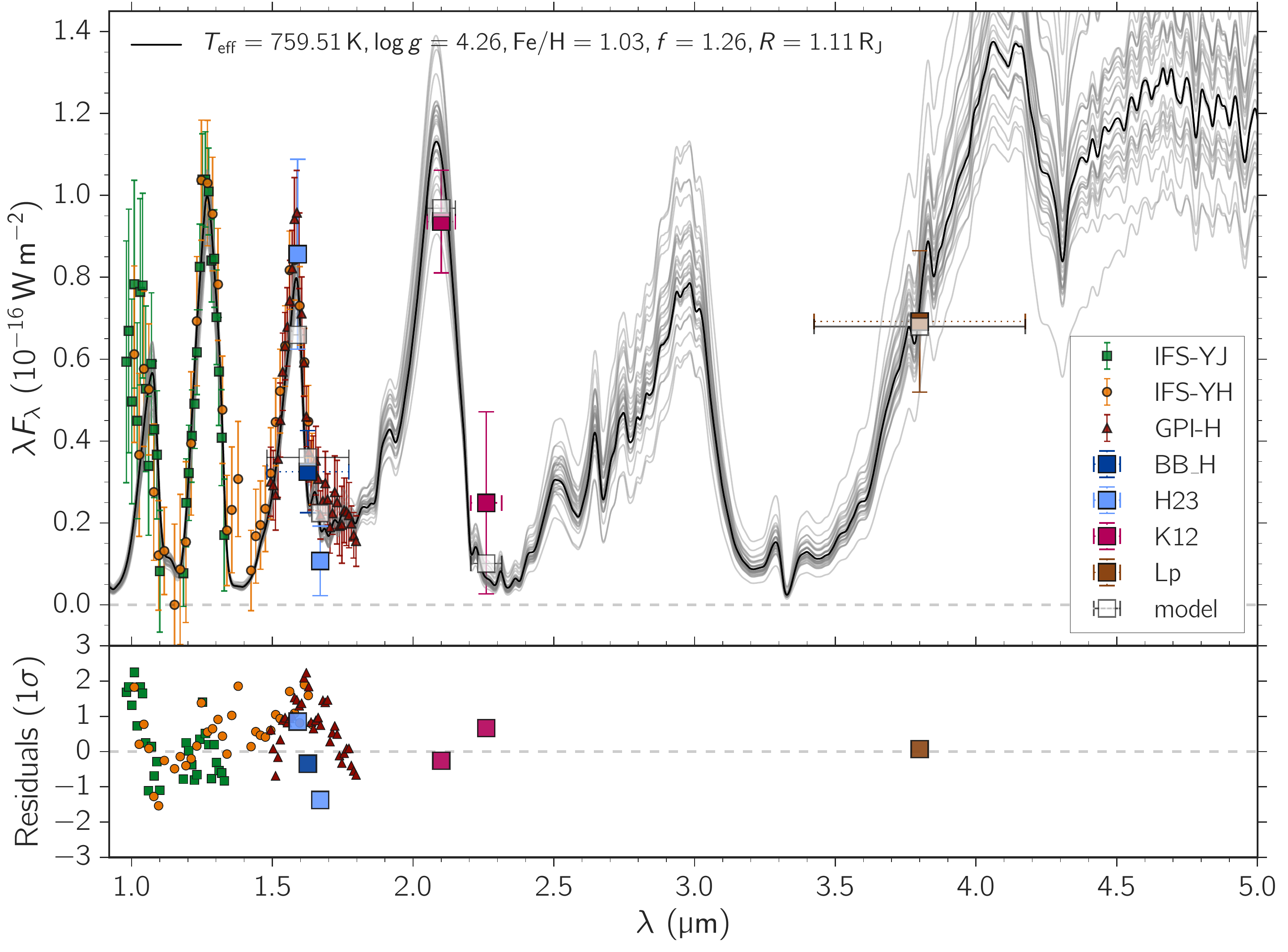} 
\caption{The plot shows the \emph{petitCODE} cloudy model interpolated to the parameters best describing the data according to the posterior probability distribution (black line), as well as the SPHERE spectrophotometric data and the GPI H-band spectrum and L' data point from \citet{Macintosh2015}. Notice that for photometric data points the x-errorbar does not reflect uncertainties, but the filter width. The gray lines represent 32 randomly drawn samples from the posterior probability distribution to reflect the spread of plausible model parameter combinations that fit the data. Photometric points describe the average flux in the respective filter, whereas the orange points describe the average flux in the respective filter for the best fitting model. The residuals are shown in multiples of $1\sigma$ uncertainties of the data.}
\label{fig:spectrum2}
\end{figure*}

\subsubsection{Discussion of physical parameters}
\label{sec:physical_parameters}
\begin{figure*}
\centering
\includegraphics[width=\textwidth]{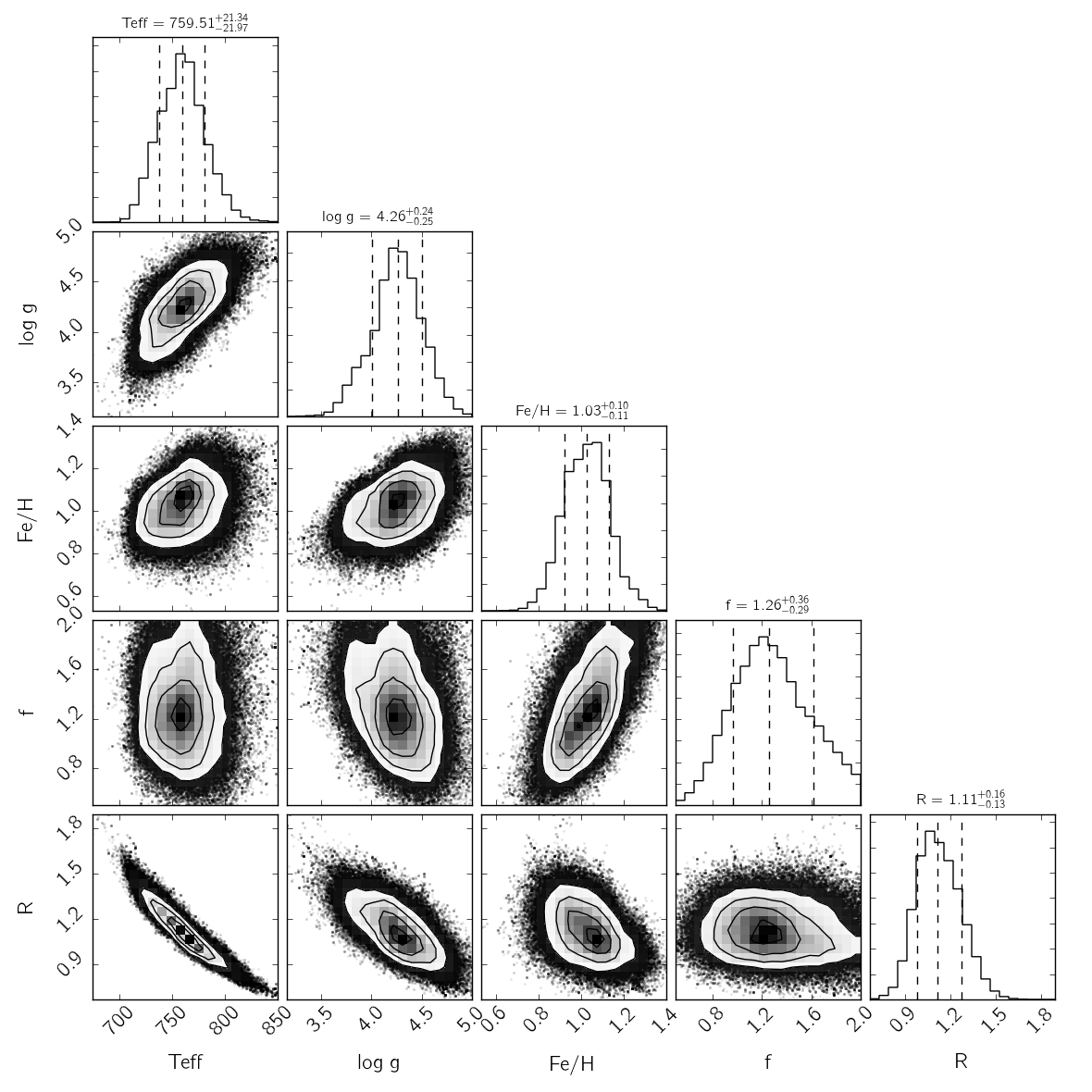}
\caption{Corner plot showing the posterior probability distribution of the cloudy \emph{petitCODE} grid with respect to each of its parameter pair as well as the marginalized distribution for each parameters. The uncertainties are given as 16\% to 84\% quantiles as commonly done for multivariate MCMC results.}
\label{fig:corner}
\end{figure*}

The best fitting models for the cloudy model grid (PTC-C) are shown in Fig.~\ref{fig:spectrum2}, with the black line representing the best fit and the gray lines showing the spectrum for 16 randomly drawn parameter combinations from the posterior probability distribution. As the most extensive model of the three, the posterior probability distribution of the PTC-C model is shown for each of the model parameters along with their marginalized values in Fig.~\ref{fig:corner}. Cloud free models are incapable of explaining all of the observed spectral features simultaneously: models which explain the Y-, J-, H-peaks are not able to explain the K1- and L'-band data (see Fig.~\ref{fig:corner_clear}). They also result in model predictions that are unphysical for young giant planets, e.g. high $\mathrm{log}\, g = 5.35^{+0.15}_{-0.12}$ and very low radius $R = 0.40 \pm 0.02\, \mathrm{R}_\mathrm{J}$ (see Fig.~\ref{fig:corner_clear}). Cloudy models vastly improve the consistency with the data over the whole spectral range for which data is available. Our discussion below will center on the results obtained on the \emph{petitCODE} cloudy models. They cover the complete parameter space relevant for 51~Eri~b, including metallicity and cloud sedimentation values ($f_\mathrm{sed}$). The results of all tested models are summarized in Table~\ref{tab:model_results}.

\begin{figure}
\centering
\includegraphics[width=\columnwidth]{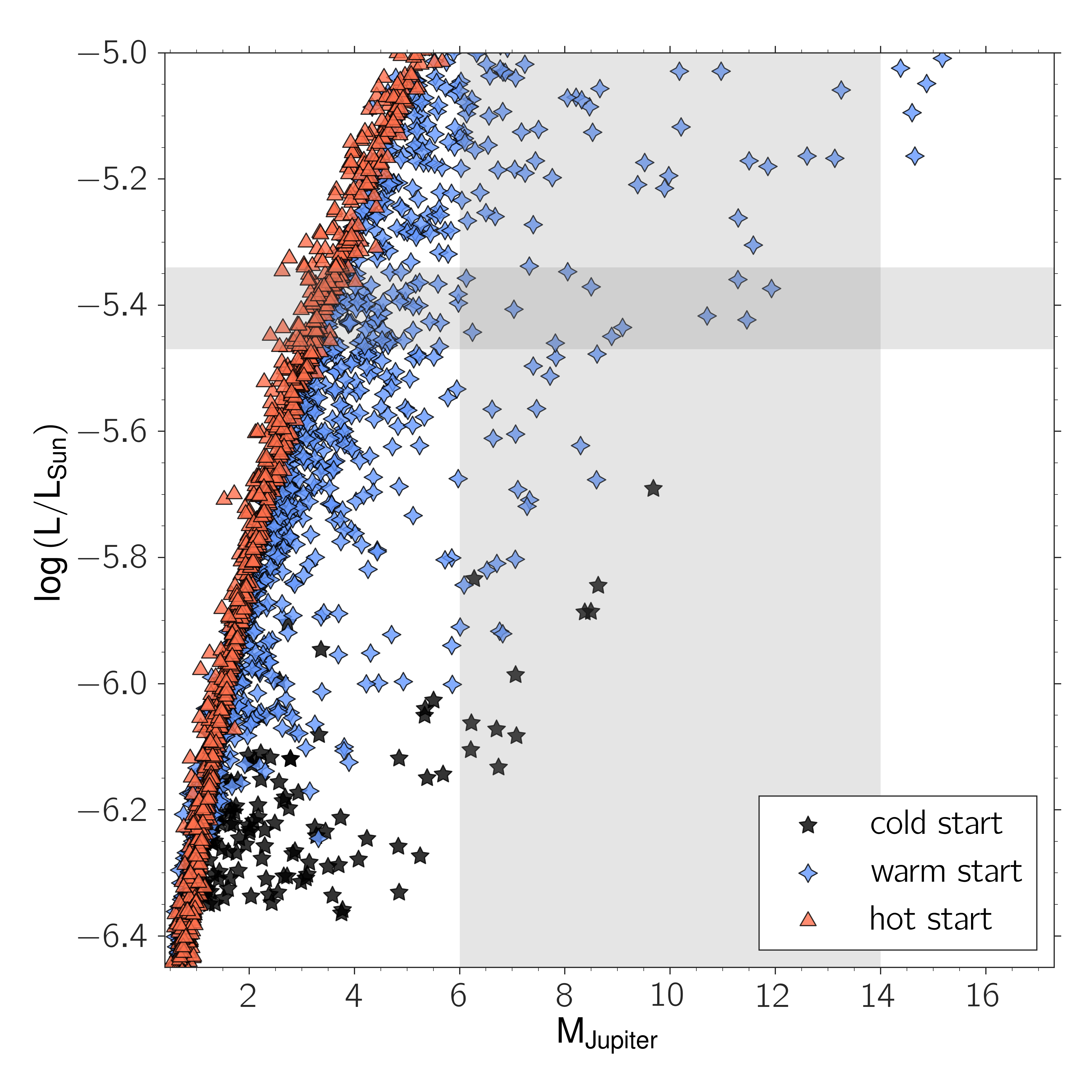}
\caption{Luminosity-Mass relationship from core-accretion population synthesis model at 20 Myr. Population using different entropy assumption are plotted, corresponding to what is traditionally referred to as cold (black) and hot start (red). The warm start (blue) model corresponds to a cold gas accretion model, but allowing for higher core-masses than $17\, \mathrm{M}_\mathrm{Earth}$. Gray shaded regions correspond to the luminosity of 51 Eri b derived in this work, and the mass range determined from surface gravity and radius.}
\label{fig:luminositymass}
\end{figure}

\paragraph{\textbf{Temperature, radius, and surface gravity}}We obtain a value of \printT{} for the effective Temperature, \printR{} for the radius and \printlogg{} (cgs-units) for the surface gravity of 51~Eri~b. The effective temperature and radius of the planet are expectedly correlated ($T_\mathrm{eff} \propto R^{-0.5}$ for black bodies) as they both relate to the luminosity of the planet. With a temperature that is likely above 700~K, it appears to be above the temperature for which sulfur chemistry becomes an important factor for 51~Eri~b as discussed in \citet{Zahnle2016}. The radius is consistent with the radius of Jupiter, possibly slightly larger as expected for young objects \citep{Chabrier2009, Mordasini2012b}.

\paragraph{\textbf{Mass}}
To check the consistency of the best-fitting solution with our physical understanding we can use multiple approaches to derive the mass of the planet. Using the posterior sampling of surface gravities and planet radii, according to $M = g/\mathrm{g}_\mathrm{J} \cdot (R/\mathrm{R}_\mathrm{J})^2$, where $\mathrm{g}_\mathrm{J} = 24.79$~m~s$^{-2}$ and $\mathrm{R}_\mathrm{J} = 6.9911 \times 10^7$~m are the surface gravity and (volumetric mean) radius of Jupiter respectively, we get a mass estimate of \printMlogg{}. Another approach is to use the luminosity of the planet derived similarly from the posterior sampling of the radius and effective temperature according to $L/\mathrm{L}_\odot \sim (R/\mathrm{R}_\odot)^2 \cdot (T/\mathrm{T}_\odot)^4$, where $R_\odot = 9.728\, \mathrm{R}_{\rm J}$ and $T_\odot = 5772$ K are the radius and effective temperature of the sun, we get a luminosity of \printLuminosity{} or $\mathrm{log} (L/\mathrm{L}_\odot)$ between \printLuminosityRange{} \citep[compared to -5.4 and -5.8,][]{Macintosh2015}, which can be converted to a mass assuming a formation (initial entropy) model. Figure~\ref{fig:luminositymass} shows the luminosity--mass relationship derived from a complete core accretion population synthesis model \citep{Mordasini2012a, Mordasini2012b}. The shaded region correspond to the above mentioned luminosity and (surface gravity derived) mass range. Three populations at 20 Myr using different assumptions are shown, where two population correspond to the traditional \textit{hot-start} and \textit{cold-start} population, and one to an intermediate \textit{warm-start} population. In the hot-start case, energy from the accretion shock is not radiated away efficiently and is deposited in the planet. In the cold-start case, all energy from gas accretion is radiated away, furthermore core-masses are restricted to $<17\, \mathrm{M}_\mathrm{Earth}$ to mimic the traditional cold-start model by \citet{Marley2007}. The warm start population is similar to the cold start population in terms of accretion physics, but allows higher core-masses, which in turn leads to more energy deposition during the growth of the planet \citep{Mordasini2013}. A similar result would be achieved by allowing for a spread in the radiation efficiency of the accretion shock. We can see that the observed luminosity range excludes the traditional cold-start model, but includes both objects of the hot- and warm-start case, with a large spread in masses. Small masses between $2.4$ and $5\, \mathrm{M}_\mathrm{J}$ are preferred in the hot-start case, whereas a big spread of masses between $2.4$ and $12\, \mathrm{M}_\mathrm{J}$ are possible in the warm-start case. While in this synthesis model objects with smaller mass are more common in this luminosity range, big masses are not excluded from the point of view of the formation model.\\
The above discussion and Fig.~\ref{fig:luminositymass} exemplify the problem of determining the mass of directly imaged exoplanets in the absence of low-mass and cool benchmark objects with independent mass measurement. In the two approaches, using the measured luminosity together with evolutionary models gives a statistical picture of the distribution of planets resulting from the planet formation synthesis modeling approach, and allows for a wide range of masses for the given age and luminosity, depending strongly on the accretion physics assumed. In principle, the mass derived from the surface gravity and radius are more constraining, but depend strongly on the atmospheric model assumptions, which in case of cold and cloudy objects have still many uncertainties. Assuming the model physics represent the real nature of the planet, the determined log g and radius is more consistent with a more massive planet that would be expected based on hot-start evolution models. It should be mentioned, however, that we can rule out the brown dwarf regime, as brown dwarfs at this age we would expect to see a significantly larger radius, due to their deuterium burning as additional heat source.\\
In conclusion it can be said that both mass estimates are highly model dependent and there are multiple big sources of uncertainty in both approaches. For atmospheric models it is possible that the cloud physics are not sufficiently well modeled, leading to big uncertainties in the surface gravity determination. The surface gravity determination is also most strongly impacted by the J-band flux for which some amount of variability cannot be completely ruled out at this point. On the other hand evolutionary models have a big intrinsic spread as they reflect the statistics of populations rather than single objects (e.g., different core-mass fractions). Initial conditions for planet formation and evolution are not well constrained. Another aspect which deserves further research is the current lack of evolutionary models that consider super-solar metallicity objects. While the composition is reflected in the core-mass of the models used here, the thermal evolution does not include the increase in opacity caused by high metallicity. This is also an issue for all other evolution models currently available.

\paragraph{\textbf{Metallicity}}The metallicity \printFEH{} is super-solar, and significantly above that of the solar metallicity host star, similar, but even more pronounced compared to what has previously been observed in the cool object GJ~504~b \citep[$\mathrm{Fe/H} = 0.60 \pm 0.12$ around a slightly metal-rich host star,][]{Skemer2016}. The metallicity determined here is in good agreement with predictions of bulk composition for giant planets formed by core-accretion \citep[e.g.,][]{Mordasini2014}. Other studies of massive directly imaged exoplanets also suggest super-solar metallicities, for example HR 8799 b \citep{2013ApJ...778...97L}.

Increasing the planetary metallicity strongly enhances the K-band brightness, redistributing a part of the flux from shorter to longer wavelengths. The reason for the metallicity-dependent K-band brightness is that a change in metallicity shifts the position of the planetary photosphere within the atmosphere, because in hydrostatic equilibrium it holds that $d\tau = (\kappa/g)\, \mathrm{d}P$, where $\tau$ is the optical depth, $\kappa$ the opacity, $g$ the planetary surface gravity and $P$ the pressure. If the pressure dependence of $\kappa$ is neglected, an increase in $\kappa$, resulting from an increase in metallicity, will shift the photosphere ($\tau_{\rm Phot}\sim 2/3$) to lower pressures. Nonetheless, it is key here to note that the opacity does vary as a function of pressure: the strength of pressure broadened molecular and atomic line wings is proportional to the pressure $P$, but for the many lines of water and methane the effect in the K-band is of secondary importance. More importantly, the continuum opacity due to CIA of H$_2$--H$_2$ and H$_2$--He pairs is linear in pressure for all wavelengths. Further, CIA exhibits a peak in opacity at 2.3~$\muup$m, i.e. very close to the K-band, whereas the CIA opacity in the neighboring H-band is lower by a factor of 100. Consequently, as the photosphere is shifted to lower pressures, due to an increased metallicity, the contribution of CIA to the total opacity in the K-band diminishes, such that the opacity minimum resulting from the scissor-like cross-over of the water and methane opacities becomes visible as an emission feature \citep{2001ApJ...556..357A}. Due to the steep decrease of the CIA opacity towards smaller wavelengths the H-band is unaffected by the increase in metallicity.
As a final test we carried out runs neglecting the CIA opacities and were unable to reproduce the effect of the metallicity-dependent K-band.

The strong influence of metallicity on key spectral features shows the importance of having a broad wavelength coverage of all features present in the near- to mid-infrared, as well as using model grids that include non-solar metallicity as a free parameter. Finally, to make sure our methodology and models are not systematically biased towards providing high metallicity results, we analyzed two benchmark brown dwarfs \citep[Gl~570D and HD~3651B, similar to][]{Line2015} and confirmed that the metallicities derived are reasonable. The details of this analysis can be found in Appendix~\ref{sec:bd_benchmark}.\\

\paragraph{\textbf{Clouds}}
For the cloud sedimentation parameter, we derive a value of \printf{}. A lower $f_{\rm sed}$ results in more vertically extended optically thicker clouds with smaller particle sizes. While the slight differences in atmospheric model implementations make it difficult to compare this result exactly with previous research, $f_{\rm sed}$ as low as this ($<2$) are unusual for self-luminous substellar objects of low temperature, especially considering that our implementation of $f_{\rm sed}$ would result in a lower value in the \citet{ackermanmarley2001} implementation (see model discussion in Sec.~\ref{sec:atmospheric_modeling}). Values of $f_{\rm sed}$ between 3 and 5 are usually reported, e.g for GJ 504 b \citep{Skemer2016} and GJ 758 B \citep[$f_{\rm sed} = 5$][]{Vigan2016}. \citet{2008ApJ...678.1372C} report values between 1 and 4 for their sample of L and T brown dwarfs, but all of these objects are significantly hotter than 51~Eri~b and only models of solar metallicity are considered. The lack of similar objects and detailed analyses including metallicity as a free parameter make a real comparison difficult. Lack of metallicity as free parameter in the model can significantly alter the cloud parameter, because it will tend to compensate for the lack or overabundance of heavy elements in the spectrum. We encourage modelers to include small $f_{\rm sed}$-values as well as metallicity in their consideration for future model grids.\\
The \printf{} we obtain for 51 Eri b reflects a particle size distribution with mean values of 1 $\muup\mathrm{m}$, and slightly below in the upper regions (below $10^{-2}$ bar). Due to the width of $\sigma=2$ of the log-normal size distribution, however, the opacities are dominated by the larger particles.

\begin{table*}[t]
\caption{Summary of model results}
\centering
\begin{tabular}{Sl Sc Sc Sc Sc Sc Sc Sc Sc}
\hline\hline
Model&$T_\mathrm{eff}$&$\mathrm{log}\, g$&[Fe/H]&$f _\mathrm{sed}$&$R$&$L$&$\log \, L$&$M_\mathrm{gravity}$\\
 &(K)&$\log_{10} \mathrm{(cgs)}$&(dex)&&$(\mathrm{R}_\mathrm{J})$&$(10^{-6} \, \mathrm{L}_\odot)$&$\log_{10} \, (\mathrm{L}_\odot)$&$(\mathrm{M}_\mathrm{J})$\\
\hline
PTC-C&$760^{+21}_{-22}$&$4.26^{+0.24}_{-0.25}$&$1.03^{+0.10}_{-0.11}$&$1.26^{+0.36}_{-0.29}$&$1.11^{+0.16}_{-0.13}$&$3.94^{+0.66}_{-0.55}$&-5.470 to -5.338&$9.1^{+4.9}_{-3.3}$\\
\hline
PTC-C w.o. Y-band&$754^{+23}_{-23}$&$4.25^{+0.32}_{-0.37}$&$1.04^{+0.11}_{-0.12}$&$1.33^{+0.38}_{-0.33}$&$1.13^{+0.17}_{-0.15}$&$3.92^{+0.68}_{-0.61}$&-5.479 to -5.337&$9.1^{+7.3}_{-4.3}$\\
PTC-Patchy\tablefootmark{a}&$757^{+24}_{-24}$&$4.47^{+0.24}_{-0.26}$&$1.25^{+0.10}_{-0.16}$&$1.07^{+0.36}_{-0.31}$&$1.11^{+0.16}_{-0.14}$&$3.84^{+0.63}_{-0.56}$&-5.484 to -5.350&$14.5^{+8.7}_{-5.6}$\\
PTC-Clear&$982^{+18}_{-15}$&$5.35^{+0.15}_{-0.12}$&$1.36^{+0.03}_{-0.06}$&-&$0.40^{+0.02}_{-0.02}$&$1.43^{+0.06}_{-0.06}$&-5.863 to -5.827&$14.5^{+4.7}_{-3.1}$\\
Morley+12&$684^{+16}_{-20}$&$5.19^{+0.10}_{-0.11}$&-&$4.16^{+0.52}_{-0.67}$&$1.01^{+0.07}_{-0.06}$&$2.12^{+0.14}_{-0.13}$&-5.700 to -5.645&$64.9^{+19.1}_{-15.6}$\\
PTC-C (Macintosh et al\tablefootmark{b})&$785^{+11}_{-17}$&$3.35^{+0.29}_{-0.21}$&$0.83^{+013}_{-0.12}$&$2.54^{+0.32}_{-0.47}$&$1.12^{+0.08}_{-0.05}$&$4.55^{+0.34}_{-0.30}$&-5.372 to -5.311&$1.2^{+1.0}_{-0.4}$
\end{tabular}
\label{tab:model_results}
\tablefoot{See Appendix~\ref{sec:corner_plots} for corner plots of the posterior probability distributions of the model parameters for each model.
\tablefoottext{a}{Effective temperature calculated by $T_\mathrm{eff}=\left[ CF \cdot T^4_\mathrm{cloudy} + (1-CF) \cdot T_\mathrm{clear}^4) \right]^{1/4}$ with $T_\mathrm{cloudy}=751^{+24}_{-25}$, $T_\mathrm{clear}=813^{+67}_{-40}$ and cloud fraction $CF=0.91 \pm 0.05$.}
\tablefoottext{b}{Using our cloudy model and the same data as \citet{Macintosh2015} without covariance treatment.}}
\end{table*}

\subsubsection{A note on patchy-cloud models}
\label{sec:patchy_cloud}
As variability has been observed in a number of brown dwarfs, the idea that for cool substellar objects cloud coverage may be less than 100\% as in our uniform cloud model should not be excluded a-priori. In \citet{Macintosh2015} such a "patchy cloud"-model, which can be expressed as a linear combination of a clear and a cloudy atmosphere, was used. They also included non-equilibrium chemistry in the cloudless model. We tested this idea with the following simple composite model using \emph{petitCODE}
\begin{equation}
F_\mathrm{patchy} = CF \cdot F_\mathrm{cloudy} + (1 - CF) \cdot F_\mathrm{clear},
\end{equation}
where CF is the cloud fraction and $F_\mathrm{cloudy}$ and $F_\mathrm{clear}$ are the flux of the cloudy and clear model, respectively. Under this model we have the following MCMC fitting parameters $\theta = (T_\mathrm{cloudy},  T_\mathrm{clear}, CF, \log\,g, \mathrm{[Fe/H]}, f_\mathrm{sed}, R)$, i.e. we now fit for the cloud fraction and allow the two models to have different temperatures, as the cloudy and clear model fluxes probe different temperatures. Because both models must, however, describe the same physical planet, we keep the metallicity, as well as the surface gravity and radius, the same for both models. Furthermore, we impose that $T_\mathrm{cloudy} < T_\mathrm{clear}$ as a prior, as the cloudy model flux is supposed to come from higher in the atmosphere than the clear flux, which in this model corresponds to holes in the clouds.\\
However, the result of this test shows no significant improvement of the fit for the resulting composite model spectrum as cloud coverage tends towards $>90$\% (see Appendix Fig.~\ref{fig:corner_composite} for corner plot). As the resulting spectra (Appendix Fig.~\ref{fig:fit_composite}) look almost the same, we conclude that a patchy-cloud model may not be necessary to explain the spectrum, and at least at this point, do not justify the increase in model complexity. It should be pointed out, that using the patchy-cloud model improves the fit marginally when only data from \citet{Macintosh2015} is used. This may be attributed to the higher J-band flux in GPI and resulting bluer spectrum.\\
To be clear, we do not wish to say that patchy-cloud models in general do not work or should be avoided, only that for this particular planet, with this data set, and this particular model, cloudy models alone seem to be capable of fitting the data quite well. It may well be that inclusion of more physics (e.g., non-equilibrium chemistry) improves the results. It is also important to keep in mind that a simple linear combination of a clear and cloudy model, as done here, is not self-consistent and strictly speaking not physically correct \citep{Marley2010}. A detailed model comparison with a more rigorous patchy-cloud model should be done in the future to test, whether further increasing the model complexity is justified by the gain in fit quality, for example by using Bayesian evidence (e.g., nested sampling).

\subsubsection{Unexplained spectral features}
\label{sec:unexplained}
There exist a number of features in the spectrum of 51~Eri~b that cannot be explained with the current model, either pointing to unaccounted systematic in the data or the need for more sophisticated atmospheric models.
\begin{enumerate}
\item The Y-band peak in the data is stronger and seems to extend to smaller wavelengths than predicted by the model. The Y-band is difficult to observe with good signal-to-noise, mostly because overall instrument performance degrades towards shorter wavelengths (e.g., worse AO correction and end-to-end instrument throughput). It is also subject to some unresolved telluric features at short wavelengths ($\lesssim 1$ $\muup\mathrm{m}$) in the Earth's atmosphere.
There are multiple plausible scenarios for the perceived discrepancy: a) residual speckle flux at planet position at these wavelengths, b) a genuine instrument systematic effect (e.g. unaccounted variations in system transmission), c) a real physical phenomenon / improper treatment of potassium wings or abundances in model. If there is residual speckle flux (i.e. speckle noise) and the noise is spectrally correlated (as it is, the treatment of which is described in Sect.~\ref{sec:correlation} and taken into account), we expect it to affect at least half of the Y-band channels consistently (as about 6 neighboring channels are correlated). Seeing visually that a number of points scatter "systematically" higher or lower than the model is actually what we should expect in this case at low SNR. It is important to remember that we can only plot 1D errorbars which looks like we have a systematic, if we work under the assumption that the measurements are uncorrelated and should scatter randomly around the true values. The proper treatment of IFS covariances is a relatively new practice in this field and bears to be kept in mind. As such the method of forced photometry should only be used in conjecture with proper covariance treatment. On the other hand it makes it challenging to distinguish residual speckle flux from other instrument systematics that may only be present in certain wavelength regions. "Systematic deviations" which conform to the correlation length are not much of an issue for the parameter fitting because we already take this effect into account. To confirm this, we performed the same fit without including the Y-band data at all. It only marginally changes the results, which shows that the relatively large and correlated uncertainties in the Y-band data, at this point are not very constraining (see Figure~\ref{fig:corner_wo_Y} for posterior distribution in the case that Y-band data is excluded), which strengthens our confidence in the robustness of the analysis and our treatment of the noise. Resolving the issue which of the three scenarios (or mixture of them) is dominant will require obtaining more high SNR Y-band measurements. If the elevated flux level in the observation are shown to be persistent and significant in upcoming observations, this raises the possibility that the model treatment of potassium wings or abundances needs to be reconsidered and improved (e.g., better alkaline profiles, mechanism for depletion of alkaline species).

\item We observe an emission feature at $\sim$1.35 $\muup\mathrm{m}$, which is not explained by the model. While it is possible that this is caused by instrument systematics or the fact, that it is in a region of strong telluric absorption, it is striking that both the GPI observations as well as SPHERE show an increase in flux. A very similar feature in the deep water bands between the J- and H-peaks at 1.35 -- 1.40 $\muup\mathrm{m}$ has been observed and discussed by \citet{King2010} in the $\epsilon$ Indi Ba and Bb brown dwarf binary members. They also list objects with descriptions of similar features, the T1 spectral standard SDSS0151+1244 \citep{Burgasser2006b}, the T8.5 and T9 dwarfs ULAS1238 and ULAS1335 \citep{Burningham2008}, and some L dwarfs \citep[e.g., 2MASS J1507--1627 (L5)][]{Burgasser2007}. \citet{King2010} argue that this feature is due to the structure of the strongest part of the water absorption bands in the higher levels of the atmosphere, possibly due to an underestimated local temperature in this region of the atmosphere. According to their toy model, raising the temperature, and therefore changing the temperature gradient can reconcile the modeled and observed flux levels, although they could not point to a reasonable physical mechanism, like back-warming due to an additional opacity source. If this feature is indeed a real feature, in the case of 51~Eri~b, it may be related to its atmospheric cloud structure, but at this point this is very speculative. The fact that both the target planet as well as Earth's atmosphere contain complex telluric features at these wavelengths makes it difficult to draw strong conclusions.
\item The H-band feature's broad width towards shorter wavelengths and extended wing towards longer wavelengths has a profound impact on the model fit. In general the H-band wings strongly favor models with higher log~g and lower metallicity, whereas the rest of the spectrum favors lower log~g and higher metallicity (especially the need for high metallicity to produce the K1-peak). Excluding the GPI H-band spectrum from the fit allows the PTC-C model as well as the \citet{Morley2012} models to fit the strength of all of the observed features well (except for the width and height of the Y-band peak and the width of the H-band peak). Including H-band wings in the fit puts strong weight on these features and the resulting best model is a compromise between fitting the H-band wings and the amplitude of the peak. This spectrum while fitting the overall shape of the H-band well, does not match the absolute amplitude of the H-peak. A zoom in on the wavelength range covered by IFS data is shown in Fig.~\ref{fig:fit_zoom}. This trade-off shows how important extensive coverage of the spectral bands is for drawing physical conclusions.
\end{enumerate}

\section{Constraints on additional companions}
\label{sec:constraints}
\subsection{VLT-NACO/SAM(L')}
To constrain the presence of any potential companions at smaller separations, we processed and analyzed archival Sparse Aperture Masking (SAM) data taken with the VLT-NACO instrument. The observations were made on 2009-12-26 using the L' filter and the 7-hole aperture mask. Calibrator stars used where HIP 22226, HIP 30034, HIP 24947 and HIP 32435, and conditions where between median and bad. Single exposures had DITs (detector integration times) of 0.2s with a total of 3200 frames (NDIT = 200, 16 cubes). The calibrators had DITs of 0.25 with the same number of frames. Data were processed using the IDL aperture masking pipeline developed at the University of Sydney. The data processing steps are described in \citet{Tuthill2000, Kraus2008} and the references therein. Briefly, the images were sky subtracted, flat fielded, cleaned of bad pixels and cosmic rays, then windowed with a super-Gaussian function. The closure phases were then measured from the Fourier transforms of the resulting cleaned cubes. The closure phases were calibrated by subtracting the average of those measured on several unresolved calibrator stars observed during the same night with the same instrument configuration.
To estimate the detection limits of the SAM data, a Monte Carlo simulation was performed. Using a Gaussian distribution, we generated 10,000 simulated datasets consistent with the measured uncertainties. For each point on a grid of separation, position angle and contrast, we defined our detection limits to be the point at which 99.9\% of the simulated datasets were fit better by a point source model than the binary model. These 3.3$\sigma$ detection limits were then scaled to 5$\sigma$ to simplify the comparison with the results from SPHERE. No additional point sources are detected.

\subsection{SPHERE}
\begin{figure}[!]
\centering
\includegraphics[width=0.45\textwidth]{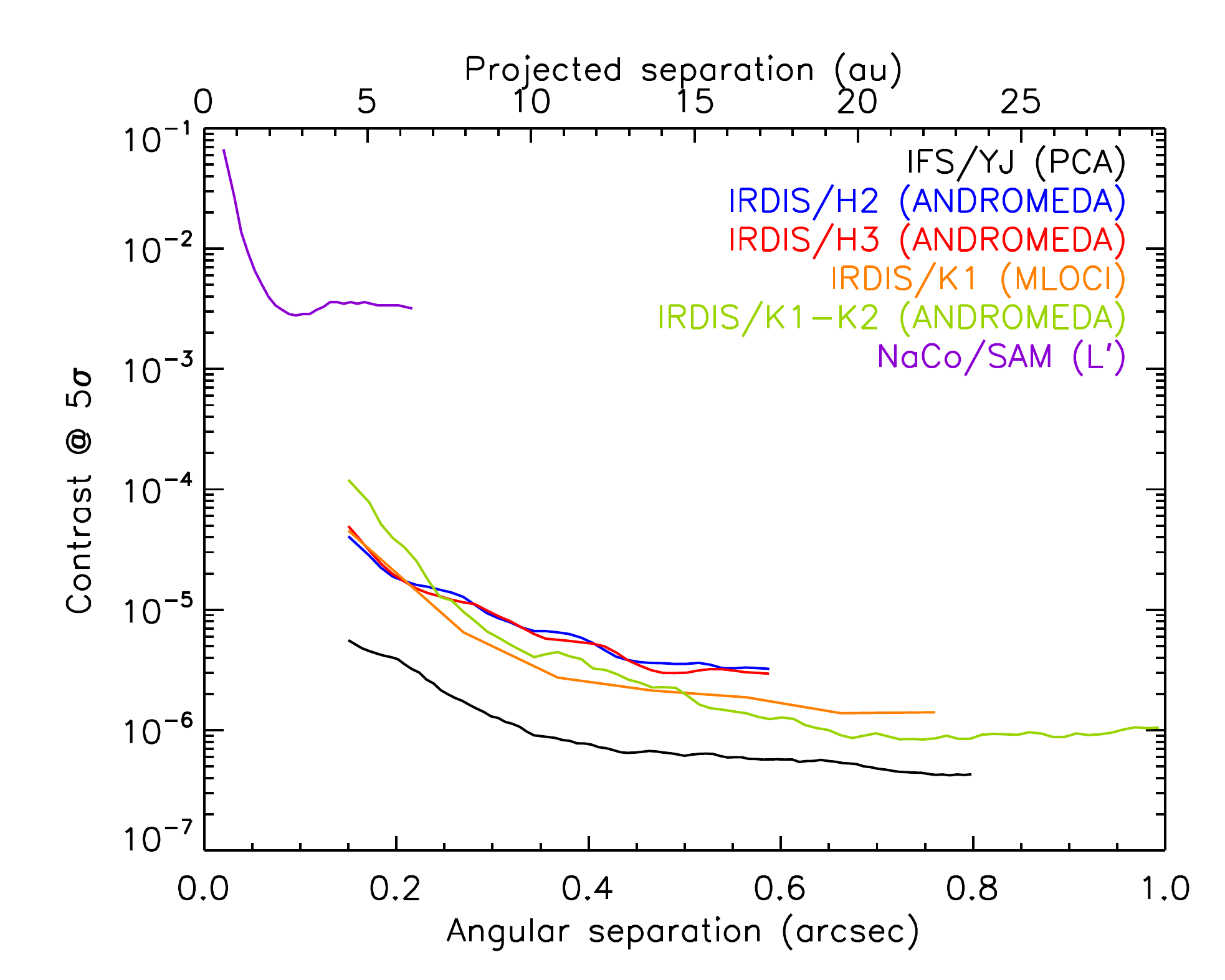}
\includegraphics[width=0.45\textwidth]{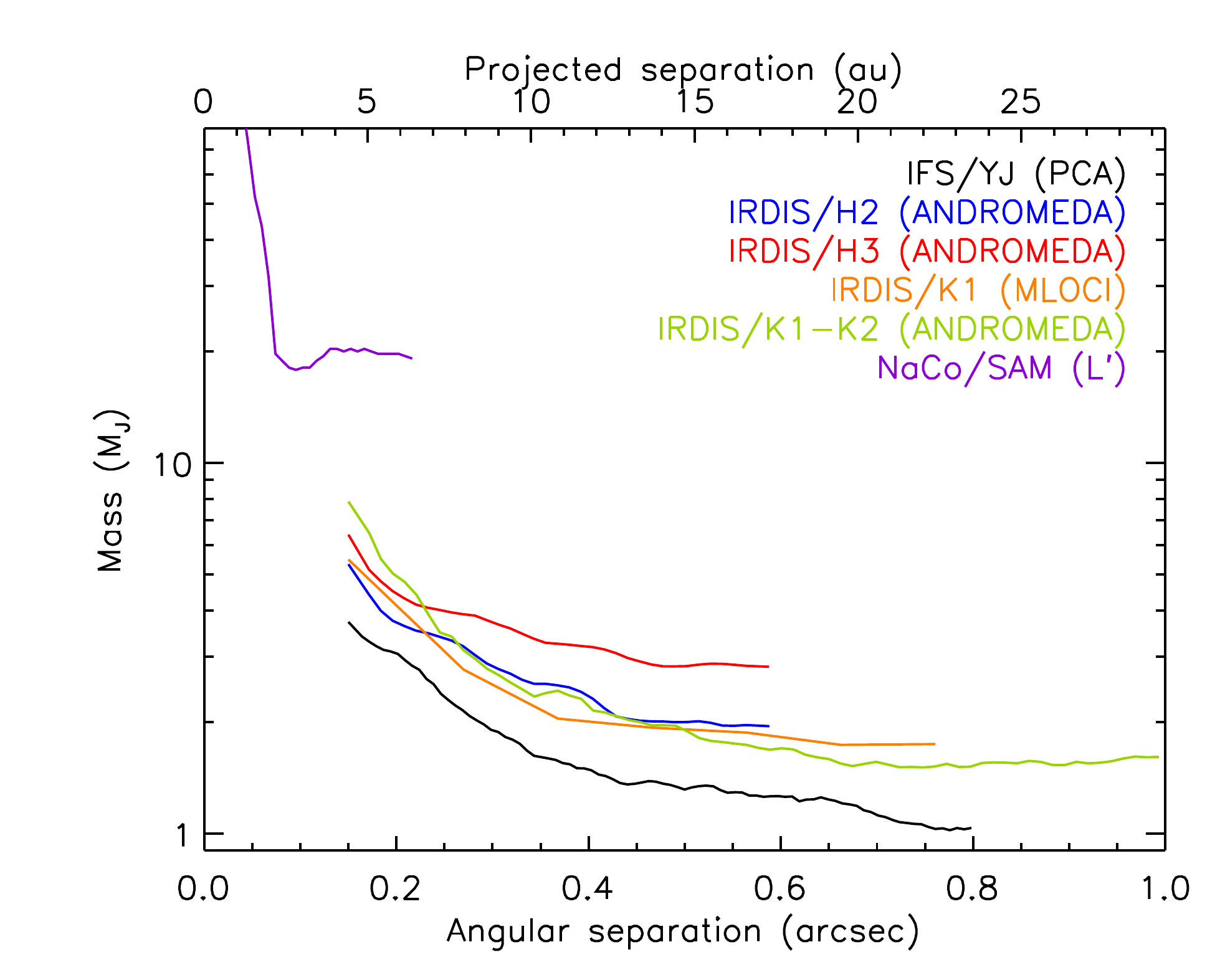}
\caption{The 5$\sigma$-contrast (top) and mass (bottom) is plotted for NACO/SAM (L') as well as the observations with the best quality for IRDIS and IFS respectively.}
\label{fig:contrast}
\end{figure}

We detect no additional point sources in the SPHERE data. Contrast curves for the more extended FoV achievable with IFS and IRDIS have been compiled using different reduction methods. The methodology for deriving the contrast was kept as similar as possible between the algorithms. For LOCI and PCA reductions they correspond to the azimuthal 5$\sigma$ self-subtraction corrected variance in the respective separation bin, whereas ANDROMEDA inherently models the detectable signal contrast and does not need self-subtraction correction. The effect of small-sample statistics at small separations \citep{Mawet2014} and the coronagraphic throughput (A. Boccaletti, priv. comm.) has been accounted for. The IFS-YJ PCA reduction was performed with the more aggressive simultaneous ADI+SDI algorithm for detection and shown is the median combination of all channels. The top panel of Fig.~\ref{fig:contrast} shows the achieved contrast with both the innermost region explored by small aperture masking as well as the exploration region of SPHERE. It is not straightforward to convert the "detection images" shown in Fig.~\ref{fig:ifs_collapse} into quantitative contrast curves and detection limits. As such they are not used for this purpose in this paper, but serve as a qualitative probe for additional candidates. However, no obvious candidates are seen.

For the conversion from contrast to mass limits, we used the same JHKL' magnitudes, distance, and age for the host star as in \citet{Macintosh2015}. We use evolutionary tracks of \citet{Baraffe2003} together with the atmosphere model of \citet{Baraffe2015} for the SPHERE data and BT-Settl models of \citet{Allard2012} for the NaCo data, because \citet{Baraffe2015} does not include NaCo L'-filters. The mass limit derived from the IFS data assumes a companion-to-star contrast which is constant with wavelength as a conservative choice, for which we already run into the lower mass limits of the used models. The mass limits (see bottom panel of Fig.~\ref{fig:contrast}) constrain quite well the presence of additional components in the system. The SAM data reject > $20 \, \mathrm{M}_\mathrm{J}$ companions at $\sim 2-4$ au, while the IFS data are sensitive to planets more massive than $4\, \mathrm{M}_\mathrm{J}$ beyond 4.5 au and $2 \, \mathrm{M}_\mathrm{J}$ beyond 9 au.

\section{Summary \& conclusions}
\label{sec:conclusion}
Our new study of 51~Eri~b provides new and improved spectra and photometry, and allow to revise the previous flux measurements and to explore new wavelength bands, especially the Y- and K-band. The obtained photometric measurements obtained with SPHERE are: $\mathrm{J}=19.74\pm0.71$ mag, $\mathrm{J3}=18.86\pm0.26$ mag, $\mathrm{H2}=18.56\pm0.28$ mag, and $\mathrm{K1}=17.55\pm0.14$ mag. The broad wavelength coverage was made possible by combining data sets from the very near infrared (Y-band) up to the L'-band allowing us to take a comprehensive view of the planet for the first time, and showed how important thorough knowledge of all features is for understanding and modeling the system. Given 51~Eri~b's separation of $\sim$ 0.5\arcsec, it is very suitable for high-contrast spectral observations, and will become a benchmark object for current and future atmospheric models, especially once further spectra and photometry at longer wavelengths are obtained and all NIR features are mapped in detail. The models produced in this work provide strong predictions on the expected flux and shape of these features and validation or rejection of these predictions will further improve our understanding.\\
In this study, we combined for the first time for SPHERE data, the use of the recently developed ANDROMEDA algorithm to extract an unbiased planetary spectrum with a proper treatment of the spectral covariance and forced photometry, with a state-of-the-art atmospheric model including clouds and varying metallicities combined with a detailed MCMC analysis.\\
We would like to advocate the use of \textit{forced photometry} together with the proper treatment of the noise covariance in the direct imaging community, to use all fluxes obtained at the known position of a planet, even if the flux values obtained are below the detection limit, which is a quantity related to the probability of a previously unknown point source to be a real detection and cannot be directly applied as a cut-off threshold for an already known object. Furthermore, the usage of empirical covariances for IFS spectra can be used to take care of the relative weighting of spectral and photometric data, without having to rely on ad-hoc weighting factors to artificially down-weight the spectral data with respect to independent photometric data.
Our best fitting planet parameters for the cloudy models are \printT{}, \printR{}, \printlogg{} (cgs-units), highly super-solar metallicity \printFEH{}, and \printf{}, indicating the presence of a vertically extended, optically thick cloud cover with small particle size. According to our models the planet seems to have an effective temperature above 700~K and thus sulfur chemistry as discussed in \citet{Zahnle2016}, probably does not play a major role. We note that the effective temperature is in general higher compared to \citet{Macintosh2015}. The new parameters are suggestive of a higher mass for the planet than previously thought. The high surface gravity at a radius slightly bigger than Jupiter's is consistent with a high mass planet \printMlogg{}, whereas the formation model that we consider is compatible with a wide range of masses depending on the initial conditions and does not strongly constrain the mass. Assuming the model atmosphere derived mass would mean that we can reject a pure hot and pure cold start model. However, if 51~Eri~b were in the brown dwarf mass regime we would expect to see a higher radius if 51~Eri~b due to deuterium burning, which makes this scenario unlikely. This discussion highlights the immense difficult of precise mass determinations using direct imaging.\\
Tests performed for patchy-cloud models showed that they do not improve the result significantly for the data used in comparison with a model of uniform cloud coverage and at this point do not seem to justify the increase in model complexity which comes with the linear combination of clear and cloudy models. Further tests should be done to explore patchiness in detail, for example using Bayesian evidence in the model comparison to account for overfitting and complexity. The consistency of the H-band flux over three independent measurements, speaks against strong variability in the J-band, but to conclusively answer this question additional data is necessary. If there truly is variability, a more complex model, like a patchy-cloud model will become necessary to explain the data.
We note that there is a strong need to consider super-solar metallicities in models of exoplanet atmospheres, beyond what is currently done. Beyond the characterization of the planet itself, neglecting super-solar metallicity will impact thermal evolution models of exoplanet, and consequently also limits placed on the occurrence rate of planets of certain mass through direct imaging. This impact will be especially noticeable if observations are performed in the K-band.\\
The empirical comparison to other sub-stellar objects confirmed the peculiarity of 51~Eri~b. It is located in a unique place in the color-color and color-magnitude diagrams, which may be related to low-surface gravity and/or young age effects, but also shares common properties with other late-T dwarfs. The empirical characterization approach is limited by a lack of comparable objects from clusters and young moving groups with spectral type later than T5.
Finally, no additional point sources have been detected in the data. However, the SPHERE/IFS observations together with the NACO/SAM data provide strong constraints on the existence of additional objects in the system, rejecting > $20 \, \mathrm{M}_\mathrm{J}$ companions at $\sim 2-4$ au and planets more massive than $4\, \mathrm{M}_\mathrm{J}$ beyond 4.5 au and $2 \, \mathrm{M}_\mathrm{J}$ beyond 9 au.\\

Future IFS observations in the Y-, K-, and L-band, with existing and upcoming instruments (e.g., SPHERE, GPI, ALES, CHARIS), as well as photometric measurements with JWST/NIRCAM and MIRI at mid-infrared wavelengths, can significantly improve the constraints on the atmospheric parameters. This will make 51~Eri~b one of the planets with the best spectral coverage available and serve as a benchmark for atmospheric model development.

\begin{acknowledgements}
We acknowledge financial support from the Programme National de Planétologie (PNP) and the Programme National de Physique Stellaire (PNPS) of CNRS-INSU. This work has also been supported by a grant from the French Labex OSUG@2020 (Investissements d’avenir – ANR10 LABX56). The project is supported by CNRS, by the Agence Nationale de la Recherche (ANR-14-CE33-0018). This work is partly based on data products produced at the SPHERE Data Centre hosted at OSUG/IPAG, Grenoble. H. A. acknowledges support from the Millennium Science Initiative (Chilean Ministry of Economy) through grant RC130007 and from FONDECYT grant 3150643. C. M. acknowledges the support of the Swiss National Science Foundation via grant BSSGI0\_155816 “PlanetsInTime”. J. Carson and D. Melnick were supported by the South Carolina Space Grant Consortium. We thank P. Delorme and E. Lagadec (SPHERE Data Centre) for their efficient help during the data reduction process. It also made use of the SPHERE Target Database developed and supported by CESAM/LAM. SPHERE is an instrument designed and built by a consortium consisting of IPAG (Grenoble, France), MPIA (Heidelberg, Germany), LAM (Marseille, France), LESIA (Paris, France), Laboratoire Lagrange (Nice, France), INAF–Osservatorio di Padova (Italy), Observatoire de Genève (Switzerland), ETH Zurich (Switzerland), NOVA (Netherlands), ONERA (France) and ASTRON (Netherlands) in collaboration with ESO. SPHERE was funded by ESO, with additional contributions from CNRS (France), MPIA (Germany), INAF (Italy), FINES (Switzerland) and NOVA (Netherlands). SPHERE also received funding from the European Commission Sixth and Seventh Framework Programmes as part of the Optical Infrared Coordination Network for Astronomy (OPTICON) under grant number RII3-Ct-2004-001566 for FP6 (2004–2008), grant number 226604 for FP7 (2009–2012) and grant number 312430 for FP7 (2013–2016).
\end{acknowledgements}

%
    \bibliographystyle{aa} 
    \bibliography{bibliography} 

\begin{thebibliography}{131}
\expandafter\ifx\csname natexlab\endcsname\relax\def\natexlab#1{#1}\fi

\bibitem[{{Ackerman} \& {Marley}(2001)}]{ackermanmarley2001}
{Ackerman}, A.~S. \& {Marley}, M.~S. 2001, \apj, 556, 872

\bibitem[{{Allard} {et~al.}(2001){Allard}, {Hauschildt}, {Alexander},
  {Tamanai}, \& {Schweitzer}}]{2001ApJ...556..357A}
{Allard}, F., {Hauschildt}, P.~H., {Alexander}, D.~R., {Tamanai}, A., \&
  {Schweitzer}, A. 2001, \apj, 556, 357

\bibitem[{{Allard} {et~al.}(2012){Allard}, {Homeier}, \&
  {Freytag}}]{Allard2012}
{Allard}, F., {Homeier}, D., \& {Freytag}, B. 2012, Philosophical Transactions
  of the Royal Society of London Series A, 370, 2765

\bibitem[{{Allers} \& {Liu}(2013)}]{2013ApJ...772...79A}
{Allers}, K.~N. \& {Liu}, M.~C. 2013, \apj, 772, 79

\bibitem[{{Amara} \& {Quanz}(2012)}]{2012MNRAS.427..948A}
{Amara}, A. \& {Quanz}, S.~P. 2012, \mnras, 427, 948

\bibitem[{{Apai} {et~al.}(2016){Apai}, {Kasper}, {Skemer}, {Hanson},
  {Lagrange}, {Biller}, {Bonnefoy}, {Buenzli}, \& {Vigan}}]{Apai2016}
{Apai}, D., {Kasper}, M., {Skemer}, A., {et~al.} 2016, \apj, 820, 40

\bibitem[{{Artigau} {et~al.}(2003){Artigau}, {Nadeau}, \&
  {Doyon}}]{2003IAUS..211..451A}
{Artigau}, E., {Nadeau}, D., \& {Doyon}, R. 2003, in IAU Symposium, Vol. 211,
  Brown Dwarfs, ed. E.~{Mart{\'{\i}}n}, 451

\bibitem[{{Bailey} {et~al.}(2014){Bailey}, {Meshkat}, {Reiter}, {Morzinski},
  {Males}, {Su}, {Hinz}, {Kenworthy}, {Stark}, {Mamajek}, {Briguglio}, {Close},
  {Follette}, {Puglisi}, {Rodigas}, {Weinberger}, \&
  {Xompero}}]{2014ApJ...780L...4B}
{Bailey}, V., {Meshkat}, T., {Reiter}, M., {et~al.} 2014, \apjl, 780, L4

\bibitem[{{Baraffe} {et~al.}(2003){Baraffe}, {Chabrier}, {Barman}, {Allard}, \&
  {Hauschildt}}]{Baraffe2003}
{Baraffe}, I., {Chabrier}, G., {Barman}, T.~S., {Allard}, F., \& {Hauschildt},
  P.~H. 2003, \aap, 402, 701

\bibitem[{{Baraffe} {et~al.}(2015){Baraffe}, {Homeier}, {Allard}, \&
  {Chabrier}}]{Baraffe2015}
{Baraffe}, I., {Homeier}, D., {Allard}, F., \& {Chabrier}, G. 2015, \aap, 577,
  A42

\bibitem[{{Baudino} {et~al.}(2015){Baudino}, {B{\'e}zard}, {Boccaletti},
  {Bonnefoy}, {Lagrange}, \& {Galicher}}]{Baudino2015}
{Baudino}, J.-L., {B{\'e}zard}, B., {Boccaletti}, A., {et~al.} 2015, \aap, 582,
  A83

\bibitem[{{Bell} {et~al.}(2015){Bell}, {Mamajek}, \& {Naylor}}]{Bell2015}
{Bell}, C.~P.~M., {Mamajek}, E.~E., \& {Naylor}, T. 2015, \mnras, 454, 593

\bibitem[{{Best} {et~al.}(2015){Best}, {Liu}, {Magnier}, {Deacon}, {Aller},
  {Redstone}, {Burgett}, {Chambers}, {Draper}, {Flewelling}, {Hodapp},
  {Kaiser}, {Metcalfe}, {Tonry}, {Wainscoat}, \&
  {Waters}}]{2015ApJ...814..118B}
{Best}, W.~M.~J., {Liu}, M.~C., {Magnier}, E.~A., {et~al.} 2015, \apj, 814, 118

\bibitem[{{Beuzit} {et~al.}(2008){Beuzit}, {Feldt}, {Dohlen}, {Mouillet},
  {Puget}, {Wildi}, {Abe}, {Antichi}, {Baruffolo}, {Baudoz}, {Boccaletti},
  {Carbillet}, {Charton}, {Claudi}, {Downing}, {Fabron}, {Feautrier},
  {Fedrigo}, {Fusco}, {Gach}, {Gratton}, {Henning}, {Hubin}, {Joos}, {Kasper},
  {Langlois}, {Lenzen}, {Moutou}, {Pavlov}, {Petit}, {Pragt}, {Rabou}, {Rigal},
  {Roelfsema}, {Rousset}, {Saisse}, {Schmid}, {Stadler}, {Thalmann}, {Turatto},
  {Udry}, {Vakili}, \& {Waters}}]{Beuzit2008}
{Beuzit}, J.-L., {Feldt}, M., {Dohlen}, K., {et~al.} 2008, in \procspie, Vol.
  7014, Ground-based and Airborne Instrumentation for Astronomy II, 701418

\bibitem[{{Biller} {et~al.}(2015){Biller}, {Vos}, {Bonavita}, {Buenzli},
  {Baxter}, {Crossfield}, {Allers}, {Liu}, {Bonnefoy}, {Deacon}, {Brandner},
  {Schlieder}, {Dupuy}, {Kopytova}, {Manjavacas}, {Allard}, {Homeier}, \&
  {Henning}}]{Biller2015}
{Biller}, B.~A., {Vos}, J., {Bonavita}, M., {et~al.} 2015, \apjl, 813, L23

\bibitem[{{Binks} \& {Jeffries}(2014)}]{Binks2014}
{Binks}, A.~S. \& {Jeffries}, R.~D. 2014, \mnras, 438, L11

\bibitem[{{Boccaletti} {et~al.}(2008){Boccaletti}, {Abe}, {Baudrand}, {Daban},
  {Douet}, {Guerri}, {Robbe-Dubois}, {Bendjoya}, {Dohlen}, \&
  {Mawet}}]{Boccaletti2008a}
{Boccaletti}, A., {Abe}, L., {Baudrand}, J., {et~al.} 2008, in Society of
  Photo-Optical Instrumentation Engineers (SPIE) Conference Series, Vol. 7015,
  Society of Photo-Optical Instrumentation Engineers (SPIE) Conference Series,
  1

\bibitem[{{Bohlin}(2007)}]{2007ASPC..364..315B}
{Bohlin}, R.~C. 2007, in Astronomical Society of the Pacific Conference Series,
  Vol. 364, The Future of Photometric, Spectrophotometric and Polarimetric
  Standardization, ed. C.~{Sterken}, 315

\bibitem[{{Bonnefoy} {et~al.}(2016){Bonnefoy}, {Zurlo}, {Baudino}, {Lucas},
  {Mesa}, {Maire}, {Vigan}, {Galicher}, {Homeier}, {Marocco}, {Gratton},
  {Chauvin}, {Allard}, {Desidera}, {Kasper}, {Moutou}, {Lagrange}, {Antichi},
  {Baruffolo}, {Baudrand}, {Beuzit}, {Boccaletti}, {Cantalloube}, {Carbillet},
  {Charton}, {Claudi}, {Costille}, {Dohlen}, {Dominik}, {Fantinel},
  {Feautrier}, {Feldt}, {Fusco}, {Gigan}, {Girard}, {Gluck}, {Gry}, {Henning},
  {Janson}, {Langlois}, {Madec}, {Magnard}, {Maurel}, {Mawet}, {Meyer},
  {Milli}, {Moeller-Nilsson}, {Mouillet}, {Pavlov}, {Perret}, {Pujet}, {Quanz},
  {Rochat}, {Rousset}, {Roux}, {Salasnich}, {Salter}, {Sauvage}, {Schmid},
  {Sevin}, {Soenke}, {Stadler}, {Turatto}, {Udry}, {Vakili}, {Wahhaj}, \&
  {Wildi}}]{2016A&A...587A..58B}
{Bonnefoy}, M., {Zurlo}, A., {Baudino}, J.~L., {et~al.} 2016, \aap, 587, A58

\bibitem[{{Borysow}(1991)}]{1991Icar...92..273B}
{Borysow}, A. 1991, \icarus, 92, 273

\bibitem[{{Burgasser}(2007)}]{Burgasser2007}
{Burgasser}, A.~J. 2007, \aj, 134, 1330

\bibitem[{{Burgasser}(2014)}]{2014ASInC..11....7B}
{Burgasser}, A.~J. 2014, in Astronomical Society of India Conference Series,
  Vol.~11, Astronomical Society of India Conference Series

\bibitem[{{Burgasser} {et~al.}(2006{\natexlab{a}}){Burgasser}, {Burrows}, \&
  {Kirkpatrick}}]{Burgasser2006}
{Burgasser}, A.~J., {Burrows}, A., \& {Kirkpatrick}, J.~D. 2006{\natexlab{a}},
  \apj, 639, 1095

\bibitem[{{Burgasser} {et~al.}(2010){Burgasser}, {Cruz}, {Cushing}, {Gelino},
  {Looper}, {Faherty}, {Kirkpatrick}, \& {Reid}}]{2010ApJ...710.1142B}
{Burgasser}, A.~J., {Cruz}, K.~L., {Cushing}, M., {et~al.} 2010, \apj, 710,
  1142

\bibitem[{{Burgasser} {et~al.}(2011){Burgasser}, {Cushing}, {Kirkpatrick},
  {Gelino}, {Griffith}, {Looper}, {Tinney}, {Simcoe}, {Bochanski}, {Skrutskie},
  {Mainzer}, {Thompson}, {Marsh}, {Bauer}, \& {Wright}}]{2011ApJ...735..116B}
{Burgasser}, A.~J., {Cushing}, M.~C., {Kirkpatrick}, J.~D., {et~al.} 2011,
  \apj, 735, 116

\bibitem[{{Burgasser} {et~al.}(2006{\natexlab{b}}){Burgasser}, {Geballe},
  {Leggett}, {Kirkpatrick}, \& {Golimowski}}]{Burgasser2006b}
{Burgasser}, A.~J., {Geballe}, T.~R., {Leggett}, S.~K., {Kirkpatrick}, J.~D.,
  \& {Golimowski}, D.~A. 2006{\natexlab{b}}, \apj, 637, 1067

\bibitem[{{Burgasser} {et~al.}(2006{\natexlab{c}}){Burgasser}, {Kirkpatrick},
  {Cruz}, {Reid}, {Leggett}, {Liebert}, {Burrows}, \&
  {Brown}}]{2006ApJS..166..585B}
{Burgasser}, A.~J., {Kirkpatrick}, J.~D., {Cruz}, K.~L., {et~al.}
  2006{\natexlab{c}}, \apjs, 166, 585

\bibitem[{{Burgasser} {et~al.}(2000){Burgasser}, {Kirkpatrick}, {Reid},
  {Liebert}, {Gizis}, \& {Brown}}]{2000AJ....120..473B}
{Burgasser}, A.~J., {Kirkpatrick}, J.~D., {Reid}, I.~N., {et~al.} 2000, \aj,
  120, 473

\bibitem[{{Burningham} {et~al.}(2011){Burningham}, {Leggett}, {Homeier},
  {Saumon}, {Lucas}, {Pinfield}, {Tinney}, {Allard}, {Marley}, {Jones},
  {Murray}, {Ishii}, {Day-Jones}, {Gomes}, \& {Zhang}}]{2011MNRAS.414.3590B}
{Burningham}, B., {Leggett}, S.~K., {Homeier}, D., {et~al.} 2011, \mnras, 414,
  3590

\bibitem[{{Burningham} {et~al.}(2008){Burningham}, {Pinfield}, {Leggett},
  {Tamura}, {Lucas}, {Homeier}, {Day-Jones}, {Jones}, {Clarke}, {Ishii},
  {Kuzuhara}, {Lodieu}, {Zapatero Osorio}, {Venemans}, {Mortlock}, {Barrado Y
  Navascu{\'e}s}, {Martin}, \& {Magazz{\`u}}}]{Burningham2008}
{Burningham}, B., {Pinfield}, D.~J., {Leggett}, S.~K., {et~al.} 2008, \mnras,
  391, 320

\bibitem[{{Cantalloube} {et~al.}(2015){Cantalloube}, {Mouillet}, {Mugnier},
  {Milli}, {Absil}, {Gomez Gonzalez}, {Chauvin}, {Beuzit}, \&
  {Cornia}}]{Cantaloube2015}
{Cantalloube}, F., {Mouillet}, D., {Mugnier}, L.~M., {et~al.} 2015, ArXiv
  e-prints [\eprint[arXiv]{1508.06406}]

\bibitem[{{Casagrande} {et~al.}(2011){Casagrande}, {Sch{\"o}nrich}, {Asplund},
  {Cassisi}, {Ram{\'{\i}}rez}, {Mel{\'e}ndez}, {Bensby}, \&
  {Feltzing}}]{Casagrande2011}
{Casagrande}, L., {Sch{\"o}nrich}, R., {Asplund}, M., {et~al.} 2011, \aap, 530,
  A138

\bibitem[{{Chabrier} {et~al.}(2009){Chabrier}, {Baraffe}, {Leconte},
  {Gallardo}, \& {Barman}}]{Chabrier2009}
{Chabrier}, G., {Baraffe}, I., {Leconte}, J., {Gallardo}, J., \& {Barman}, T.
  2009, in American Institute of Physics Conference Series, Vol. 1094, 15th
  Cambridge Workshop on Cool Stars, Stellar Systems, and the Sun, ed.
  E.~{Stempels}, 102--111

\bibitem[{{Chilcote} {et~al.}(2015){Chilcote}, {Barman}, {Fitzgerald},
  {Graham}, {Larkin}, {Macintosh}, {Bauman}, {Burrows}, {Cardwell}, {De Rosa},
  {Dillon}, {Doyon}, {Dunn}, {Erikson}, {Gavel}, {Goodsell}, {Hartung},
  {Hibon}, {Ingraham}, {Kalas}, {Konopacky}, {Maire}, {Marchis}, {Marley},
  {Marois}, {Millar-Blanchaer}, {Morzinski}, {Norton}, {Oppenheimer}, {Palmer},
  {Patience}, {Perrin}, {Poyneer}, {Pueyo}, {Rantakyr{\"o}}, {Sadakuni},
  {Saddlemyer}, {Savransky}, {Serio}, {Sivaramakrishnan}, {Song}, {Soummer},
  {Thomas}, {Wallace}, {Wiktorowicz}, \& {Wolff}}]{Chilcote2015}
{Chilcote}, J., {Barman}, T., {Fitzgerald}, M.~P., {et~al.} 2015, \apjl, 798,
  L3

\bibitem[{{Claudi} {et~al.}(2008){Claudi}, {Turatto}, {Gratton}, {Antichi},
  {Bonavita}, {Bruno}, {Cascone}, {De Caprio}, {Desidera}, {Giro}, {Mesa},
  {Scuderi}, {Dohlen}, {Beuzit}, \& {Puget}}]{Claudi2008}
{Claudi}, R.~U., {Turatto}, M., {Gratton}, R.~G., {et~al.} 2008, in \procspie,
  Vol. 7014, Ground-based and Airborne Instrumentation for Astronomy II, 70143E

\bibitem[{{Cushing} {et~al.}(2008){Cushing}, {Marley}, {Saumon}, {Kelly},
  {Vacca}, {Rayner}, {Freedman}, {Lodders}, \& {Roellig}}]{2008ApJ...678.1372C}
{Cushing}, M.~C., {Marley}, M.~S., {Saumon}, D., {et~al.} 2008, \apj, 678, 1372

\bibitem[{{Cutri} {et~al.}(2003){Cutri}, {Skrutskie}, {van Dyk}, {Beichman},
  {Carpenter}, {Chester}, {Cambresy}, {Evans}, {Fowler}, {Gizis}, {Howard},
  {Huchra}, {Jarrett}, {Kopan}, {Kirkpatrick}, {Light}, {Marsh}, {McCallon},
  {Schneider}, {Stiening}, {Sykes}, {Weinberg}, {Wheaton}, {Wheelock}, \&
  {Zacarias}}]{Cutri2003}
{Cutri}, R.~M., {Skrutskie}, M.~F., {van Dyk}, S., {et~al.} 2003, {2MASS All
  Sky Catalog of point sources.}

\bibitem[{{Cutri} {et~al.}(2013){Cutri}, {Wright}, {Conrow}, {Fowler},
  {Eisenhardt}, {Grillmair}, {Kirkpatrick}, {Masci}, {McCallon}, {Wheelock},
  {Fajardo-Acosta}, {Yan}, {Benford}, {Harbut}, {Jarrett}, {Lake}, {Leisawitz},
  {Ressler}, {Stanford}, {Tsai}, {Liu}, {Helou}, {Mainzer}, {Gettings},
  {Gonzalez}, {Hoffman}, {Marsh}, {Padgett}, {Skrutskie}, {Beck}, {Papin}, \&
  {Wittman}}]{Cutri2013}
{Cutri}, R.~M., {Wright}, E.~L., {Conrow}, T., {et~al.} 2013, {Explanatory
  Supplement to the AllWISE Data Release Products}, Tech. rep.

\bibitem[{{De Rosa} {et~al.}(2015){De Rosa}, {Nielsen}, {Blunt}, {Graham},
  {Konopacky}, {Marois}, {Pueyo}, {Rameau}, {Ryan}, {Wang}, {Bailey},
  {Chontos}, {Fabrycky}, {Follette}, {Macintosh}, {Marchis}, {Ammons},
  {Arriaga}, {Chilcote}, {Cotten}, {Doyon}, {Duch{\^e}ne}, {Esposito},
  {Fitzgerald}, {Gerard}, {Goodsell}, {Greenbaum}, {Hibon}, {Ingraham},
  {Johnson-Groh}, {Kalas}, {Lafreni{\`e}re}, {Maire}, {Metchev},
  {Millar-Blanchaer}, {Morzinski}, {Oppenheimer}, {Patel}, {Patience},
  {Perrin}, {Rajan}, {Rantakyr{\"o}}, {Ruffio}, {Schneider},
  {Sivaramakrishnan}, {Song}, {Tran}, {Vasisht}, {Ward-Duong}, \&
  {Wolff}}]{DeRosa2015}
{De Rosa}, R.~J., {Nielsen}, E.~L., {Blunt}, S.~C., {et~al.} 2015, \apjl, 814,
  L3

\bibitem[{{Delorme} {et~al.}(2017){Delorme}, {Dupuy}, {Gagn{\'e}}, {Reyl{\'e}},
  {Forveille}, {Liu}, {Artigau}, {Albert}, {Delfosse}, {Allard}, {Homeier},
  {Malo}, {Morley}, {Naud}, \& {Bonnefoy}}]{Delorme2017}
{Delorme}, P., {Dupuy}, T., {Gagn{\'e}}, J., {et~al.} 2017, ArXiv e-prints
  [\eprint[arXiv]{1703.00843}]

\bibitem[{{Delorme} {et~al.}(2012){Delorme}, {Gagn{\'e}}, {Malo}, {Reyl{\'e}},
  {Artigau}, {Albert}, {Forveille}, {Delfosse}, {Allard}, \&
  {Homeier}}]{2012A&A...548A..26D}
{Delorme}, P., {Gagn{\'e}}, J., {Malo}, L., {et~al.} 2012, \aap, 548, A26

\bibitem[{{Dohlen} {et~al.}(2008){Dohlen}, {Langlois}, {Saisse}, {Hill},
  {Origne}, {Jacquet}, {Fabron}, {Blanc}, {Llored}, {Carle}, {Moutou}, {Vigan},
  {Boccaletti}, {Carbillet}, {Mouillet}, \& {Beuzit}}]{Dohlen2008}
{Dohlen}, K., {Langlois}, M., {Saisse}, M., {et~al.} 2008, in \procspie, Vol.
  7014, Ground-based and Airborne Instrumentation for Astronomy II, 70143L

\bibitem[{{Faherty} {et~al.}(2012){Faherty}, {Burgasser}, {Walter}, {Van der
  Bliek}, {Shara}, {Cruz}, {West}, {Vrba}, \&
  {Anglada-Escud{\'e}}}]{2012ApJ...752...56F}
{Faherty}, J.~K., {Burgasser}, A.~J., {Walter}, F.~M., {et~al.} 2012, \apj,
  752, 56

\bibitem[{{Faherty} {et~al.}(2010){Faherty}, {Burgasser}, {West}, {Bochanski},
  {Cruz}, {Shara}, \& {Walter}}]{2010AJ....139..176F}
{Faherty}, J.~K., {Burgasser}, A.~J., {West}, A.~A., {et~al.} 2010, \aj, 139,
  176

\bibitem[{{Foreman-Mackey} {et~al.}(2013){Foreman-Mackey}, {Hogg}, {Lang}, \&
  {Goodman}}]{Foreman2013}
{Foreman-Mackey}, D., {Hogg}, D.~W., {Lang}, D., \& {Goodman}, J. 2013, \pasp,
  125, 306

\bibitem[{{Fusco} {et~al.}(2014){Fusco}, {Sauvage}, {Petit}, {Costille},
  {Dohlen}, {Mouillet}, {Beuzit}, {Kasper}, {Suarez}, {Soenke}, {Fedrigo},
  {Downing}, {Baudoz}, {Sevin}, {Perret}, {Barrufolo}, {Salasnich}, {Puget},
  {Feautrier}, {Rochat}, {Moulin}, {Deboulb{\'e}}, {Hugot}, {Vigan}, {Mawet},
  {Girard}, \& {Hubin}}]{Fusco2014}
{Fusco}, T., {Sauvage}, J.-F., {Petit}, C., {et~al.} 2014, in \procspie, Vol.
  9148, Adaptive Optics Systems IV, 91481U

\bibitem[{{Gagn{\'e}} {et~al.}(2015){Gagn{\'e}}, {Burgasser}, {Faherty},
  {Lafreni{\'e}re}, {Doyon}, {Filippazzo}, {Bowsher}, \&
  {Nicholls}}]{2015ApJ...808L..20G}
{Gagn{\'e}}, J., {Burgasser}, A.~J., {Faherty}, J.~K., {et~al.} 2015, \apjl,
  808, L20

\bibitem[{{Gagn{\'e}} {et~al.}(2014){Gagn{\'e}}, {Lafreni{\`e}re}, {Doyon},
  {Malo}, \& {Artigau}}]{2014ApJ...783..121G}
{Gagn{\'e}}, J., {Lafreni{\`e}re}, D., {Doyon}, R., {Malo}, L., \& {Artigau},
  {\'E}. 2014, \apj, 783, 121

\bibitem[{Goodman \& Weare(2010)}]{Goodman2010}
Goodman, J. \& Weare, J. 2010, Communications in Applied Mathematics and
  Computational Science, 5, 65

\bibitem[{{Greco} \& {Brandt}(2016)}]{Greco2016}
{Greco}, J.~P. \& {Brandt}, T.~D. 2016, ArXiv e-prints
  [\eprint[arXiv]{1602.00691}]

\bibitem[{{Helou} \& {Walker}(1988)}]{Helou1988}
{Helou}, G. \& {Walker}, D.~W., eds. 1988, {Infrared astronomical satellite
  (IRAS) catalogs and atlases. Volume 7: The small scale structure catalog},
  Vol.~7, 1--265

\bibitem[{{Henning} \& {Stognienko}(1996)}]{henningstognienko1996}
{Henning}, T. \& {Stognienko}, R. 1996, \aap, 311, 291

\bibitem[{{Hiranaka} {et~al.}(2016){Hiranaka}, {Cruz}, {Douglas}, {Marley}, \&
  {Baldassare}}]{2016arXiv160609485H}
{Hiranaka}, K., {Cruz}, K.~L., {Douglas}, S.~T., {Marley}, M.~S., \&
  {Baldassare}, V.~F. 2016, ArXiv e-prints [\eprint[arXiv]{1606.09485}]

\bibitem[{{Hoeg} {et~al.}(1997){Hoeg}, {B{\"a}ssgen}, {Bastian}, {Egret},
  {Fabricius}, {Gro{\ss}mann}, {Halbwachs}, {Makarov}, {Perryman},
  {Schwekendiek}, {Wagner}, \& {Wicenec}}]{Hoeg1997}
{Hoeg}, E., {B{\"a}ssgen}, G., {Bastian}, U., {et~al.} 1997, \aap, 323

\bibitem[{{Ingraham} {et~al.}(2014){Ingraham}, {Marley}, {Saumon}, {Marois},
  {Macintosh}, {Barman}, {Bauman}, {Burrows}, {Chilcote}, {De Rosa}, {Dillon},
  {Doyon}, {Dunn}, {Erikson}, {Fitzgerald}, {Gavel}, {Goodsell}, {Graham},
  {Hartung}, {Hibon}, {Kalas}, {Konopacky}, {Larkin}, {Maire}, {Marchis},
  {McBride}, {Millar-Blanchaer}, {Morzinski}, {Norton}, {Oppenheimer},
  {Palmer}, {Patience}, {Perrin}, {Poyneer}, {Pueyo}, {Rantakyr{\"o}},
  {Sadakuni}, {Saddlemyer}, {Savransky}, {Soummer}, {Sivaramakrishnan}, {Song},
  {Thomas}, {Wallace}, {Wiktorowicz}, \& {Wolff}}]{Ingraham2014}
{Ingraham}, P., {Marley}, M.~S., {Saumon}, D., {et~al.} 2014, \apjl, 794, L15

\bibitem[{{Jaeger} {et~al.}(1998){Jaeger}, {Molster}, {Dorschner}, {Henning},
  {Mutschke}, \& {Waters}}]{jaegermolster1998}
{Jaeger}, C., {Molster}, F.~J., {Dorschner}, J., {et~al.} 1998, \aap, 339, 904

\bibitem[{{Jones} {et~al.}(2013){Jones}, {Noll}, {Kausch}, {Szyszka}, \&
  {Kimeswenger}}]{2013A&A...560A..91J}
{Jones}, A., {Noll}, S., {Kausch}, W., {Szyszka}, C., \& {Kimeswenger}, S.
  2013, \aap, 560, A91

\bibitem[{{Kao} {et~al.}(2016){Kao}, {Hallinan}, {Pineda}, {Escala},
  {Burgasser}, {Bourke}, \& {Stevenson}}]{2016ApJ...818...24K}
{Kao}, M.~M., {Hallinan}, G., {Pineda}, J.~S., {et~al.} 2016, \apj, 818, 24

\bibitem[{{King} {et~al.}(2010){King}, {McCaughrean}, {Homeier}, {Allard},
  {Scholz}, \& {Lodieu}}]{King2010}
{King}, R.~R., {McCaughrean}, M.~J., {Homeier}, D., {et~al.} 2010, \aap, 510,
  A99

\bibitem[{{Kraus} {et~al.}(2008){Kraus}, {Ireland}, {Martinache}, \&
  {Lloyd}}]{Kraus2008}
{Kraus}, A.~L., {Ireland}, M.~J., {Martinache}, F., \& {Lloyd}, J.~P. 2008,
  \apj, 679, 762

\bibitem[{{Lagrange} {et~al.}(2010){Lagrange}, {Bonnefoy}, {Chauvin}, {Apai},
  {Ehrenreich}, {Boccaletti}, {Gratadour}, {Rouan}, {Mouillet}, {Lacour}, \&
  {Kasper}}]{Lagrange2010}
{Lagrange}, A.-M., {Bonnefoy}, M., {Chauvin}, G., {et~al.} 2010, Science, 329,
  57

\bibitem[{{Lang} {et~al.}(2016){Lang}, {Hogg}, \& {Schlegel}}]{Lang2016}
{Lang}, D., {Hogg}, D.~W., \& {Schlegel}, D.~J. 2016, \aj, 151, 36

\bibitem[{{Langlois} {et~al.}(2013){Langlois}, {Vigan}, {Moutou}, {Sauvage},
  {Dohlen}, {Costille}, {Mouillet}, \& {Le Mignant}}]{Langlois2013}
{Langlois}, M., {Vigan}, A., {Moutou}, C., {et~al.} 2013, in Proceedings of the
  Third AO4ELT Conference, ed. S.~{Esposito} \& L.~{Fini}, 63

\bibitem[{{Lee} {et~al.}(2013){Lee}, {Heng}, \& {Irwin}}]{2013ApJ...778...97L}
{Lee}, J.-M., {Heng}, K., \& {Irwin}, P.~G.~J. 2013, \apj, 778, 97

\bibitem[{{Leggett} {et~al.}(2000){Leggett}, {Allard}, {Dahn}, {Hauschildt},
  {Kerr}, \& {Rayner}}]{2000ApJ...535..965L}
{Leggett}, S.~K., {Allard}, F., {Dahn}, C., {et~al.} 2000, \apj, 535, 965

\bibitem[{{Line} {et~al.}(2016){Line}, {Marley}, {Liu}, {Morley}, {Burningham},
  {Hinkel}, {Teske}, \& {Fortney}}]{Line2016}
{Line}, M.~R., {Marley}, M.~S., {Liu}, M.~C., {et~al.} 2016, ArXiv e-prints
  [\eprint[arXiv]{1612.02809}]

\bibitem[{{Line} {et~al.}(2015){Line}, {Teske}, {Burningham}, {Fortney}, \&
  {Marley}}]{Line2015}
{Line}, M.~R., {Teske}, J., {Burningham}, B., {Fortney}, J.~J., \& {Marley},
  M.~S. 2015, \apj, 807, 183

\bibitem[{{Liu} {et~al.}(2016){Liu}, {Dupuy}, \& {Allers}}]{Liu2016}
{Liu}, M.~C., {Dupuy}, T.~J., \& {Allers}, K.~N. 2016, \apj, 833, 96

\bibitem[{{Liu} {et~al.}(2007){Liu}, {Leggett}, \& {Chiu}}]{Liu2007}
{Liu}, M.~C., {Leggett}, S.~K., \& {Chiu}, K. 2007, \apj, 660, 1507

\bibitem[{{Mace} {et~al.}(2013){Mace}, {Kirkpatrick}, {Cushing}, {Gelino},
  {Griffith}, {Skrutskie}, {Marsh}, {Wright}, {Eisenhardt}, {McLean},
  {Thompson}, {Mix}, {Bailey}, {Beichman}, {Bloom}, {Burgasser}, {Fortney},
  {Hinz}, {Knox}, {Lowrance}, {Marley}, {Morley}, {Rodigas}, {Saumon},
  {Sheppard}, \& {Stock}}]{2013ApJS..205....6M}
{Mace}, G.~N., {Kirkpatrick}, J.~D., {Cushing}, M.~C., {et~al.} 2013, \apjs,
  205, 6

\bibitem[{{Macintosh} {et~al.}(2015){Macintosh}, {Graham}, {Barman}, {De Rosa},
  {Konopacky}, {Marley}, {Marois}, {Nielsen}, {Pueyo}, {Rajan}, {Rameau},
  {Saumon}, {Wang}, {Patience}, {Ammons}, {Arriaga}, {Artigau}, {Beckwith},
  {Brewster}, {Bruzzone}, {Bulger}, {Burningham}, {Burrows}, {Chen}, {Chiang},
  {Chilcote}, {Dawson}, {Dong}, {Doyon}, {Draper}, {Duch{\^e}ne}, {Esposito},
  {Fabrycky}, {Fitzgerald}, {Follette}, {Fortney}, {Gerard}, {Goodsell},
  {Greenbaum}, {Hibon}, {Hinkley}, {Cotten}, {Hung}, {Ingraham},
  {Johnson-Groh}, {Kalas}, {Lafreniere}, {Larkin}, {Lee}, {Line}, {Long},
  {Maire}, {Marchis}, {Matthews}, {Max}, {Metchev}, {Millar-Blanchaer},
  {Mittal}, {Morley}, {Morzinski}, {Murray-Clay}, {Oppenheimer}, {Palmer},
  {Patel}, {Perrin}, {Poyneer}, {Rafikov}, {Rantakyr{\"o}}, {Rice}, {Rojo},
  {Rudy}, {Ruffio}, {Ruiz}, {Sadakuni}, {Saddlemyer}, {Salama}, {Savransky},
  {Schneider}, {Sivaramakrishnan}, {Song}, {Soummer}, {Thomas}, {Vasisht},
  {Wallace}, {Ward-Duong}, {Wiktorowicz}, {Wolff}, \&
  {Zuckerman}}]{Macintosh2015}
{Macintosh}, B., {Graham}, J.~R., {Barman}, T., {et~al.} 2015, Science, 350, 64

\bibitem[{{Macintosh} {et~al.}(2014){Macintosh}, {Graham}, {Ingraham},
  {Konopacky}, {Marois}, {Perrin}, {Poyneer}, {Bauman}, {Barman}, {Burrows},
  {Cardwell}, {Chilcote}, {De Rosa}, {Dillon}, {Doyon}, {Dunn}, {Erikson},
  {Fitzgerald}, {Gavel}, {Goodsell}, {Hartung}, {Hibon}, {Kalas}, {Larkin},
  {Maire}, {Marchis}, {Marley}, {McBride}, {Millar-Blanchaer}, {Morzinski},
  {Norton}, {Oppenheimer}, {Palmer}, {Patience}, {Pueyo}, {Rantakyro},
  {Sadakuni}, {Saddlemyer}, {Savransky}, {Serio}, {Soummer},
  {Sivaramakrishnan}, {Song}, {Thomas}, {Wallace}, {Wiktorowicz}, \&
  {Wolff}}]{Macintosh2014}
{Macintosh}, B., {Graham}, J.~R., {Ingraham}, P., {et~al.} 2014, Proceedings of
  the National Academy of Science, 111, 12661

\bibitem[{{Maire} {et~al.}(2014){Maire}, {Boccaletti}, {Rameau}, {Chauvin},
  {Lagrange}, {Bonnefoy}, {Desidera}, {Sylvestre}, {Baudoz}, {Galicher}, \&
  {Mouillet}}]{Maire2014}
{Maire}, A.-L., {Boccaletti}, A., {Rameau}, J., {et~al.} 2014, \aap, 566, A126

\bibitem[{{Maire} {et~al.}(2016{\natexlab{a}}){Maire}, {Bonnefoy}, {Ginski},
  {Vigan}, {Messina}, {Mesa}, {Galicher}, {Gratton}, {Desidera}, {Kopytova},
  {Millward}, {Thalmann}, {Claudi}, {Ehrenreich}, {Zurlo}, {Chauvin},
  {Antichi}, {Baruffolo}, {Bazzon}, {Beuzit}, {Blanchard}, {Boccaletti}, {de
  Boer}, {Carle}, {Cascone}, {Costille}, {De Caprio}, {Delboulb{\'e}},
  {Dohlen}, {Dominik}, {Feldt}, {Fusco}, {Girard}, {Giro}, {Gisler}, {Gluck},
  {Gry}, {Henning}, {Hubin}, {Hugot}, {Jaquet}, {Kasper}, {Lagrange},
  {Langlois}, {Le Mignant}, {Llored}, {Madec}, {Martinez}, {Mawet}, {Milli},
  {M{\"o}ller-Nilsson}, {Mouillet}, {Moulin}, {Moutou}, {Orign{\'e}}, {Pavlov},
  {Petit}, {Pragt}, {Puget}, {Ramos}, {Rochat}, {Roelfsema}, {Salasnich},
  {Sauvage}, {Schmid}, {Turatto}, {Udry}, {Vakili}, {Wahhaj}, {Weber}, \&
  {Wildi}}]{Maire2016}
{Maire}, A.-L., {Bonnefoy}, M., {Ginski}, C., {et~al.} 2016{\natexlab{a}},
  \aap, 587, A56

\bibitem[{{Maire} {et~al.}(2016{\natexlab{b}}){Maire}, {Langlois}, {Dohlen},
  {Lagrange}, {Gratton}, {Chauvin}, {Desidera}, {Girard}, {Milli}, {Vigan},
  {Zins}, {Delorme}, {Beuzit}, {Claudi}, {Feldt}, {Mouillet}, {Puget},
  {Turatto}, \& {Wildi}}]{Maire2016b}
{Maire}, A.-L., {Langlois}, M., {Dohlen}, K., {et~al.} 2016{\natexlab{b}}, in
  SPIE Conf. Ser., Vol. 9908, 990834

\bibitem[{{Mamajek} \& {Bell}(2014)}]{Mamajek2014}
{Mamajek}, E.~E. \& {Bell}, C.~P.~M. 2014, \mnras, 445, 2169

\bibitem[{{Mancini} {et~al.}(2016{\natexlab{a}}){Mancini}, {Giordano},
  {Molli{\`e}re}, {Southworth}, {Brahm}, {Ciceri}, \& {Henning}}]{mancini2016b}
{Mancini}, L., {Giordano}, M., {Molli{\`e}re}, P., {et~al.} 2016{\natexlab{a}},
  \mnras, 461, 1053

\bibitem[{{Mancini} {et~al.}(2016{\natexlab{b}}){Mancini}, {Kemmer},
  {Southworth}, {Bott}, {Molli{\`e}re}, {Ciceri}, {Chen}, \&
  {Henning}}]{mancinikemmer2016a}
{Mancini}, L., {Kemmer}, J., {Southworth}, J., {et~al.} 2016{\natexlab{b}},
  \mnras [\eprint[arXiv]{1603.08031}]

\bibitem[{{Marley} {et~al.}(2007){Marley}, {Fortney}, {Hubickyj},
  {Bodenheimer}, \& {Lissauer}}]{Marley2007}
{Marley}, M.~S., {Fortney}, J.~J., {Hubickyj}, O., {Bodenheimer}, P., \&
  {Lissauer}, J.~J. 2007, \apj, 655, 541

\bibitem[{{Marley} {et~al.}(2010){Marley}, {Saumon}, \&
  {Goldblatt}}]{Marley2010}
{Marley}, M.~S., {Saumon}, D., \& {Goldblatt}, C. 2010, \apjl, 723, L117

\bibitem[{{Marocco} {et~al.}(2014){Marocco}, {Day-Jones}, {Lucas}, {Jones},
  {Smart}, {Zhang}, {Gomes}, {Burningham}, {Pinfield}, {Raddi}, \&
  {Smith}}]{2014MNRAS.439..372M}
{Marocco}, F., {Day-Jones}, A.~C., {Lucas}, P.~W., {et~al.} 2014, \mnras, 439,
  372

\bibitem[{{Marois} {et~al.}(2014){Marois}, {Correia}, {Galicher}, {Ingraham},
  {Macintosh}, {Currie}, \& {De Rosa}}]{Marois2014}
{Marois}, C., {Correia}, C., {Galicher}, R., {et~al.} 2014, in Society of
  Photo-Optical Instrumentation Engineers (SPIE) Conference Series, Vol. 9148,
  Society of Photo-Optical Instrumentation Engineers (SPIE) Conference Series,
  0

\bibitem[{{Marois} {et~al.}(2006){Marois}, {Lafreni{\`e}re}, {Doyon},
  {Macintosh}, \& {Nadeau}}]{Marois2006}
{Marois}, C., {Lafreni{\`e}re}, D., {Doyon}, R., {Macintosh}, B., \& {Nadeau},
  D. 2006, \apj, 641, 556

\bibitem[{{Marois} {et~al.}(2008){Marois}, {Macintosh}, {Barman}, {Zuckerman},
  {Song}, {Patience}, {Lafreni{\`e}re}, \& {Doyon}}]{Marois2008b}
{Marois}, C., {Macintosh}, B., {Barman}, T., {et~al.} 2008, Science, 322, 1348

\bibitem[{{Marois} {et~al.}(2010{\natexlab{a}}){Marois}, {Macintosh}, \&
  {V{\'e}ran}}]{Marois2010}
{Marois}, C., {Macintosh}, B., \& {V{\'e}ran}, J.-P. 2010{\natexlab{a}}, in
  Society of Photo-Optical Instrumentation Engineers (SPIE) Conference Series,
  Vol. 7736, Society of Photo-Optical Instrumentation Engineers (SPIE)
  Conference Series, 1

\bibitem[{{Marois} {et~al.}(2010{\natexlab{b}}){Marois}, {Zuckerman},
  {Konopacky}, {Macintosh}, \& {Barman}}]{Marois2010b}
{Marois}, C., {Zuckerman}, B., {Konopacky}, Q.~M., {Macintosh}, B., \&
  {Barman}, T. 2010{\natexlab{b}}, \nat, 468, 1080

\bibitem[{{Mawet} {et~al.}(2014){Mawet}, {Milli}, {Wahhaj}, {Pelat}, {Absil},
  {Delacroix}, {Boccaletti}, {Kasper}, {Kenworthy}, {Marois}, {Mennesson}, \&
  {Pueyo}}]{Mawet2014}
{Mawet}, D., {Milli}, J., {Wahhaj}, Z., {et~al.} 2014, \apj, 792, 97

\bibitem[{{Mermilliod}(2006)}]{Mermilliod2006}
{Mermilliod}, J.~C. 2006, VizieR Online Data Catalog, 2168

\bibitem[{{Mesa} {et~al.}(2015){Mesa}, {Gratton}, {Zurlo}, {Vigan}, {Claudi},
  {Alberi}, {Antichi}, {Baruffolo}, {Beuzit}, {Boccaletti}, {Bonnefoy},
  {Costille}, {Desidera}, {Dohlen}, {Fantinel}, {Feldt}, {Fusco}, {Giro},
  {Henning}, {Kasper}, {Langlois}, {Maire}, {Martinez}, {Moeller-Nilsson},
  {Mouillet}, {Moutou}, {Pavlov}, {Puget}, {Salasnich}, {Sauvage}, {Sissa},
  {Turatto}, {Udry}, {Vakili}, {Waters}, \& {Wildi}}]{Mesa2015}
{Mesa}, D., {Gratton}, R., {Zurlo}, A., {et~al.} 2015, \aap, 576, A121

\bibitem[{{Molli{\`e}re} {et~al.}(2017){Molli{\`e}re}, {van Boekel}, {Bouwman},
  {Henning}, {Lagage}, \& {Min}}]{Molliere2017}
{Molli{\`e}re}, P., {van Boekel}, R., {Bouwman}, J., {et~al.} 2017, \aap, 600,
  A10

\bibitem[{{Molli{\`e}re} {et~al.}(2015){Molli{\`e}re}, {van Boekel},
  {Dullemond}, {Henning}, \& {Mordasini}}]{mollierevanboekel2015}
{Molli{\`e}re}, P., {van Boekel}, R., {Dullemond}, C., {Henning}, T., \&
  {Mordasini}, C. 2015, \apj, 813, 47

\bibitem[{{Monet} {et~al.}(1992){Monet}, {Dahn}, {Vrba}, {Harris}, {Pier},
  {Luginbuhl}, \& {Ables}}]{1992AJ....103..638M}
{Monet}, D.~G., {Dahn}, C.~C., {Vrba}, F.~J., {et~al.} 1992, \aj, 103, 638

\bibitem[{{Montet} {et~al.}(2015){Montet}, {Bowler}, {Shkolnik}, {Deck},
  {Wang}, {Horch}, {Liu}, {Hillenbrand}, {Kraus}, \&
  {Charbonneau}}]{Montet2015}
{Montet}, B.~T., {Bowler}, B.~P., {Shkolnik}, E.~L., {et~al.} 2015, \apjl, 813,
  L11

\bibitem[{{Mordasini}(2013)}]{Mordasini2013}
{Mordasini}, C. 2013, \aap, 558, A113

\bibitem[{{Mordasini} {et~al.}(2012{\natexlab{a}}){Mordasini}, {Alibert},
  {Georgy}, {Dittkrist}, {Klahr}, \& {Henning}}]{Mordasini2012b}
{Mordasini}, C., {Alibert}, Y., {Georgy}, C., {et~al.} 2012{\natexlab{a}},
  \aap, 547, A112

\bibitem[{{Mordasini} {et~al.}(2012{\natexlab{b}}){Mordasini}, {Alibert},
  {Klahr}, \& {Henning}}]{Mordasini2012a}
{Mordasini}, C., {Alibert}, Y., {Klahr}, H., \& {Henning}, T.
  2012{\natexlab{b}}, \aap, 547, A111

\bibitem[{{Mordasini} {et~al.}(2014){Mordasini}, {Klahr}, {Alibert}, {Miller},
  \& {Henning}}]{Mordasini2014}
{Mordasini}, C., {Klahr}, H., {Alibert}, Y., {Miller}, N., \& {Henning}, T.
  2014, \aap, 566, A141

\bibitem[{{Morley} {et~al.}(2012){Morley}, {Fortney}, {Marley}, {Visscher},
  {Saumon}, \& {Leggett}}]{Morley2012}
{Morley}, C.~V., {Fortney}, J.~J., {Marley}, M.~S., {et~al.} 2012, \apj, 756,
  172

\bibitem[{{Naud} {et~al.}(2014){Naud}, {Artigau}, {Malo}, {Albert}, {Doyon},
  {Lafreni{\`e}re}, {Gagn{\'e}}, {Saumon}, {Morley}, {Allard}, {Homeier},
  {Beichman}, {Gelino}, \& {Boucher}}]{2014ApJ...787....5N}
{Naud}, M.-E., {Artigau}, {\'E}., {Malo}, L., {et~al.} 2014, \apj, 787, 5

\bibitem[{{Noll} {et~al.}(2012){Noll}, {Kausch}, {Barden}, {Jones}, {Szyszka},
  {Kimeswenger}, \& {Vinther}}]{2012A&A...543A..92N}
{Noll}, S., {Kausch}, W., {Barden}, M., {et~al.} 2012, \aap, 543, A92

\bibitem[{Palik(2012)}]{palik2012}
Palik, E. 2012, Handbook of Optical Constants of Solids No. Bd. 1 (Elsevier
  Science)

\bibitem[{{Patel} {et~al.}(2014){Patel}, {Metchev}, \&
  {Heinze}}]{2014ApJS..212...10P}
{Patel}, R.~I., {Metchev}, S.~A., \& {Heinze}, A. 2014, \apjs, 212, 10

\bibitem[{{Pavlov} {et~al.}(2008){Pavlov}, {M{\"o}ller-Nilsson}, {Feldt},
  {Henning}, {Beuzit}, \& {Mouillet}}]{Pavlov2008}
{Pavlov}, A., {M{\"o}ller-Nilsson}, O., {Feldt}, M., {et~al.} 2008, in
  \procspie, Vol. 7019, Advanced Software and Control for Astronomy II, 701939

\bibitem[{{Prugniel} {et~al.}(2007){Prugniel}, {Soubiran}, {Koleva}, \& {Le
  Borgne}}]{Prugniel2007}
{Prugniel}, P., {Soubiran}, C., {Koleva}, M., \& {Le Borgne}, D. 2007, ArXiv
  Astrophysics e-prints [\eprint{astro-ph/0703658}]

\bibitem[{{Racine} {et~al.}(1999){Racine}, {Walker}, {Nadeau}, {Doyon}, \&
  {Marois}}]{1999PASP..111..587R}
{Racine}, R., {Walker}, G.~A.~H., {Nadeau}, D., {Doyon}, R., \& {Marois}, C.
  1999, \pasp, 111, 587

\bibitem[{{Radigan} {et~al.}(2012){Radigan}, {Jayawardhana}, {Lafreni{\`e}re},
  {Artigau}, {Marley}, \& {Saumon}}]{Radigan2012}
{Radigan}, J., {Jayawardhana}, R., {Lafreni{\`e}re}, D., {et~al.} 2012, \apj,
  750, 105

\bibitem[{{Rameau} {et~al.}(2013){Rameau}, {Chauvin}, {Lagrange}, {Boccaletti},
  {Quanz}, {Bonnefoy}, {Girard}, {Delorme}, {Desidera}, {Klahr}, {Mordasini},
  {Dumas}, \& {Bonavita}}]{Rameau2013}
{Rameau}, J., {Chauvin}, G., {Lagrange}, A.-M., {et~al.} 2013, \apjl, 772, L15

\bibitem[{{Ram{\'{\i}}rez} {et~al.}(2013){Ram{\'{\i}}rez}, {Allende Prieto}, \&
  {Lambert}}]{Ramirez2013}
{Ram{\'{\i}}rez}, I., {Allende Prieto}, C., \& {Lambert}, D.~L. 2013, \apj,
  764, 78

\bibitem[{{Riviere-Marichalar} {et~al.}(2014){Riviere-Marichalar}, {Barrado},
  {Montesinos}, {Duch{\^e}ne}, {Bouy}, {Pinte}, {Menard}, {Donaldson}, {Eiroa},
  {Krivov}, {Kamp}, {Mendigut{\'{\i}}a}, {Dent}, \&
  {Lillo-Box}}]{2014A&A...565A..68R}
{Riviere-Marichalar}, P., {Barrado}, D., {Montesinos}, B., {et~al.} 2014, \aap,
  565, A68

\bibitem[{{Saumon} {et~al.}(2006){Saumon}, {Marley}, {Cushing}, {Leggett},
  {Roellig}, {Lodders}, \& {Freedman}}]{Saumon2006}
{Saumon}, D., {Marley}, M.~S., {Cushing}, M.~C., {et~al.} 2006, \apj, 647, 552

\bibitem[{{Scott} \& {Duley}(1996)}]{scottduley1996}
{Scott}, A. \& {Duley}, W.~W. 1996, \apjs, 105, 401

\bibitem[{{Servoin} \& {Piriou}(1973)}]{servoinpiriou1973}
{Servoin}, J.~L. \& {Piriou}, B. 1973, Physica Status Solidi B Basic Research,
  55, 677

\bibitem[{{Simon} \& {Schaefer}(2011)}]{Simon2011}
{Simon}, M. \& {Schaefer}, G.~H. 2011, \apj, 743, 158

\bibitem[{{Skemer} {et~al.}(2016){Skemer}, {Morley}, {Zimmerman}, {Skrutskie},
  {Leisenring}, {Buenzli}, {Bonnefoy}, {Bailey}, {Hinz}, {Defr{\'e}re},
  {Esposito}, {Apai}, {Biller}, {Brandner}, {Close}, {Crepp}, {De Rosa},
  {Desidera}, {Eisner}, {Fortney}, {Freedman}, {Henning}, {Hofmann},
  {Kopytova}, {Lupu}, {Maire}, {Males}, {Marley}, {Morzinski}, {Oza},
  {Patience}, {Rajan}, {Rieke}, {Schertl}, {Schlieder}, {Stone}, {Su}, {Vaz},
  {Visscher}, {Ward-Duong}, {Weigelt}, \& {Woodward}}]{Skemer2016}
{Skemer}, A.~J., {Morley}, C.~V., {Zimmerman}, N.~T., {et~al.} 2016, \apj, 817,
  166

\bibitem[{{Soummer}(2005)}]{Soummer2005}
{Soummer}, R. 2005, \apjl, 618, L161

\bibitem[{{Soummer} {et~al.}(2012){Soummer}, {Pueyo}, \&
  {Larkin}}]{Soummer2012}
{Soummer}, R., {Pueyo}, L., \& {Larkin}, J. 2012, \apjl, 755, L28

\bibitem[{{Stephens} {et~al.}(2009){Stephens}, {Leggett}, {Cushing}, {Marley},
  {Saumon}, {Geballe}, {Golimowski}, {Fan}, \& {Noll}}]{2009ApJ...702..154S}
{Stephens}, D.~C., {Leggett}, S.~K., {Cushing}, M.~C., {et~al.} 2009, \apj,
  702, 154

\bibitem[{{Thalmann} {et~al.}(2008){Thalmann}, {Schmid}, {Boccaletti},
  {Mouillet}, {Dohlen}, {Roelfsema}, {Carbillet}, {Gisler}, {Beuzit}, {Feldt},
  {Gratton}, {Joos}, {Keller}, {Kragt}, {Pragt}, {Puget}, {Rigal}, {Snik},
  {Waters}, \& {Wildi}}]{Thalmann2008}
{Thalmann}, C., {Schmid}, H.~M., {Boccaletti}, A., {et~al.} 2008, in \procspie,
  Vol. 7014, Ground-based and Airborne Instrumentation for Astronomy II, 70143F

\bibitem[{Thi\'ebaut {et~al.}(2016)Thi\'ebaut, Denis, Mugnier, Ferrari, Mary,
  Langlois, Cantalloube, \& Devaney}]{Thiebaut-p-16}
Thi\'ebaut, E., Denis, L., Mugnier, L.~M., {et~al.} 2016, in Adaptive Optics
  Systems {V}, ed. E.~Marchetti, L.~M. Close, \& J.-P. V\'eran, Vol. 9909,
  Proc.\ Soc.\ Photo-Opt.\ Instrum.\ Eng., 990957--990957--10, conference date
  Jun. 2016, Edinburgh, UK

\bibitem[{{Tuthill} {et~al.}(2000){Tuthill}, {Monnier}, {Danchi}, {Wishnow}, \&
  {Haniff}}]{Tuthill2000}
{Tuthill}, P.~G., {Monnier}, J.~D., {Danchi}, W.~C., {Wishnow}, E.~H., \&
  {Haniff}, C.~A. 2000, \pasp, 112, 555

\bibitem[{{Venemans} {et~al.}(2015){Venemans}, {Ba{\~n}ados}, {Decarli},
  {Farina}, {Walter}, {Chambers}, {Fan}, {Rix}, {Schlafly}, {McMahon},
  {Simcoe}, {Stern}, {Burgett}, {Draper}, {Flewelling}, {Hodapp}, {Kaiser},
  {Magnier}, {Metcalfe}, {Morgan}, {Price}, {Tonry}, {Waters}, {AlSayyad},
  {Banerji}, {Chen}, {Gonz{\'a}lez-Solares}, {Greiner}, {Mazzucchelli},
  {McGreer}, {Miller}, {Reed}, \& {Sullivan}}]{Venemans2015}
{Venemans}, B.~P., {Ba{\~n}ados}, E., {Decarli}, R., {et~al.} 2015, \apjl, 801,
  L11

\bibitem[{{Vigan} {et~al.}(2016{\natexlab{a}}){Vigan}, {Bonnefoy}, {Ginski},
  {Beust}, {Galicher}, {Janson}, {Baudino}, {Buenzli}, {Hagelberg}, {D'Orazi},
  {Desidera}, {Maire}, {Gratton}, {Sauvage}, {Chauvin}, {Thalmann}, {Malo},
  {Salter}, {Zurlo}, {Antichi}, {Baruffolo}, {Baudoz}, {Blanchard},
  {Boccaletti}, {Beuzit}, {Carle}, {Claudi}, {Costille}, {Delboulb{\'e}},
  {Dohlen}, {Dominik}, {Feldt}, {Fusco}, {Gluck}, {Girard}, {Giro}, {Gry},
  {Henning}, {Hubin}, {Hugot}, {Jaquet}, {Kasper}, {Lagrange}, {Langlois}, {Le
  Mignant}, {Llored}, {Madec}, {Martinez}, {Mawet}, {Mesa}, {Milli},
  {Mouillet}, {Moulin}, {Moutou}, {Orign{\'e}}, {Pavlov}, {Perret}, {Petit},
  {Pragt}, {Puget}, {Rabou}, {Rochat}, {Roelfsema}, {Salasnich}, {Schmid},
  {Sevin}, {Siebenmorgen}, {Smette}, {Stadler}, {Suarez}, {Turatto}, {Udry},
  {Vakili}, {Wahhaj}, {Weber}, \& {Wildi}}]{Vigan2016}
{Vigan}, A., {Bonnefoy}, M., {Ginski}, C., {et~al.} 2016{\natexlab{a}}, \aap,
  587, A55

\bibitem[{{Vigan} {et~al.}(2016{\natexlab{b}}){Vigan}, {Bonnefoy}, {Ginski},
  {Beust}, {Galicher}, {Janson}, {Baudino}, {Buenzli}, {Hagelberg}, {D'Orazi},
  {Desidera}, {Maire}, {Gratton}, {Sauvage}, {Chauvin}, {Thalmann}, {Malo},
  {Salter}, {Zurlo}, {Antichi}, {Baruffolo}, {Baudoz}, {Blanchard},
  {Boccaletti}, {Beuzit}, {Carle}, {Claudi}, {Costille}, {Delboulb{\'e}},
  {Dohlen}, {Dominik}, {Feldt}, {Fusco}, {Gluck}, {Girard}, {Giro}, {Gry},
  {Henning}, {Hubin}, {Hugot}, {Jaquet}, {Kasper}, {Lagrange}, {Langlois}, {Le
  Mignant}, {Llored}, {Madec}, {Martinez}, {Mawet}, {Mesa}, {Milli},
  {Mouillet}, {Moulin}, {Moutou}, {Orign{\'e}}, {Pavlov}, {Perret}, {Petit},
  {Pragt}, {Puget}, {Rabou}, {Rochat}, {Roelfsema}, {Salasnich}, {Schmid},
  {Sevin}, {Siebenmorgen}, {Smette}, {Stadler}, {Suarez}, {Turatto}, {Udry},
  {Vakili}, {Wahhaj}, {Weber}, \& {Wildi}}]{2016A&A...587A..55V}
{Vigan}, A., {Bonnefoy}, M., {Ginski}, C., {et~al.} 2016{\natexlab{b}}, \aap,
  587, A55

\bibitem[{{Vigan} {et~al.}(2010){Vigan}, {Moutou}, {Langlois}, {Allard},
  {Boccaletti}, {Carbillet}, {Mouillet}, \& {Smith}}]{Vigan2010}
{Vigan}, A., {Moutou}, C., {Langlois}, M., {et~al.} 2010, \mnras, 407, 71

\bibitem[{{Wagner} {et~al.}(2016){Wagner}, {Apai}, {Kasper}, {Kratter},
  {McClure}, {Robberto}, \& {Beuzit}}]{2016Sci...353..673W}
{Wagner}, K., {Apai}, D., {Kasper}, M., {et~al.} 2016, Science, 353, 673

\bibitem[{{Wahhaj} {et~al.}(2015){Wahhaj}, {Cieza}, {Mawet}, {Yang}, {Canovas},
  {de Boer}, {Casassus}, {M{\'e}nard}, {Schreiber}, {Liu}, {Biller}, {Nielsen},
  \& {Hayward}}]{Wahhaj2015}
{Wahhaj}, Z., {Cieza}, L.~A., {Mawet}, D., {et~al.} 2015, \aap, 581, A24

\bibitem[{{Zahnle} {et~al.}(2016){Zahnle}, {Marley}, {Morley}, \&
  {Moses}}]{Zahnle2016}
{Zahnle}, K., {Marley}, M.~S., {Morley}, C.~V., \& {Moses}, J.~I. 2016, \apj,
  824, 137

\bibitem[{{Zahnle} \& {Marley}(2014)}]{zahnlemarley2014}
{Zahnle}, K.~J. \& {Marley}, M.~S. 2014, \apj, 797, 41

\bibitem[{{Zapatero Osorio} {et~al.}(2014){Zapatero Osorio}, {B{\'e}jar},
  {Miles-P{\'a}ez}, {Pe{\~n}a Ram{\'{\i}}rez}, {Rebolo}, \&
  {Pall{\'e}}}]{2014A&A...568A...6Z}
{Zapatero Osorio}, M.~R., {B{\'e}jar}, V.~J.~S., {Miles-P{\'a}ez}, P.~A.,
  {et~al.} 2014, \aap, 568, A6

\bibitem[{{Zuckerman} {et~al.}(2001){Zuckerman}, {Song}, {Bessell}, \&
  {Webb}}]{Zuckerman2001}
{Zuckerman}, B., {Song}, I., {Bessell}, M.~S., \& {Webb}, R.~A. 2001, \apjl,
  562, L87

\bibitem[{{Zurlo} {et~al.}(2016){Zurlo}, {Vigan}, {Galicher}, {Maire}, {Mesa},
  {Gratton}, {Chauvin}, {Kasper}, {Moutou}, {Bonnefoy}, {Desidera}, {Abe},
  {Apai}, {Baruffolo}, {Baudoz}, {Baudrand}, {Beuzit}, {Blancard},
  {Boccaletti}, {Cantalloube}, {Carle}, {Cascone}, {Charton}, {Claudi},
  {Costille}, {de Caprio}, {Dohlen}, {Dominik}, {Fantinel}, {Feautrier},
  {Feldt}, {Fusco}, {Gigan}, {Girard}, {Gisler}, {Gluck}, {Gry}, {Henning},
  {Hugot}, {Janson}, {Jaquet}, {Lagrange}, {Langlois}, {Llored}, {Madec},
  {Magnard}, {Martinez}, {Maurel}, {Mawet}, {Meyer}, {Milli},
  {Moeller-Nilsson}, {Mouillet}, {Orign{\'e}}, {Pavlov}, {Petit}, {Puget},
  {Quanz}, {Rabou}, {Ramos}, {Rousset}, {Roux}, {Salasnich}, {Salter},
  {Sauvage}, {Schmid}, {Soenke}, {Stadler}, {Suarez}, {Turatto}, {Udry},
  {Vakili}, {Wahhaj}, {Wildi}, \& {Antichi}}]{Zurlo2016}
{Zurlo}, A., {Vigan}, A., {Galicher}, R., {et~al.} 2016, \aap, 587, A57

\end{thebibliography}
%

\begin{appendix}
\section{Alternative reductions}
\label{sec:alternative_reductions}
\begin{figure}
\centering
\includegraphics[width=\columnwidth]{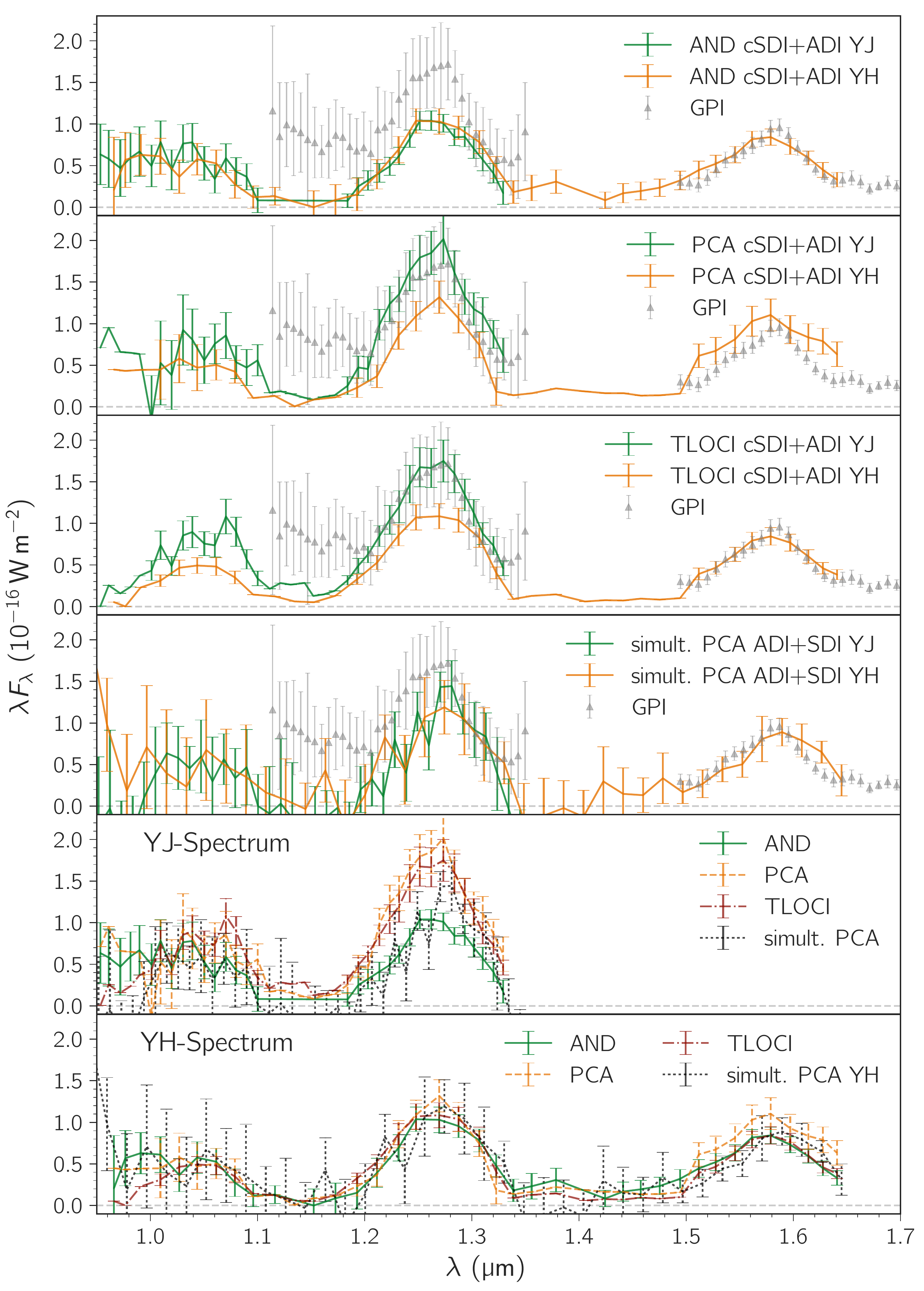}
\caption{Shown in the panels are (from top to bottom) the extracted spectra with: ANDROMEDA (panel 1), PCA (panel 2), TLOCI (panel 3), and simultaneous ADI+SDI with PCA (panel 4), all reduction for YJ (panel 5), and all reductions for YH (panel 6). The GPI spectra are plotted for comparison in the first 4 panels. The PCA and TLOCI pipeline used automatically exchange not significant detections with the upper-limit, in which case no uncertainty is displayed on the data point.}
\label{fig:alternative_reductions}
\end{figure}

Shown in Figure~\ref{fig:alternative_reductions} are the extracted spectra with the different algorithms that were tested (see Sect.~\ref{sec:IFS_reduction}): ANDROMEDA (panel 1), PCA (panel 2), TLOCI (panel 3, Specal), and additionally PCA with simultaneous use of ADI and SDI reference images (panel 4). Additionally, panel 5 shows all reductions for the YJ data set and panel 6 all reductions for the YH data set. The ANDROMEDA, PCA and TLOCI reduction all rely on the same cSDI pre-reduced data cube, whereas the simultaneous PCA ADI+SDI reduction is completely independent based on the pipeline introduced in \citet{Mesa2015}. The YH spectrum is fully consistent between all reduction methods, the only exception being the low quality of the H-band extraction using PCA.\\
We notice more uncertainties in the absolute calibration of our YJ-spectral data, which changes depending on the exact algorithm used to reduce the data. The ANDROMEDA code yields compatible fluxes between the two observations, as does the independent reduction using spectral PCA with simultaneous ADI and SDI references. Whereas the PCA and TLOCI reduction using the same classical SDI pre-reduced frames, show higher peak fluxes in the J-band. Although these reductions show comparable peak values to the GPI J-band spectrum, they do not follow the same spectral shape over the entirety of the GPI J-band spectrum.

\section{Testing metallicity determination with benchmark brown dwarfs}
\label{sec:bd_benchmark}
\begin{figure*}
\centering
\includegraphics[width=\textwidth]{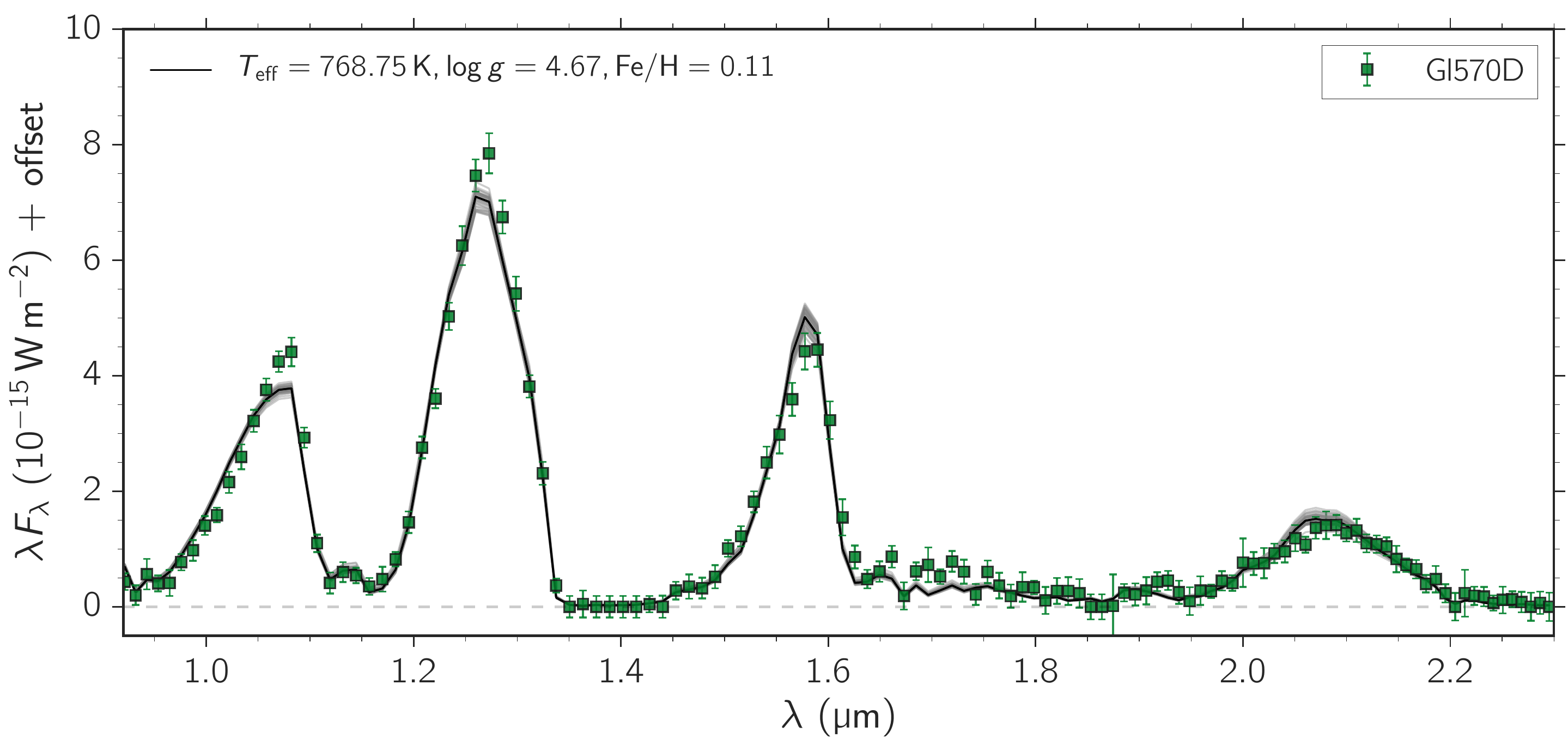} \\
\includegraphics[width=\textwidth]{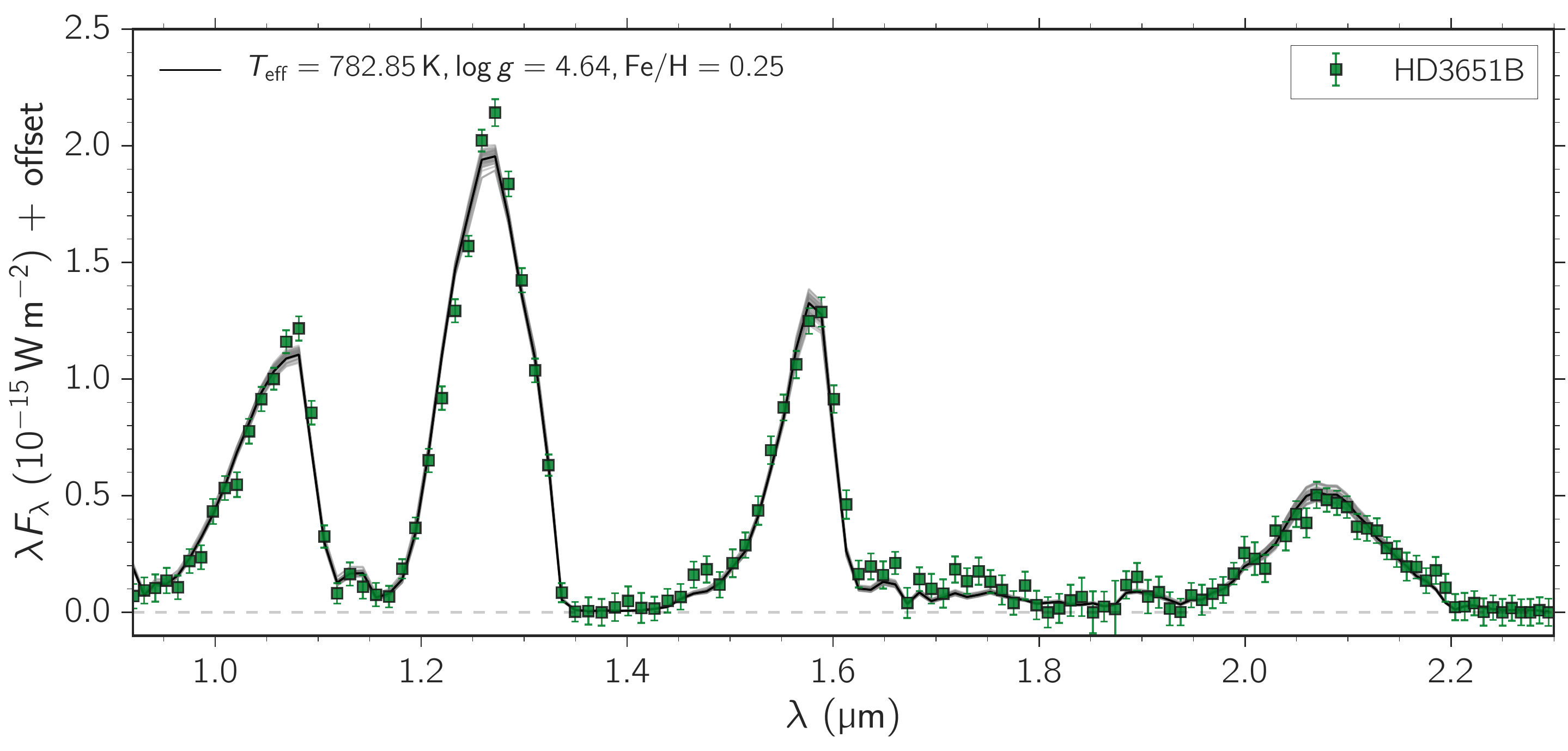}
\caption{The upper panel shows the best-fit \emph{petitCODE} clear spectrum for Gl~570D, the lower panel the same for HD~3651B. The overplotted gray lines represent the model scatter with spectra generated from randomly drawn samples of the posterior parameter distribution. Errorbars plotted include the best-fit value of the $b$-parameter, correcting for the underestimated data uncertainty.}
\label{fig:spec_BD}
\end{figure*}

\begin{figure*}
\centering
\includegraphics[width=\textwidth]{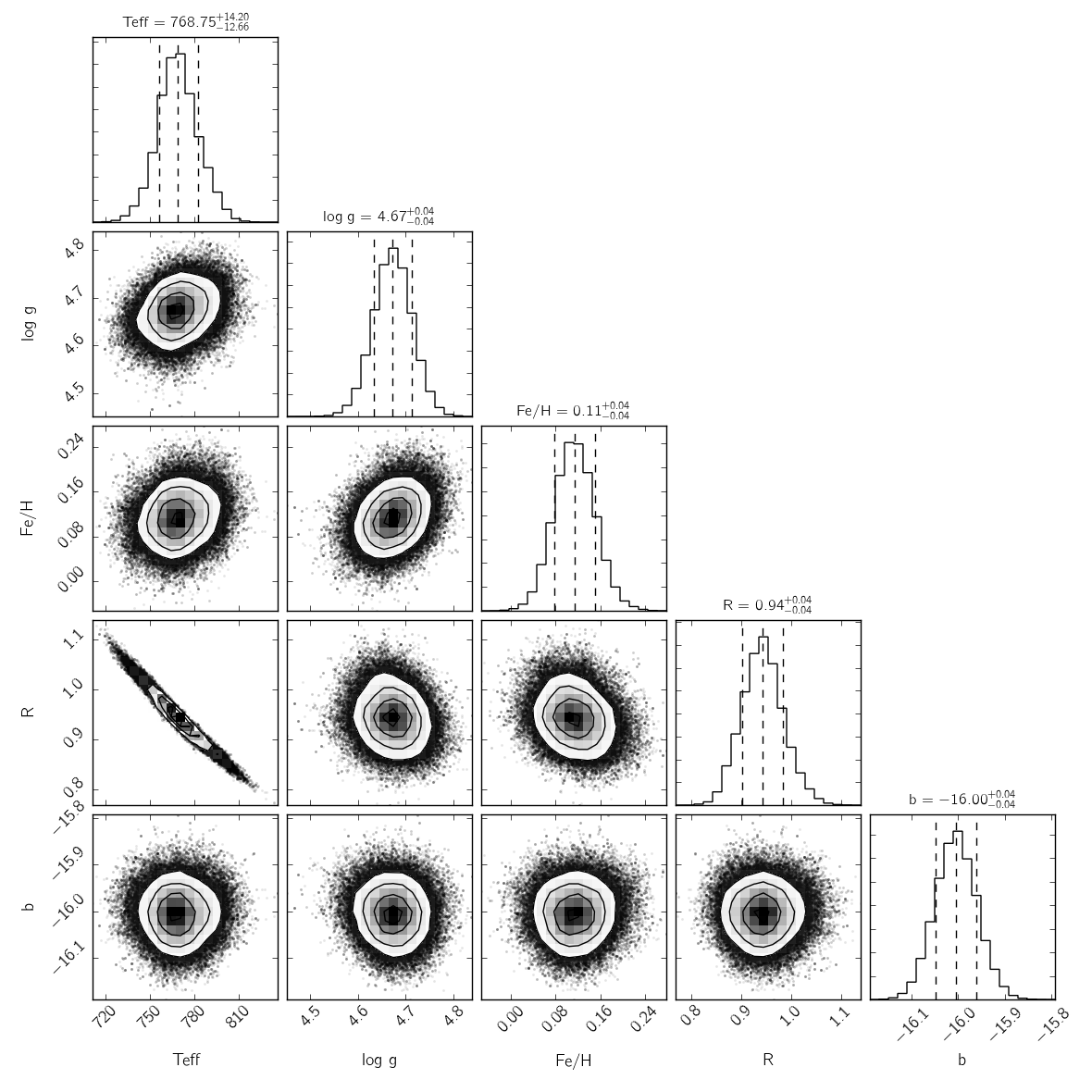}
\caption{Corner plot showing the posterior probability distribution of the clear \emph{petitCODE} fitted to Gl~570D, including a further scale parameter $b$ as additive term to the flux uncertainty.}
\label{fig:corner_Gl570D}
\end{figure*}

\begin{figure*}
\centering
\includegraphics[width=\textwidth]{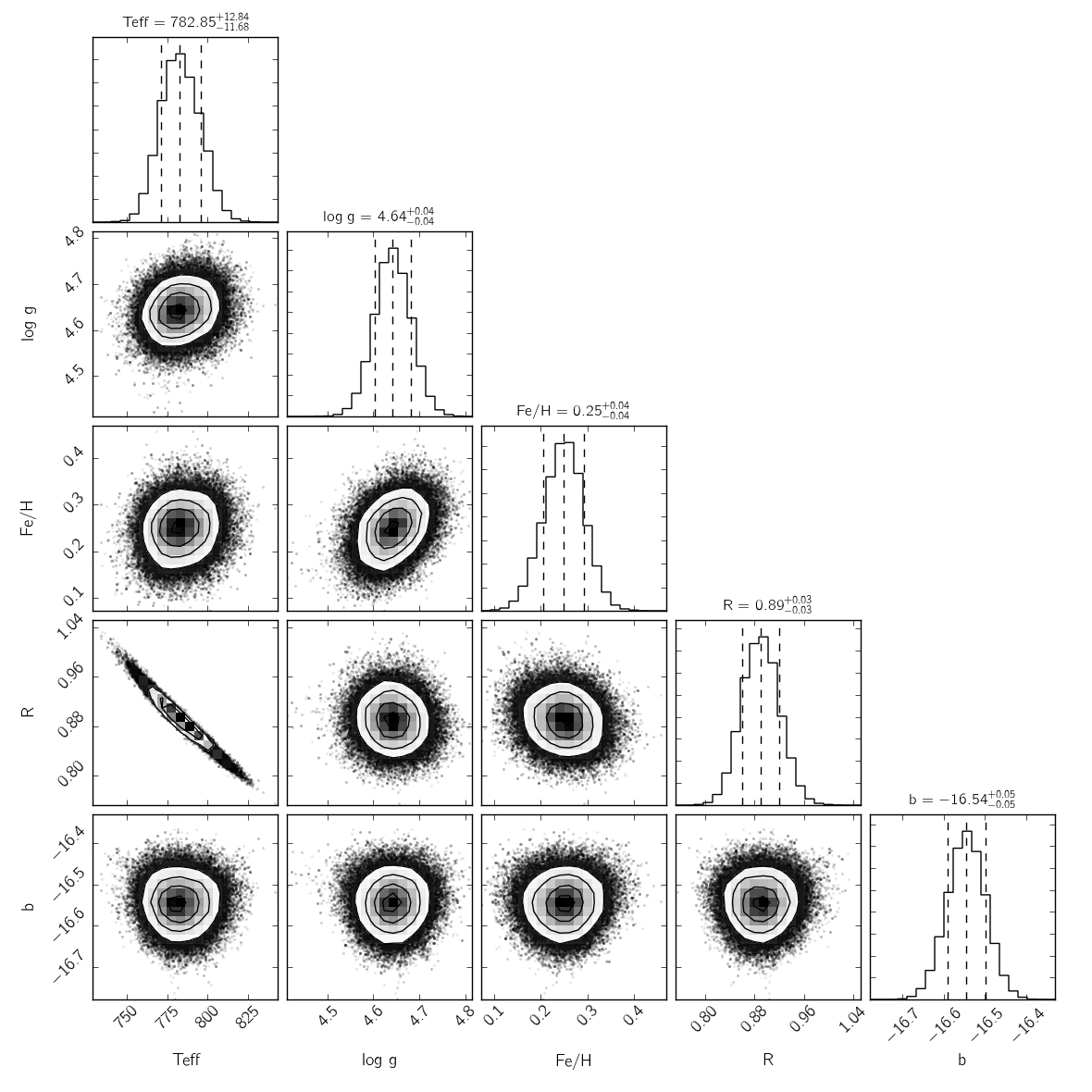}
\caption{Corner plot showing the posterior probability distribution of the clear \emph{petitCODE} fitted to HD~3651B, including a further scale parameter $b$ as additive term to the flux uncertainty.}
\label{fig:corner_HD3651B}
\end{figure*}

The two T7.5 brown dwarfs \object{Gl 570D} and \object{HD 3651B} are considered to be benchmark objects, because they are on wide orbits around extensively studied K stars with known properties \citep{Line2015, Line2016}. Having formed from the same cloud as their host stars, these brown dwarfs provide the opportunity to compare the derived parameters for the brown dwarfs, especially their composition, with their respective host star. Given the high metallicity inferred by our model for 51~Eri~b, we want to make sure that our methodology and models are not biased towards obtaining high metallicity results. Below we compare the host star metallicities with the brown dwarf metallicities obtained with our self-consistent equilibrium \emph{petitCODE} models. We further compare the parameters with the parameters derived by \citet{Line2015}.\\
To make a comparison with \citet{Line2015} easier, we follow the same methodology, using every third pixel to avoid correlations between neighboring data points and the same additional free fit-parameter $b$ in the likelihood function, which accounts for the underestimated uncertainties in the data by adding a constant $10^b$ term to the flux uncertainties. A flat prior is assumed for this parameter. Notice that all systematics in the absolute photometric calibration and distance are absorbed into the brown dwarf "radius"-parameter $R$, because with a flat prior it allows the spectrum to freely float up and down. As also pointed out in \citet{Line2015}, the absolute calibration is not necessarily reliable, so the radius should \emph{not} be seen as a physical quantity, but rather as a data scaling parameter. On the positive side, this means that the determination of $T_\mathrm{eff}$, $\log g$, and [Fe/H] is independent of the absolute photometry and distance of the objects and purely determined by the model shape and relative strength of the features.
A summary of the derived parameters for Gl~570D and HD~3651B for the \citet{Line2015} retrieval as well as the \emph{petitCODE} clear models is shown in Table~\ref{tab:benchmark}. The best-fit \emph{petitCODE} model spectra are shown in Fig.~\ref{fig:spec_BD} and the respective posterior probability distributions in Fig.~\ref{fig:corner_Gl570D}~and~\ref{fig:corner_HD3651B}, respectively. 
\citet{Line2016} gives a summary of literature metallicity values for the host star Gl~570A and \citet{Line2015} derived the metallicity for HD~3651A. For Gl~570D, \citet{Saumon2006} gave an averaged metallicity data based on recent literature of $\mathrm{[Fe/H]}=0.09 \pm 0.04$, \citet{Casagrande2011} a more recent value of $0.31$ and $-0.05 \pm 0.17$ from \citet{Line2015}. The preponderance of evidence seems to suggest a slightly super-solar metallicity, whereas HD~3651A has a super-solar metallicity in the range of $\mathrm{[Fe/H]}=0.18 \pm 0.07$ \citep{Ramirez2013}.\\
The metallicity we determined using the \emph{petitCODE} models is within the given range of host star metallicities, showing that our model and fitting approach can be used to reliably estimate metallicities. Comparing the results to the free retrieval performed by \citet{Line2015}, their metallicities fall on the lower end, whereas ours fall on the higher end of the metallicity range, which may reflect a difference in the free retrieval versus self-consistent model approach. For example, their retrieval requires an additional step to compute the chemical bulk metallicity from the retrieved abundances. On the other hand, our models assume a fixed solar C/O ratio. We note that both objects share very similar properties according to our fits, except for HD~3651B being more metal rich. This is consistent with our expectations as they are both classified as T7.5 dwarfs and the normalized spectra are virtually indistinguishable at the resolution of the SpeX instrument. Only the strength of the K-band flux, which is an indicator of metallicity mainly due to its sensitivity to collision-induced absorption (CIA) of H$_2$--H$_2$ and H$_2$--He pairs (see discussion in Sec.~\ref{sec:atmospheric_modeling}), is stronger in HD~3651B. Compared to \citet{Line2015} our models are about 50~K hotter in effective temperature. The derived surface gravity for Gl~570D is almost the same, whereas \citet{Line2015} arrives at a significantly higher surface gravity for HD~3651B.

\begin{table*}
\caption{Benchmark Brown Dwarfs}
\centering
\begin{tabular}{l c c c c c}
\hline\hline
Object&$T_\mathrm{eff}$&$\log \, g$&Chemically derived bulk [Fe/H]&[Fe/H]$_\mathrm{BD}$ - [Fe/H]$_\mathrm{host}$\tablefootmark{a}\\
\hline
\multicolumn{1}{c}{petitCODE clear model}\\
\hline
Gl 570D&$769^{+14}_{-13}$&$4.67 \pm 0.04$&$0.11 \pm 0.04$&$+0.16 \pm 0.18$\\
HD 3651B&$783^{+13}_{-12}$&$4.64 \pm 0.04$&$0.25 \pm 0.04$&$+0.07 \pm 0.08$\\
\hline
\multicolumn{1}{c}{Line et al. 2015 retrieval}\\
\hline
Gl 570D&$714^{+20}_{-23}$&$4.76^{+0.27}_{-0.28}$&$-0.15$&$-0.10 \pm 0.17$\\
HD 3651B&$726^{+22}_{-21}$&$5.12^{+0.09}_{-0.17}$&$+0.08$&$-0.10 \pm 0.07$\\
\end{tabular}
\label{tab:benchmark}
\tablefoot{Summary of modeling results for Gl~570D and HD~3651B using \emph{petitCODE} clear models and comparison to atmospheric retrieval results by \citet{Line2015}.
\tablefoottext{a}{Difference of best fit model to middle of metallicity range of host star. Uncertainty includes the width of host star metallicity range in all cases, as well as uncertainty for brown dwarf metallicity in the \emph{petitCODE} case. Host star metallicities used here: Gl 570D: -0.22 -- 0.12 or $-0.05 \pm 0.17$ \citep{Line2015};  HD 3651B: 0.11 -- 0.25 or $0.18 \pm 0.07$ \citep{Ramirez2013}.}}
\end{table*}

\section{Model spectra}
\begin{figure}[h]
\centering
\includegraphics[width=\columnwidth]{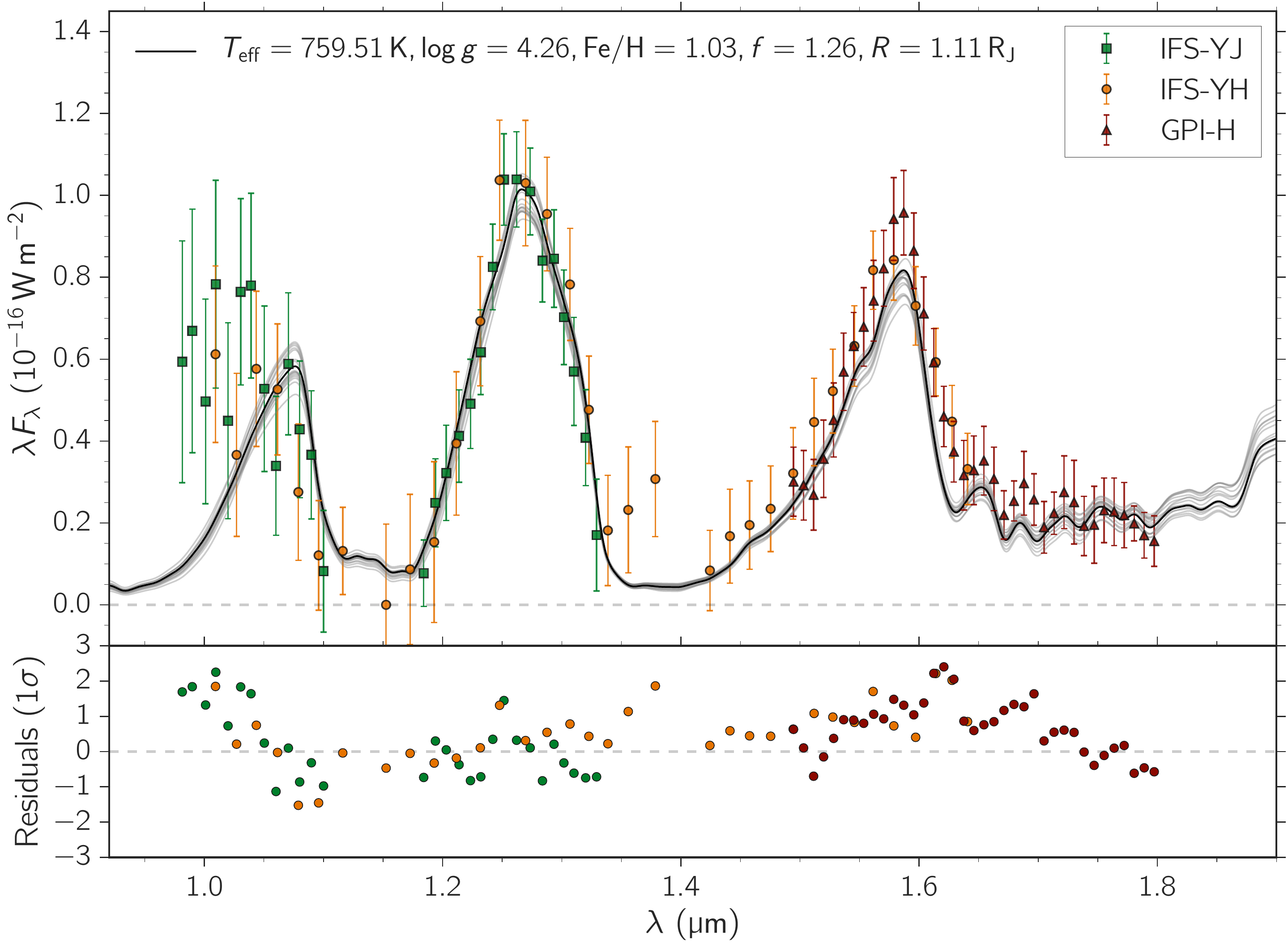}
\caption{Same plot as Fig.~\ref{fig:spectrum2}, but zooming in on the wavelength range covered by spectral data.}
\label{fig:fit_zoom}
\end{figure}

\begin{figure}[h]
\centering
\includegraphics[width=\columnwidth]{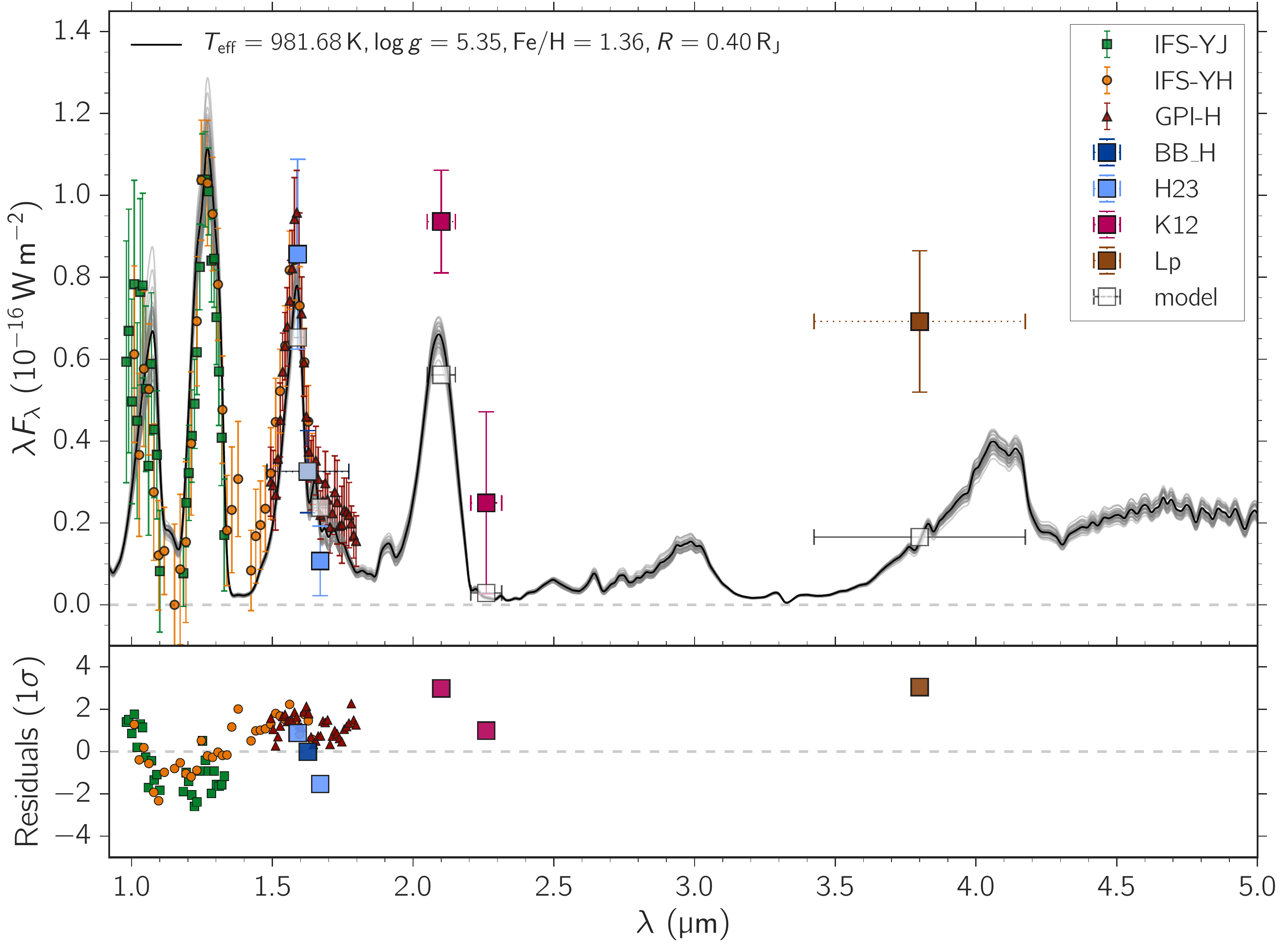}
\caption{Same plot as Fig.~\ref{fig:spectrum2}, but using the clear model.}
\label{fig:fit_clear}
\end{figure}

\begin{figure}[h]
\centering
\includegraphics[width=\columnwidth]{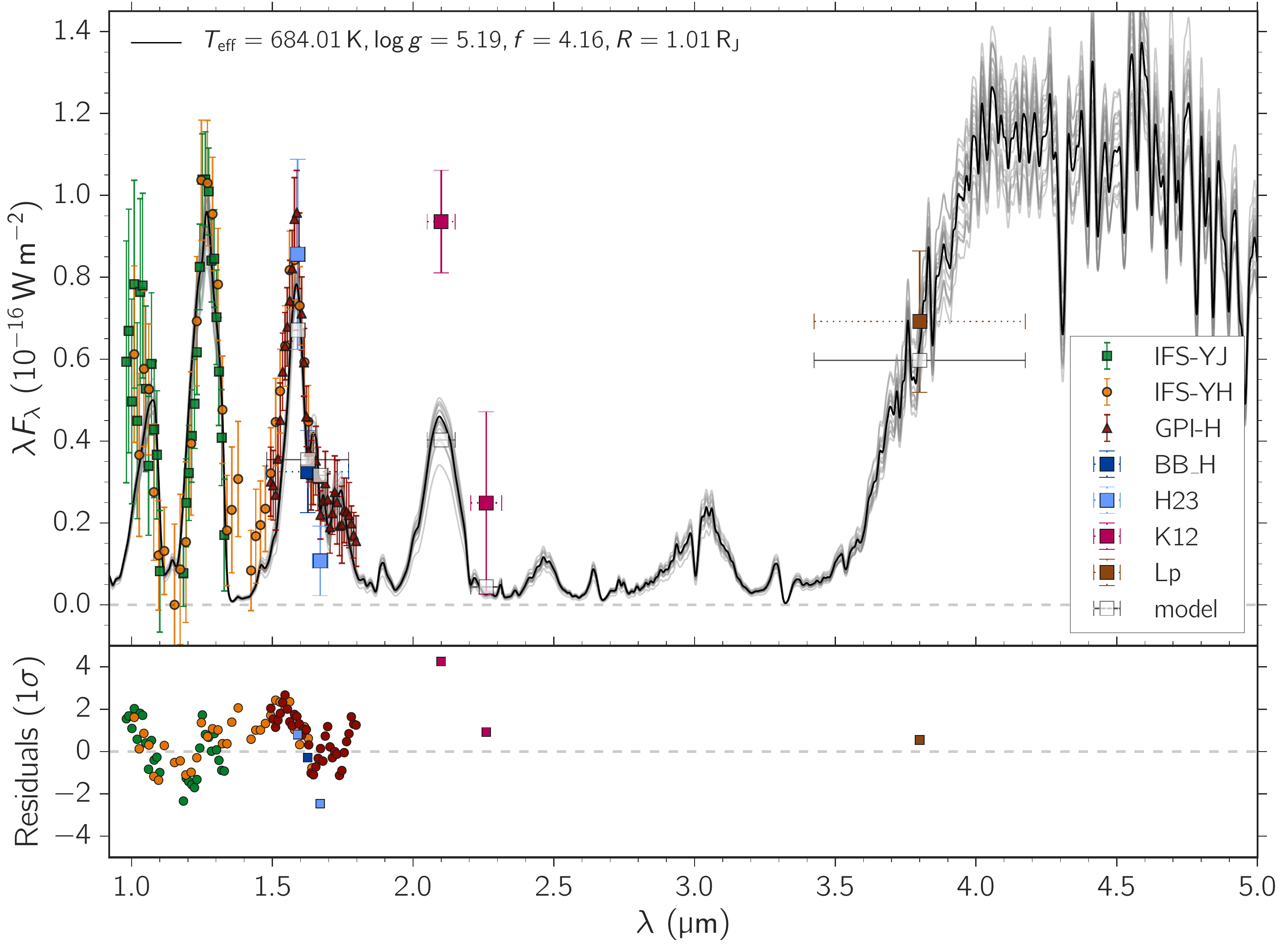}
\caption{Same plot as Fig.~\ref{fig:spectrum2}, but using the \citet{Morley2012} model.}
\label{fig:fit_morley}
\end{figure}

\begin{figure}[h]
\centering
\includegraphics[width=\columnwidth]{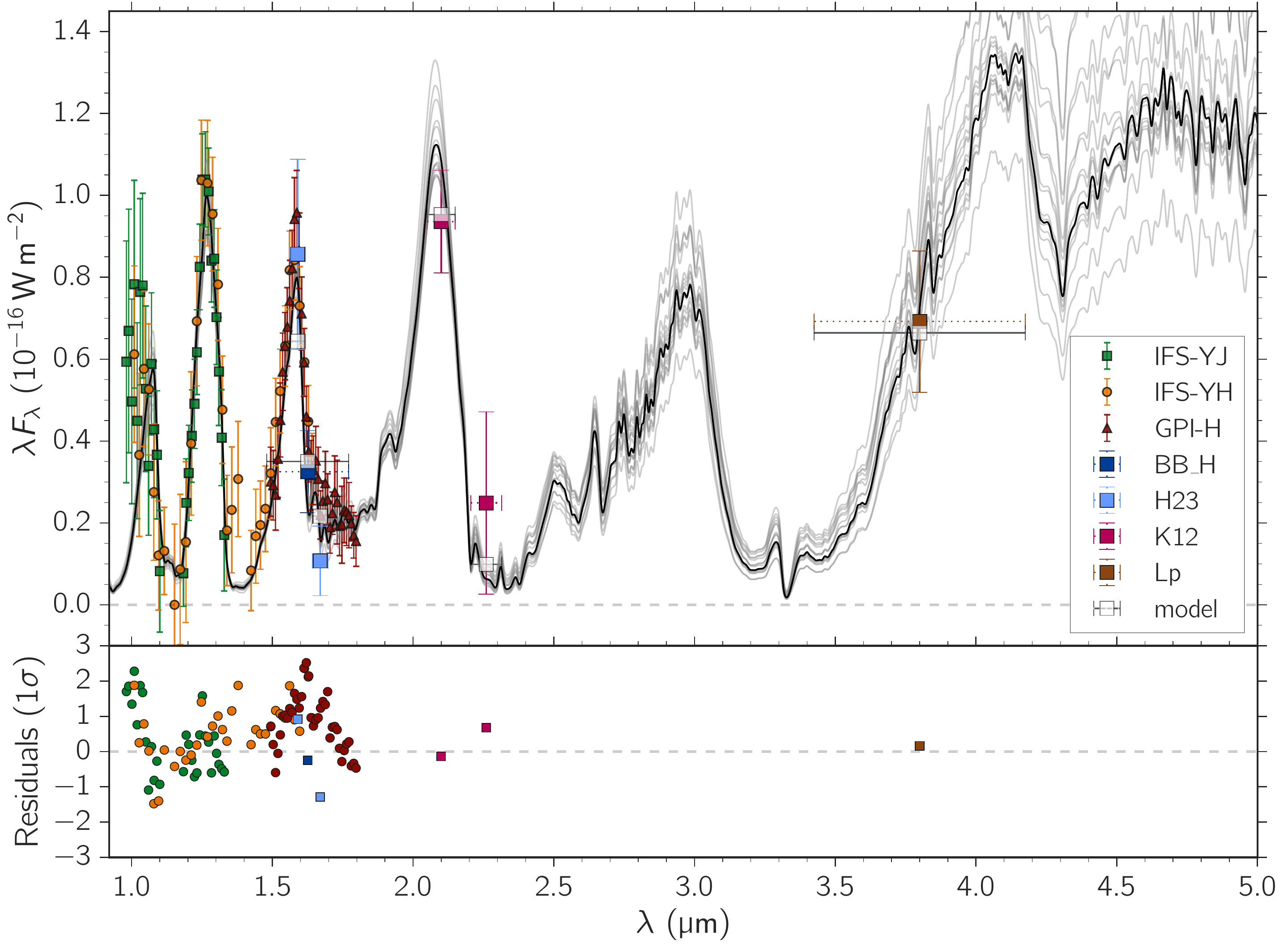}
\caption{Same plot as Fig.~\ref{fig:spectrum2}, but using the patchy-cloud model described in Sec.~\ref{sec:patchy_cloud}. Best fitting parameters are: $T_{cloud} = 750 \pm 25$ K,  $T_{clear} = 815^{+70}_{-40}$ K, $CF = 0.9 \pm 0.05$, $\log \, g = 4.5 \pm 0.3$, $\mathrm{[Fe/H]} = 1.25$, $f_{\rm sed} = 1.10 \pm 0.15$, $R = 1.10 \pm 0.15$ $\mathrm{R}_\mathrm{J}$}
\label{fig:fit_composite}
\end{figure}

Figure~\ref{fig:fit_zoom} shows a zoom in on the IFS spectra in the best-fit cloudy \emph{petitCODE} model. It can be seen that there is very good agreement between the model and the data in shape and amplitude, except for a systematic offset in the amplitude of the H-band which still exists and could not be modeled without negatively impacting the overall fit to the rest of the spectrum.\\
Figure~\ref{fig:fit_clear} shows the the same plot as Fig.~\ref{fig:spectrum2}, but with the cloud-free \emph{petitCODE} model. It is immediately apparent that the cloud-free model is incapable of explaining the long wavelengths of the spectrum (K1- and L'-band), which results in unphysical parameters in temperature and radius and the lack of clouds in the model is compensated with extremely high metallicities.\\
Figure~\ref{fig:fit_morley} of the \citet{Morley2012} shows that the lack of metallicity as a free-parameter (only solar metallicity was available) also skews the overall parameters, especially in order to fit the K1-peak. Since high metallicities are not allowed, the $f_\mathrm{sed}$ parameter increases in an attempt to compensate, again the resulting physical parameters are unreliable. All of this shows that a complex model which allows coverage of at least the basic physics (e.g., a cloud model with free parameters, non-solar metallicity), is a bare minimum to model these cold giant planets.\\
Figure~\ref{fig:fit_composite} shows the spectrum resulting from the patchy-cloud model introduced in Sec.~\ref{sec:patchy_cloud}. The spectrum is almost indistinguishable from a pure cloudy model and does not improve the result significantly.

\section{Corner plots}
\label{sec:corner_plots}
\begin{figure*}
\centering
\includegraphics[width=\textwidth]{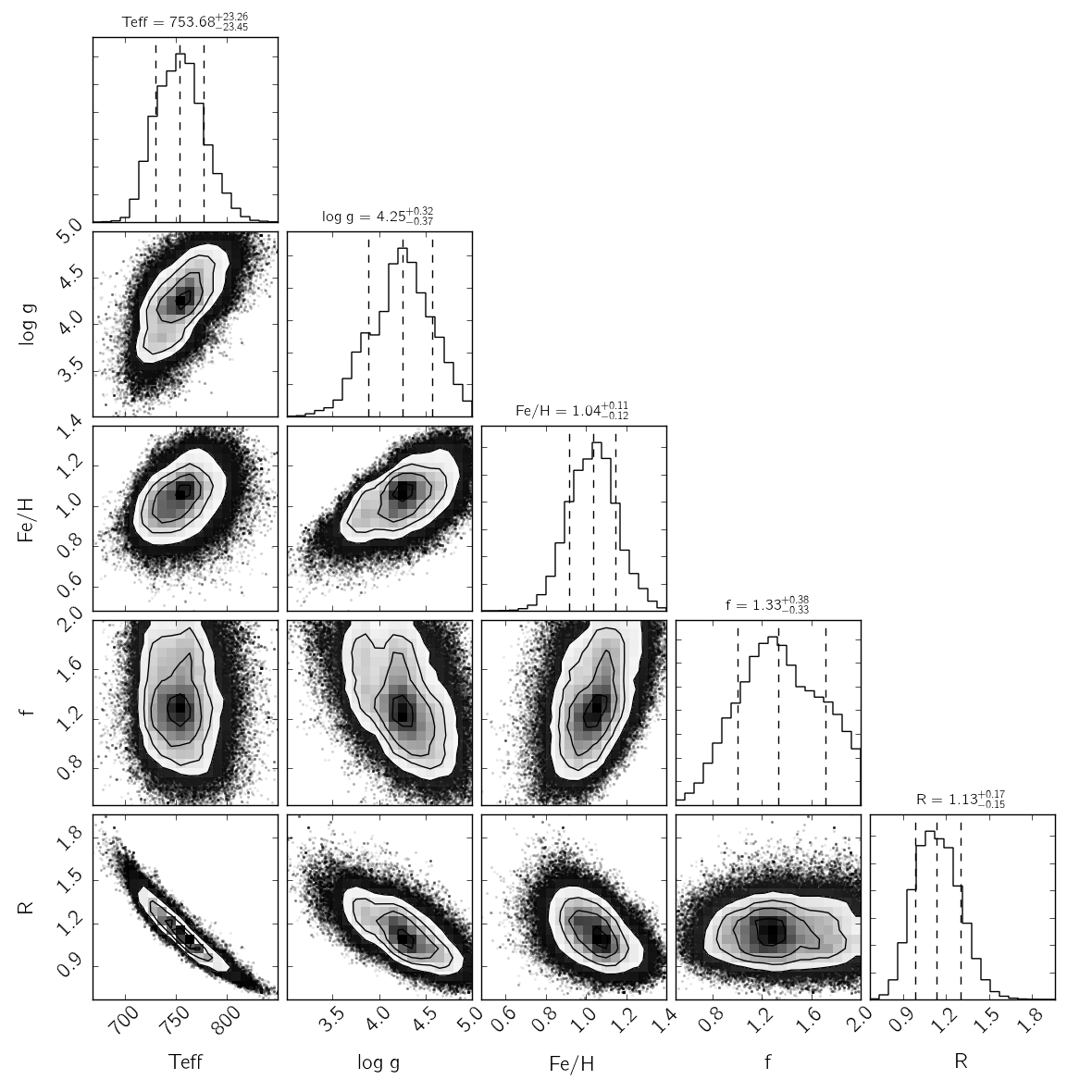}
\caption{The same as Figure~\ref{fig:corner}, but excluding the Y-band data. Corner plot showing the posterior probability distribution of the cloudy \emph{petitCODE} grid with respect to each of its parameter pair as well as the marginalized distribution for each parameters. The uncertainties are given as 16\% to 84\% quantiles as commonly done for multivariate MCMC results.}
\label{fig:corner_wo_Y}
\end{figure*}

\begin{figure*}
\centering
\includegraphics[width=\textwidth]{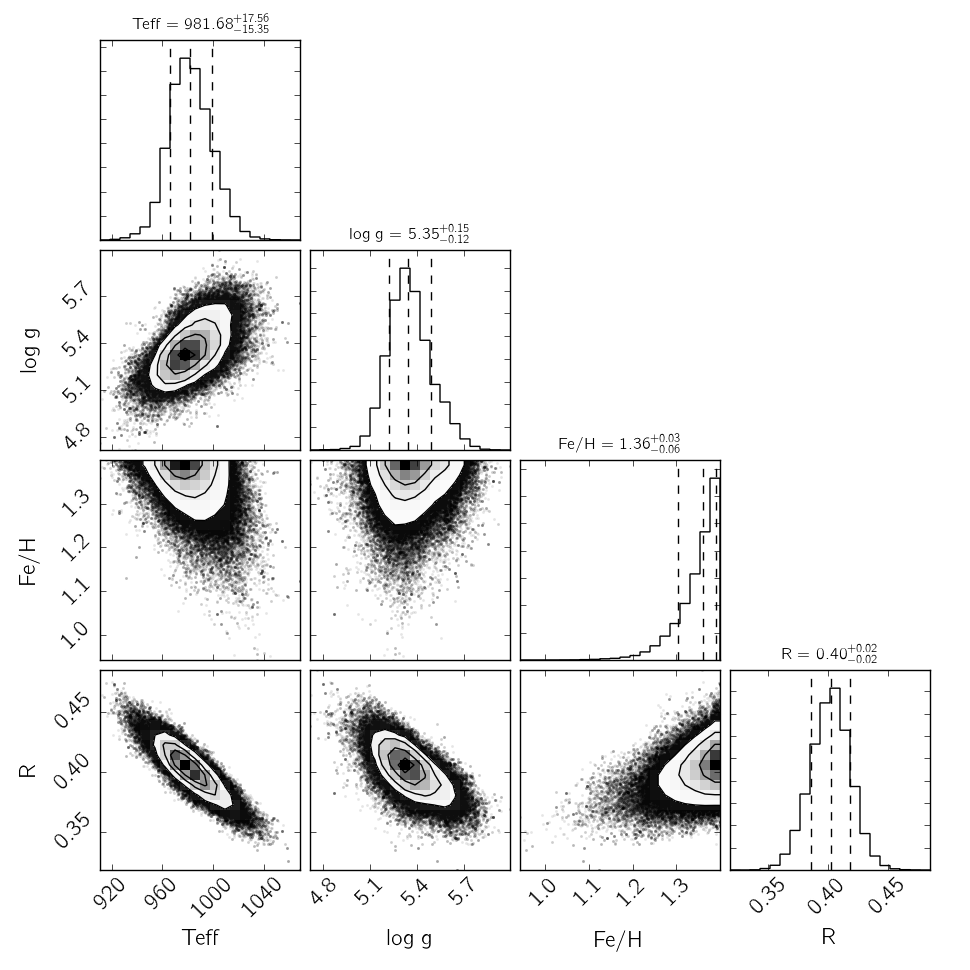}
\caption{Corner plot showing the posterior probability distribution of the clear \emph{petitCODE} grid, with respect to each of its parameter pair as well as the marginalized distribution for each parameters. The uncertainties are given as 16\% to 84\% quantiles as commonly done for multivariate MCMC results. Note that a clear model atmosphere requires a small radius, which speaks against this model.}
\label{fig:corner_clear}
\end{figure*}

\begin{figure*}
\centering
\includegraphics[width=\textwidth]{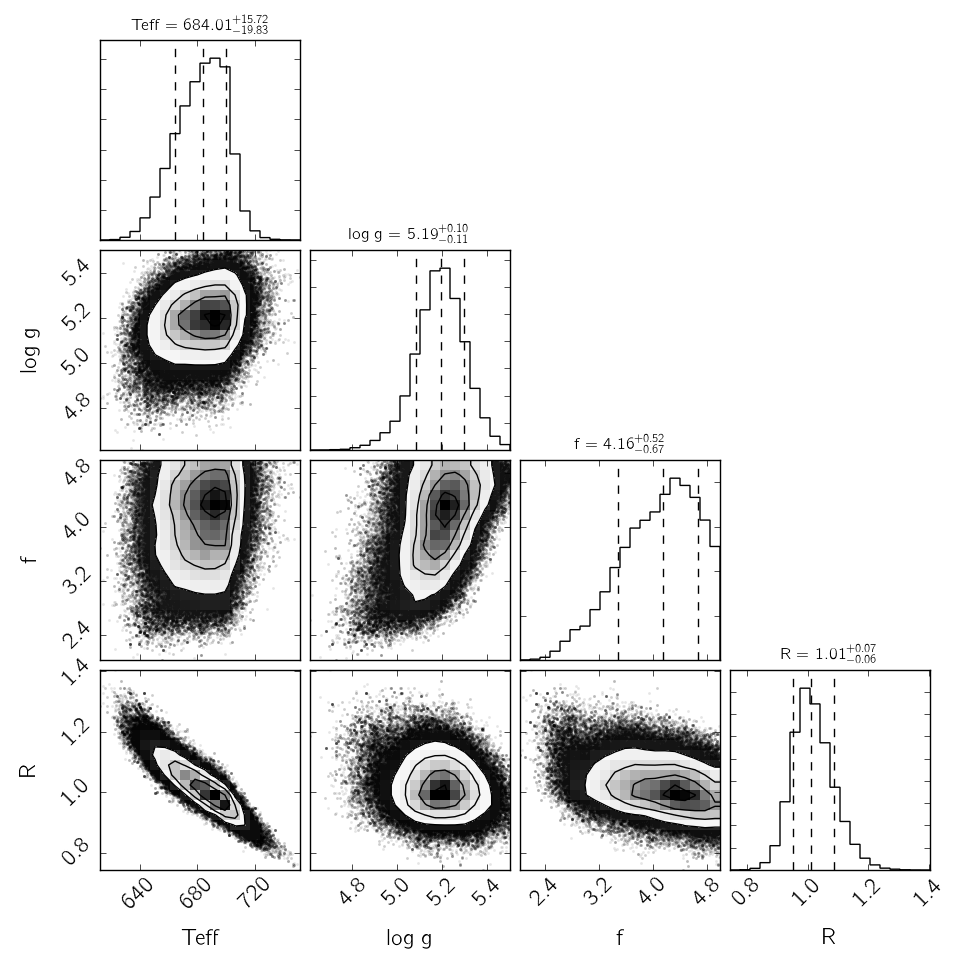}
\caption{Corner plot showing the posterior probability distribution of the \citet{Morley2012} grid, with respect to each of its parameter pair as well as the marginalized distribution for each parameters. The uncertainties are given as 16\% to 84\% quantiles as commonly done for multivariate MCMC results.}
\label{fig:corner_morley}
\end{figure*}

\begin{figure*}
\centering
\includegraphics[width=\textwidth]{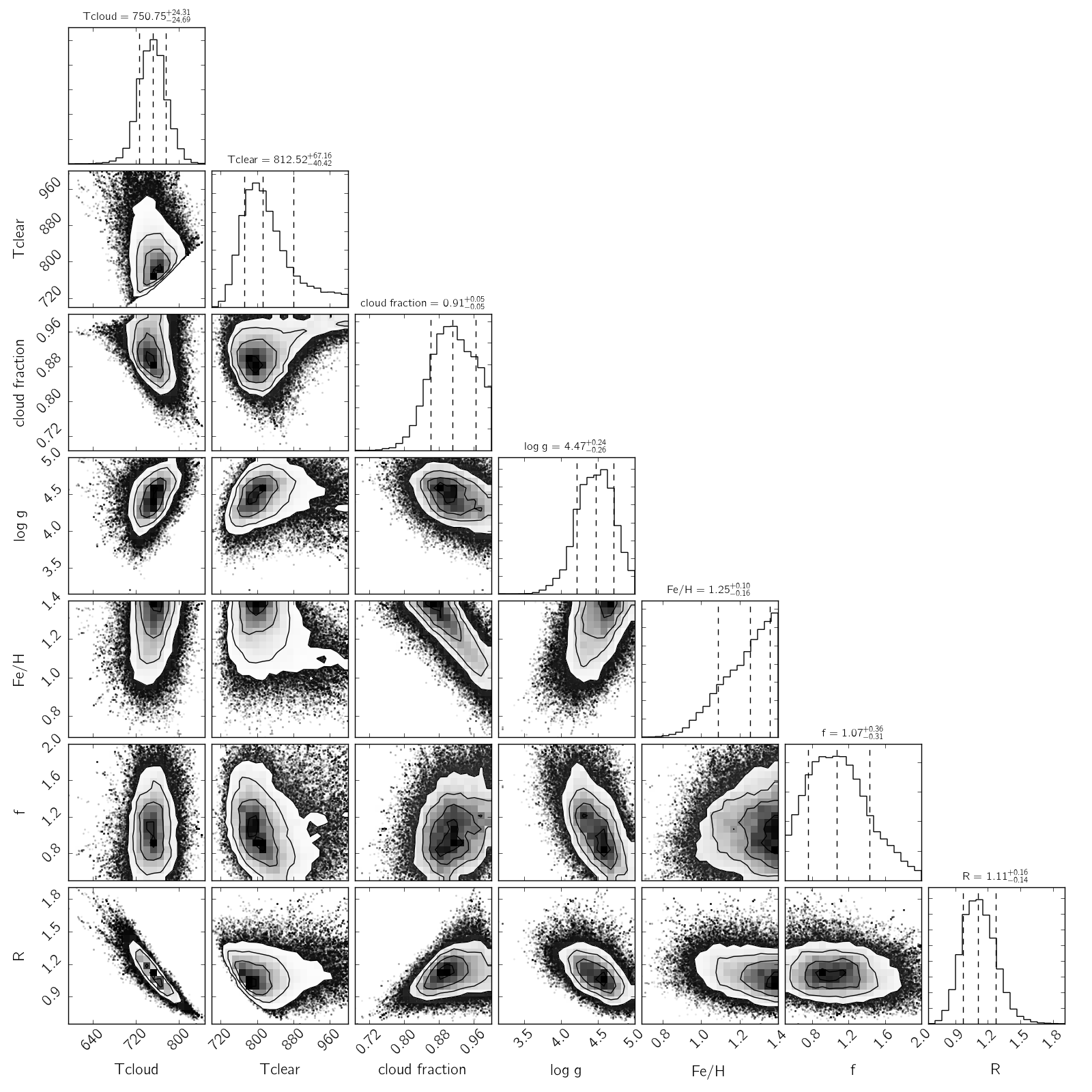}
\caption{Corner plot showing the posterior probability distribution of the patchy-cloud model described in Sec.~\ref{sec:patchy_cloud}, with respect to each of its parameter pair as well as the marginalized distribution for each parameters. The uncertainties are given as 16\% to 84\% quantiles as commonly done for multivariate MCMC results.}
\label{fig:corner_composite}
\end{figure*}

\begin{figure*}
\centering
\includegraphics[width=\textwidth]{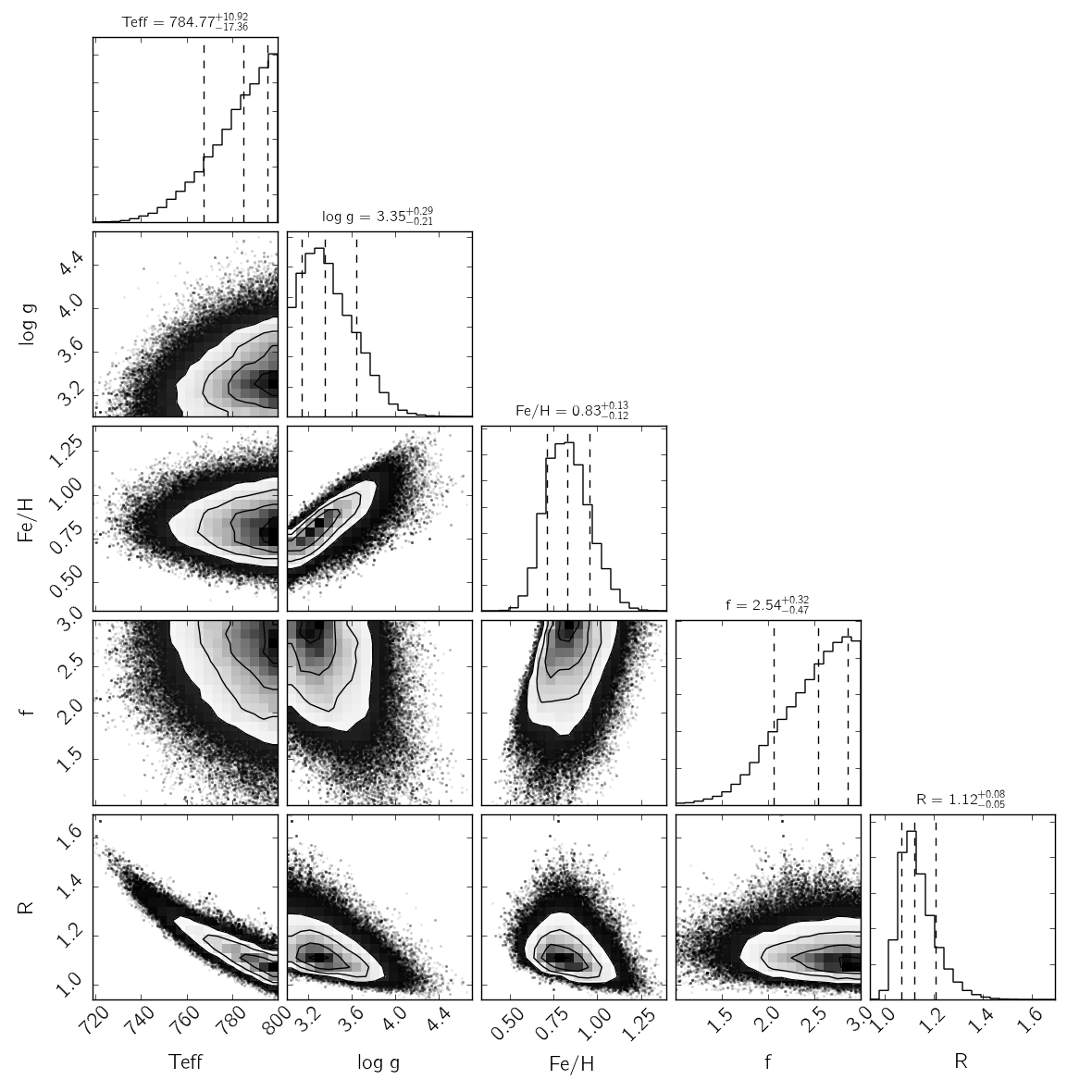}
\caption{Corner plot showing the posterior probability distribution using only data published in \citet{Macintosh2015} (without taking the covariance into account), with respect to each of its parameter pair as well as the marginalized distribution for each parameters. The uncertainties are given as 16\% to 84\% quantiles as commonly done for multivariate MCMC results.}
\label{fig:corner_macintosh}
\end{figure*}

Shown are the corner plots for additional models and data sets. Figure~\ref{fig:corner_wo_Y} shows the parameter distribution for the cloudy \emph{petitCODE} model in case we exclude the Y-band data complete. We see that the Y-band does not significantly constrain the models. Figure~\ref{fig:corner_clear} shows the corner plot for the clear \emph{petitCODE} model and Figure~\ref{fig:corner_morley} for the \citet{Morley2012} model. As discussed above, both lead to skewed results, because important physics is missing. Figure~\ref{fig:corner_composite} shows the corner plot for the patchy-cloud model, a linear combination of cloudy and cloud-free models, which share the same parameters except for temperature and are linked by a cloud fraction parameter. Cloud fractions are very high and, as pointed out above, the resulting spectrum does not improve the fit significantly.\\
As additional experiment Figure~\ref{fig:corner_macintosh} shows the posterior distribution for the cloudy model grid when only data from \citet{Macintosh2015} is used. This fit does not include the covariance matrices and should reduce roughly to a straightforward fit as the discovery paper described (except that the model can vary in metallicity). With the $\sim 40$\% higher J-band flux and missing K-band, we retrieve a very low surface gravity (same as the discovery paper), but significantly higher temperature outside of our model grid. In the a patchy-cloud model of the original paper a higher J-band contribution can come from a clear model, but this is more difficult to explain in a pure cloudy model.

\end{appendix}

\end{document}